\documentclass[11 pt, a4paper]{article}
\usepackage{array,epsfig}
\usepackage{amsthm}
\usepackage{amsmath}
\usepackage{color}
\usepackage{enumitem}
\usepackage{booktabs}
\usepackage{sidecap}
\usepackage{setspace}
\usepackage{comment}
\usepackage{url}
\usepackage{float}
\usepackage{apacite} 
\usepackage{parskip}
\usepackage{standalone}
\usepackage{tabu}
\usepackage[font=footnotesize,labelfont=bf]{caption}
\usepackage{bm, bbm, mdframed, mathtools, mathabx, multirow, subfigure, tikz, multirow, tcolorbox, soul, caption}
\usepackage[left=2.5cm, right=2.5cm, top=3cm, bottom=3cm]{geometry}
\usepackage[english]{babel}
\usepackage[mathscr]{euscript}
\newcommand{\E}{\mathrm{E}}

\numberwithin{equation}{section} 
\tcbuselibrary{breakable}
\usepackage{mathdots}
\usepackage{lscape}
\usepackage{pdflscape}

\usepackage[round, longnamesfirst]{natbib}
\usepackage[T1]{fontenc} 
\usepackage[autostyle]{csquotes}  
\usepackage[citecolor=black, linkcolor = black, colorlinks=true, allcolors=black]{hyperref}
\usepackage{glossaries}
\usepackage{appendix}
\usepackage{verbatim}
\usepackage{tcolorbox}
\usepackage{xcolor}
\usepackage{listings}
\usepackage{rotating}
\usepackage{comment}
\usepackage{graphicx}
\usepackage{dsfont}
\usepackage{algorithm2e}
\usepackage{bookmark}
\usepackage{setspace}
\usepackage{xr}
\usepackage{threeparttable}

\DeclareMathOperator{\Var}{Var}

\RestyleAlgo{ruled}
\SetKwComment{Comment}{/* }{ */}
\usepackage{tikz}
\usetikzlibrary{shapes.geometric, arrows, positioning, arrows.meta}
\tikzstyle{c_solid} = [circle, minimum width=1cm, minimum height=1cm,text centered,draw=black, fill=white]
\tikzstyle{c_dashed} = [circle, minimum width=1cm, minimum height=1cm,text centered,draw=black, dashed]
\tikzstyle{arrow} = [thick,->,>={Stealth[scale=1.3]}]
\newtheorem{assumption}{Assumption}
\newtheorem{theorem}{Theorem}
\newtheorem{lemma}{Lemma}
\newtheorem{definition}{Definition}



\title{\textbf{Causal Machine Learning for Moderation Effects}}
\author{Nora Bearth\thanks{Swiss Institute for Empirical Economic Research of the University of St. Gallen (SEW-HSG), Varnb\"uelstrasse 14, 9000 St. Gallen, CH, E-mail: \texttt{nora.bearth@unisg.ch}}\hspace{0.2cm} \& Michael Lechner\thanks{Swiss Institute for Empirical Economic Research of the University of St. Gallen (SEW-HSG), Varnb\"uelstrasse 14, 9000 St. Gallen, CH, E-mail: \texttt{michael.lechner@unisg.ch}, Michael Lechner is also affiliated with CEPR, London, CESIfo, Munich, IAB, Nuremberg and IZA, Bonn.\vspace{0.1cm} \\  Financial support from the Swiss National Science Foundation (SNSF) is gratefully acknowledged. The study is part of the project "Chances and risks of data-driven decision making for labour market policy" (grant number SNSF 407740\_187301) of the Swiss National Research Program "Digital Transformation" (NRP 77). We thank Daniele Ballinari, Hannah Busshoff, Jonathan Chassot, Riccardo Di Francesco, and Jana Mareckova and three anonymous referees for comments and suggestions on a previous version of this paper. The paper was presented at the annual meeting of the Verein für Socialpolitik (2023), the COMPIE conference (2024) and at the University of St. Gallen. We thank participants for helpful comments and suggestions. Last, we thank GPT-4 and Grammarly for editorial support.}}
\date{} 
\begin{document}
\begingroup
\let\newpage\relax
\maketitle
\endgroup

\begin{center}
 \vspace{0.5cm}
  \textbf{Abstract}\vspace{0.3cm}   \\
\end{center}

\begin{minipage}{\textwidth}
    \small
It is valuable for any decision maker to know the impact of decisions (treatments) on average and for subgroups. The causal machine learning literature has recently provided tools for estimating group average treatment effects (GATE) to better describe treatment heterogeneity. This paper addresses the challenge of interpreting such differences in treatment effects between groups while accounting for variations in other covariates. We propose a new parameter, the \textit{balanced group average treatment effect} (BGATE), which measures a GATE with a specific distribution of a priori-determined covariates. By taking the difference between two BGATEs, we can analyze heterogeneity more meaningfully than by comparing two GATEs, as we can separate the difference due to the different distributions of other variables and the difference due to the variable of interest. The main estimation strategy for this parameter is based on double/debiased machine learning for discrete treatments in an unconfoundedness setting, and the estimator is shown to be $\sqrt{N}$-consistent and asymptotically normal under standard conditions. We propose two additional estimation strategies: automatic debiased machine learning and a specific reweighting procedure. Last, we demonstrate the usefulness of these parameters in a small-scale simulation study and in an empirical example.
\end{minipage}

\thispagestyle{empty}
\newpage
\newpage
\setcounter{page}{1}

\section{Introduction} \label{2_introduction}

Detecting and interpreting heterogeneity in treatment effects is crucial for understanding the impact of interventions and (policy) decisions. Researchers have recently developed many methods to estimate heterogeneous treatment effects. However, there are still limitations in interpreting these effects. For example, suppose that the effect of a particular training program for the unemployed is larger for women than for men. By comparing the average treatment effects for these two groups, without taking into account the different distribution of other covariates of men and women, such as education or labor market experience, we might implicitly compare a group with more extended labor market experience (men) with a group with shorter labor market experience (women). Therefore, obtaining a balanced distribution of relevant characteristics across different groups may be crucial to ensure proper comparisons and draw meaningful conclusions. In the specific example, we might want to ensure that both groups have the same average years of labor market experience. This approach allows us to isolate the differences in causal effects between groups in a way that is not confounded by certain other covariates. However, the difference may still be due to differences in unobservable characteristics that vary between the two groups of interest. Determining whether a particular variable causes differences in treatment effects is often of interest. In the literature, these variables are called causal moderator variables. In addition to balancing distributions of all covariates that confound the moderation effect across groups, additional assumptions must hold to interpret the differences in treatment effects causally, i.e., so that the group variable can be considered an unconfounded moderator.

\defcitealias{Chernozhukov:2018}{Chernozhukov, Chetverikov, Demirer, Duflo, Hansen, Newey, \& Robins, 2018}  

This paper discusses how to estimate differences in treatment effects between groups in an unconfoundedness setting. First, a new parameter called \textit{balanced group average treatment effect (BGATE)} is introduced. This parameter is a group average treatment effect (GATE) with a specific distribution of a-priori-determined covariates. It is beneficial for comparing the treatment effects of two groups with each other. This results in the difference of two BGATEs called $\Delta$BGATE. We demonstrate how it relates to a difference of two group average treatment effects ($\Delta$GATE), discuss its identification and propose different estimators for discrete moderators (for simplicity, we refer to the variable for which we find heterogeneous treatment effects as moderator) and discrete treatments. For the estimator based on double/debiased machine learning (subsequently abbreviated as DML) \citepalias{Chernozhukov:2018}, we show that it is asymptotically normal centred at the true value, which allows valid inference. Additionally, we show how the $\Delta$BGATEs can be estimated using automatic debiased machine learning (Auto-DML) and a reweighting method. DML relies on using a double-robust score function that depends on the propensity score, which can be problematic if the propensity score is extreme \citep[][]{Lechner:2024,Busso:2014,Frolich:2004}. Auto-DML uses Riesz representers instead of the propensity score; hence, it does not suffer from the problem of extreme propensity scores \citep{Chernozhukov:2022a,Chernozhukov:2018b}. The reweighting method allows to estimate the $\Delta$BGATE with whatever method the user prefers for the GATE estimation, as the $\Delta$BGATE equals the $\Delta$GATE on the reweighted data. A small-scale simulation study and an empirical example using administrative labor market data from Switzerland demonstrate the practicability of the different estimators. 

The paper proceeds as follows: In Section \ref{2_effects_of_interest}, we define the parameter of interest for the case of a binary treatment and a binary moderator and discuss its interpretation and identification under unconfoundedness. Section \ref{2_estimation} proposes different estimation strategies and shows the asymptotic properties of the DML estimator. Section \ref{Simulation_study} depicts the design and results of a small-scale Monte Carlo simulation. In Section \ref{Empirical_application}, we demonstrate how the new parameter can be used in practice and Section \ref{Conclusion} concludes. Appendix \ref{Extension_Causal_Moderation} shows how the $\Delta$BGATE can be interpreted in a causal way. The identification and asymptotic proofs are depicted in Appendix \ref{Appendix_A}. The different algorithms for the estimation are explained in Appendix \ref{Estimation_procedures}. Appendix \ref{Appendix_C} shows more details about the simulation study, and Appendix \ref{appendix Empiric example} provides more details about the empirical example.

\subsection{Related Literature}

Several authors have recently contributed to the topic of effect heterogeneity \citep*[e.g.,][]{Tian:2014, Wager:2018, Athey:2019, Kuenzel:2019, Nie:2021, Semenova:2021, Knaus:2022b, DiFrancesco:2022, Foster:2023, Kennedy:2023}. The proposed methods make it possible to estimate heterogeneities between fine-grained subgroups more accurately. For a recent review of the various methods and their performance, see \cite*{Knaus:2021}. As a result, many applied papers now use such methods to detect heterogeneities \citep*[e.g.,][]{Davis:2017, Knaus:2022, Cockx:2023}. However, decision-makers are often more interested in heterogeneities for a small subset of covariates than at the finest possible granularity. Therefore, several papers show how to detect (low-dimensional) heterogeneities at the group level, called ``Group Average Treatment Effects'' (GATEs) (a GATE is a conditional average treatment effect (CATE) with a small number of conditional variables, often only a single one). Several approaches have been developed to estimate GATEs. \cite*{Abrevaya:2015} show how to identify GATEs nonparametrically under unconfoundedness and estimate them using inverse probability weighting estimators. 
\cite*{Athey:2019} and \cite*{Lechner:2018} modify a random forest algorithm to adjust for confounding to estimate heterogeneous treatment effects. \cite*{Semenova:2021} use the DML framework to find heterogeneity based on linear models. \cite*{Zimmert:2019} and \cite*{Fan:2022} develop a two-step estimator that allows estimating GATEs nonparametrically. The first step is estimated using machine learning methods, and in the second step, they apply a nonparametric local constant regression.

We add to this literature by proposing a new parameter of interest: the difference between two GATEs with (partially) balanced characteristics. By taking the difference, we can disentangle the difference in treatment effects from the difference in the distribution of covariates between the two groups. Hence, this paper is related to the Kitagawa-Oaxaca-Blinder decomposition (KOB), often used to decompose differences in outcomes between two groups into one part that is due to the difference in the distribution of other covariates and another part that is due to the variable of interest \citep{Kitagawa:1955, Oaxaca:1973, Blinder:1973}. \cite*{Vafa:2024} show that such decompositions suffer from omitted variable bias if the model for estimating them is not complex enough. They suggest a methodology for wage gap decompositions using foundation models, such as large language models. \cite{Yu:2023} introduce a new decomposition method that allows to identify how a treatment variable contributes to a group disparity in an outcome variable. They decompose it into different prevalence of treatment, different treatment effects across groups, and different patterns of selection into treatment, and they show how the decomposition can be estimated non-parametrically. Similarly, \cite*{Chernozhukov:2013} show how to model counterfactual distributions based on regression methods. They consider two scenarios: making changes to either the distribution of covariates associated with the outcome or the conditional distribution of the outcome given the covariates. The $\Delta$BGATE fits into the second scenario. Additionally, \cite{Chernozhukov:2023} introduce automatic debiased machine learning for covariate shifts. This approach is related as the $\Delta$BGATE can be interpreted as a $\Delta$GATE on a population with a shifted covariate distribution. While \cite{Chernozhukov:2023} consider the case of two distinct datasets that do not overlap and are statistically independent, we have one dataset and shift the distribution of covariates within this dataset.

\cite*{Chernozhukov:2018b} consider the problem of interpreting heterogeneous treatment effects from another angle. They suggest sorting the estimated partial effects by percentiles and comparing the covariate means of observations falling in the different percentiles to see which individuals with which characteristics are most positively and negatively affected. In contrast, we are in this paper not interested in finding the characteristics of the individuals with the most positive and negative effects but in comparing the treatment effects of the two groups while balancing the distribution of some covariates. If the goal is to to learn the characteristics of (relative) winners and losers from a treatment, their approach suits well. However, if the goal is to understand how much a variable drives the differences in treatment effects between two groups, our approach is more appropriate.

\defcitealias{Chernozhukov:2018}{Chernozhukov et al. (2018)}  

As has already become apparent from the discussion of the first part of the relevant literature, the DML literature is closely related to our work. \citetalias{Chernozhukov:2018} developed this framework, which allows using machine learning methods for causal analysis. Machine learning algorithms may introduce two biases: a regularization and an overfitting bias. The main idea of DML is that by using Neyman-orthogonal score functions, we can overcome the regularization bias, and by using cross-fitting, an efficient form of sample splitting, we can overcome the overfitting bias. Many papers have adapted the general DML framework for different settings, for example, for continuous treatments \citep*[e.g.,][]{Kennedy:2017, Semenova:2021}, for mediation analysis \citep*{Farbmacher:2022}, for panel data \citep*[e.g.][]{Clarke:2023} or for difference-in-difference estimation \citep*[e.g.,][]{Zimmert:2018, SantAnna:2020}. We contribute to this literature by using this highly flexible framework to estimate the new parameter of interest.

Last, we add to the literature on moderation effects. Discussions of moderation effects are primarily found in the psychology and political science literature \citep*[e.g.,][]{Gogineni:1995, Frazier:2004, Fairchild:2010, Marsh:2013, Bansak:2021a, Blackwell:2021}. Common approaches to analyzing moderation effects specify interactions in a regression or subgroup analysis. Without additional assumptions, these parameters cannot be interpreted causally. \cite{Bansak:2021} studies causal moderation effects in an experimental setting by showing what identifying assumptions are needed. Because of the randomization of treatment, it is possible to estimate the causal effect of the moderator on the outcome separately for the treated and control units and then subtract both estimates to obtain the moderator effect. Such differences cannot be interpreted causally if the moderator variable influences some covariates, which is often the case. In addition, \cite*{Bansak:2022} have show how to identify and estimate subgroup differences in a regression discontinuity design.

 
\section{Effects of Interest} \label{2_effects_of_interest}

\subsection{Definition}\label{2_definition}

The causal moderation framework used in this paper is based on the potential outcome framework of \cite{Rubin:1974}. A causal effect is defined as the difference between two potential outcomes, whereas for a unit, we only observe one of these potential outcomes. Therefore, finding a credible counterfactual is problematic. We observe $N$ i.i.d. observations of the independent random variables $H_i = (D_i, Y_i, Z_i, X_i)$ according to an unknown probability distribution $\mathds{P}$. Here, the focus is on a treatment $D_i$ and a moderator $Z_i$ which, for simplicity, are assumed to be binary (realizations of the treatment variable are $d \in \{0,1\}$, and of the moderator variable $z \in \{0,1\}$). The formal theory presented in Appendix \ref{Appendix_A} is based on fixed numbers of discrete values for $Z_i$ and $D_i$. As usual, the potential outcomes are indexed by the treatment variable: $(Y_i^{0}, Y_i^{1})$. Finally, a set of $k \in \{1, \dots, p\}$ covariates $X_{i,k}$ might simultaneously affect treatment allocation and potential outcomes, where $X_i = (X_{i,1}, \dots, X_{i,p})$. Additionally, potential covariates and potential moderators are defined as: $(X_i^{0}, X_i^{1}, Z_i^{0}, Z_i^{1})$.

Since only realizations of one of the potential outcomes are observed, we can never consistently estimate realizations of the individual treatment effect (ITE), $\xi_i = Y_i^1- Y_i^0$. However, under suitable assumptions, the identification of, for example, the average treatment effect (ATE) $\theta = \E[Y_i^1- Y_i^0]$ is possible \citep{Imbens:2009}.
It is often interesting to additionally investigate different aspects of the heterogeneity of the $\xi_i$ which can be captured by so-called conditional average treatment effects (CATE). A CATE measures the average treatment effect conditional on a (sub-) set of covariates $X_i$. The individualized average treatment effect (IATE) and the group average treatment effect (GATE) are specific CATEs. The IATE measures the treatment effect at the most granular aggregation level. Namely, it compares the average effect of the treatment for all individuals with a specific value of all relevant covariates used. Formally, the IATE is defined as follows:
\begin{equation*}
    \tau(x,z) = \E[Y_i^1- Y_i^0 | Z_i=z, X_i=x]
\end{equation*}
The GATE measures the treatment effect at the group level, i.e. at a more aggregated level than the IATE, but still at a finer level than the ATE. Formally, the GATE is defined as follows:
\begin{equation*}
    \theta^G(z) = \E[Y_i^1- Y_i^0| Z_i=z]  = \E[\tau(X_i, z) | Z_i = z]
\end{equation*}
As long as the interest lies only in describing effect heterogeneity, IATEs and GATEs are sufficient. However, if the interest lies in the difference in treatment effects between the two groups, the difference between the two GATEs ($\Delta$GATE), i.e.
\begin{align*}
    \theta^{\Delta G} &= \E[Y_i^1- Y_i^0 | Z_i = 1] - \E[Y_i^1- Y_i^0| Z_i = 0] \\ &= \E[\tau(X_i, 1) | Z_i = 1] - \E[\tau(X_i, 0) | Z_i = 0],
\end{align*}
may be difficult to interpret because the two groups may differ in the distribution of other covariates $X_i$.

Thus, a new parameter is introduced, the \textit{balanced group average treatment effect} (BGATE). This parameter is called BGATE, as it is designed to balance the distribution of other variables within the groups (defined by the different values of $Z_i$) we want to compare to each other. The variables used to balance the GATEs are denoted as $W_i$. $W_i$ is part of $X_i$. If $W_i$ is empty, or $W_i$ is independent of $Z_i$, the BGATE reduces to the GATE. 
Thus, the new parameter of interest, denoted by $\theta^B(z)$, is defined as
\begin{align*}
    \theta^B(z)&= \E[\E[Y_i^1- Y_i^0 | Z_i = z, W_i]] = \E[\E[\tau(X_i, z) |  Z_i = z, W_i]], 
\end{align*}
and its difference, $\theta^{\Delta B}$, as
\begin{align*}
    \theta^{\Delta B} &= \E[\E[Y_i^1- Y_i^0 | Z_i = 1, W_i] - \E[Y_i^1- Y_i^0 | Z_i = 0, W_i]] \\ &= \E[\E[\tau(X_i, 1) | Z_i = 1, W_i] - \E[\tau(X_i, 0) | Z_i = 0, W_i]] 
\end{align*}
A $\Delta$BGATE ($\theta^{\Delta B}$) represents the difference between two groups, adjusting the distribution of some other covariates ($W_i$) in both groups to the overall population distribution. The $\Delta$BGATE usually shows associational moderation effects. Under certain assumptions, we define a causal balanced group average treatment effect ($\Delta$CBGATE) that can be interpreted causally. The discussion of this parameter is referred to Appendix \ref{Extension_Causal_Moderation}.

\subsection{Decomposition of Effects} \label{2_Oaxaca_Blinder}

To clarify the distinction between a $\Delta$GATE and a $\Delta$BGATE, the $\Delta$GATE is decomposed into two components: the $\Delta$BGATE, representing the direct effect of the moderator variable, and the compositional effect, arising from differences in the distributions of $W_i$ across the groups. The compositional components capture the differences in the distribution of $W_i$ in both $Z_i$ subsamples weighted by their relative importance:
\begin{align*}
    &\underbrace{\E[Y_i^1- Y_i^0 | Z_i = 1] - \E[Y_i^1- Y_i^0| Z_i = 0]}_{\Delta\text{GATE}} &\\ &= \underbrace{\E[\E[Y_i^1- Y_i^0 | W_i, Z_i = 1] - \E[Y_i^1- Y_i^0 | W_i, Z_i = 0]]}_{\text{direct effect: $\Delta$BGATE}} \\&+ \underbrace{\frac{P(Z_i = 0)}{P(Z_i = 1)}\E[\E[Y_i^1- Y_i^0 |W_i, Z_i = 1]] - \E[\E[Y_i^1- Y_i^0 |W_i, Z_i = 1]| Z_i = 0]}_{\text{compositional effect (1)}} \\&- \underbrace{\frac{P(Z_i = 1)}{P(Z_i = 0)}\E[\E[Y_i^1- Y_i^0 |W_i, Z_i = 0]] - \E[\E[Y_i^1- Y_i^0 |W_i, Z_i = 0]| Z_i = 1]}_{\text{compositional effect (2)}}
\end{align*}
The derivation of this decomposition is shown in Appendix \ref{Decomposition_GATE}.
This decomposition is similar to the KOB-decomposition \citep{Kitagawa:1955, Oaxaca:1973, Blinder:1973} used to decompose differences in outcomes between two groups into one part that is due to the difference in the distribution of other covariates and one part that is due to the variable of interest. It is often used in the context of analyzing gender wage differences \citep[e.g.][]{Blau:2017}.

The  $\Delta$BGATE is similar to the \textit{direct effect} in the KOB-decomposition. However, in a KOB-decomposition, the distribution of the balancing variables $W_i$ is not adjusted to the overall population distribution but to the distribution of a specific group. If the distribution of $W_i$ does not differ across groups, the direct effect in the KOB-decomposition and the $\Delta$BGATE are equal. In that case, the $\Delta$BGATE also equals the $\Delta$GATE. Adjusting the distribution of the covariates to the overall population distribution is intuitive if we are interested in the population and not in a specific part of the population.

\subsection{Identification} \label{2_identifying_assumptions}

To identify the GATE, BGATE, $\Delta$GATE or $\Delta$BGATE in an unconfoundedness setting, usual identifying assumptions are needed \citep[e.g.,][]{Imbens:2004}:

\begin{assumption}\label{assumption_CIA} \textnormal{(Identifying assumptions)}
\begin{enumerate}[label=(\alph*)]
        \item Conditional Independence (CIA): $(Y_i^1, Y_i^0) \perp D_i| X_i = x, Z_i = z, \quad \forall x \in \mathcal{X}, \forall z \in \mathcal{Z}$
        \item Common support (CS): $0 < P(D_i = d | X_i = x, Z_i = z) < 1, \quad \forall d \in \{0,1\}, \forall x \in \mathcal{X}, \forall z \in \mathcal{Z}$
        \item Exogeneity of confounder: $X_i^0 = X_i^1, \quad Z_i^0 = Z_i^1$
        \item Stable Unit Treatment Value Assumption (SUTVA): $Y_i = D_iY_i^1 + (1-D_i)Y_i^0$
\end{enumerate}
\end{assumption}

\begin{lemma} \label{BGATE_lemma}
    Under Assumptions \ref{assumption_CIA}, the parameter $\theta^B(z) = \E[\E[Y_i^1- Y_i^0 | Z_i = z, W_i]]$ is identified as $\E[\E[\mu_1(Z_i, X_i) - \mu_0(Z_i, X_i)| Z_i = z, W_i]]$ with $\mu_d(z,x) = \E[Y_i | D_i = d, Z_i = z, X_i = x]$. Hence, $\theta^{\Delta B} = \E[\E[Y_i^1- Y_i^0 | Z_i = 1, W_i]  - \E[Y_i^1- Y_i^0 | Z_i = 0, W_i]]$ is identified as $\E[\E[\mu_1(Z_i, X_i) - \mu_0(Z_i, X_i)| Z_i = 1, W_i] - \E[\mu_1(Z_i, X_i) - \mu_0(Z_i, X_i)| Z_i = 0, W_i]]$.
\end{lemma}

For the proof of Lemma \ref{BGATE_lemma} see Appendix \ref{_BGATE}.

\section{Estimation and Inference} \label{2_estimation}
\defcitealias{Chernozhukov:2018}{(e.g. Chernozhukov et al., 2018, 2022)} 
Since we are interested in differences in treatment effects, the estimation strategy focuses on the $\Delta$BGATE. The estimator can easily be adapted to estimate the BGATE and can be estimated using different estimators, from causal machine learning \citetalias[e.g.][]{Chernozhukov:2018} to other semiparametric estimators \citep[e.g.][]{Ai:2007, Ai:2012}. In this paper, we explain three methods based on causal machine learning and prove analytically that the DML estimator is $\sqrt{N}$-consistent and asymptotically normal.

\defcitealias{Chernozhukov:2018}{Chernozhukov et al. (2018)}  
\subsection{Double/Debiased Machine Learning} \label{2_DML}

\subsubsection{Estimator}
The identification results suggest a three-step estimation strategy. To obtain a flexible estimator that allows for a potentially high-dimensional vector of covariates, the first suggested estimator relies on the methodology of DML as proposed by \citetalias{Chernozhukov:2018}. In the first step, the usual double robust score function is estimated. In the second step, the score function is regressed on the two indicator variables defined by the different values of the moderator variable $Z_i$ and the covariates we want to balance $W_i$. Last, we take the difference between the two groups defined by the moderator and average over the variables we balance. This approach is close to the approach of \cite{Kennedy:2023} for estimating CATEs with the DR-learner. However, instead of estimating a CATE, we average over $W_i$.

The estimated doubly robust score function is given by the following expression:
\begin{align*}
 \hat \phi^{\Delta B}(h; \theta^{\Delta B}, \hat \eta) = & \hat g_1(w) -
 \hat g_0(w) + \frac{z\left( \hat \delta(h)  -  \hat g_1(w)\right)}{ \hat \lambda_1(w)}   - \frac{(1-z)\left( \hat \delta(h) -  \hat g_0(w) \right)}{1-  \hat \lambda_1(w)} - \theta^{\Delta B} 
\end{align*}
 with
\begin{align*} \delta(h) &= \mu_1(z,x) - \mu_0(z,x) + \frac{d(y - \mu_1(z,x))}{\pi_1(z,x)} - \frac{(1-d)(y - \mu_0(z,x))}{1-\pi_1(z,x)} \\
\lambda_z(w) &= P(Z_i = z | W_i =w), \quad g_z(w) = \E[\delta(H_i)| Z_i = z, W_i = w], \\
 \mu_d(z,x) &= \E[Y_i | D_i = d, Z_i = z, X_i = x], \quad
 \pi_d(z,x) = P(D_i = d | Z_i = z ,X_i = x). 
\end{align*}
$\hat \delta(h)$, $\hat \mu_d(z, x)$, $\hat \lambda_z(w)$ and $\hat \pi_d(z,x)$ denote the estimated values of $\delta(h)$, $ \mu_d(z,x)$, $\lambda_z(w)$ and $ \pi_d(z,x)$, respectively. Furthermore, notice that $\hat g_z(w) = \hat \E[\hat \delta(H_i)| Z_i = z, W_i = w]$ is the regression of $\hat \delta(H_i)$ on $Z_i$ and $W_i$ and that $\tilde g_z(w) = \hat \E[\delta(H_i)| Z_i = z, W_i = w]$ is the corresponding oracle regression of $\delta(H_i)$ on $Z_i$ and $W_i$. Hence, $\hat \E [\dots | \dots]$ denotes a generic regression estimator, which can be linear or non-linear, depending on the presumed data-generating process. Last, the estimated nuisance parameters are $\hat\eta = (\hat\mu_d(z,x)$, $\hat \pi_d(z,x)$, $\hat \lambda_z(w),$ $\hat g_z(w))$.

As explained above, the score function $\hat \delta(h)$ has to be estimated. The product of the nuisance function errors for $\hat \delta(h)$ must converge faster than or equal to $\sqrt N$, and cross-fitting with K-folds ($K > 1$) must be used. In the second estimation step, the product of the nuisance function errors has to converge with $\sqrt N$ and cross-fitting with J-folds ($J > 1$) has to be used. If certain conditions are met, the estimator is $\sqrt N$-consistent and asymptotically normal (see Subsection \ref{asymptotic_results}).
Because $E[\phi^{\Delta B}(H_i; \theta^{\Delta B}, \eta)] = 0$, the variance of $\hat \theta^{\Delta B}$ is given by
\begin{align*} 
 \Var(\hat \theta^{\Delta B}) &= \Var(\phi^{\Delta B}(H_i; \theta^{\Delta B}, \eta)) \\&=\E[ \phi^{\Delta B}(H_i; \theta^{\Delta B}, \eta)^2] - \underbrace{\E[ \phi^{\Delta B}(H_i; \theta^{\Delta B}, \eta)]^2}_{=0} \\ &= \E[ \phi^{\Delta B}(H_i; \theta^{\Delta B}, \eta)^2]
\end{align*}
and is estimated by $\widehat{\Var}(\hat \theta^{\Delta B}) = \frac{1}{N}\sum_{j = 1}^J \sum_{i \in S_j}[ \hat \phi^{\Delta B}(H_i; \theta^{\Delta B}, \hat\eta)]^2$ with $S_j$ being a random fold in the second estimation step. Algorithm \ref{alg:BGATE-Learner_ap} in Appendix \ref{Estimation_procedures} depicts the implementation of this DML estimator.

\subsubsection{Asymptotic Properties} \label{asymptotic_results}
We investigate the asymptotic properties of the estimator and impose the following assumptions:

\begin{assumption}
    \textnormal{(Overlap)}\label{assumption_overlap}\\ 
    The propensity scores $\lambda_z(w)$ and $\pi_d(z,x)$ are bounded away from 0 and 1:
    \begin{align*}
        \kappa < \lambda_z(w), \pi_d(z,x) < 1- \kappa \quad \forall x \in \mathcal{X}, z \in \mathcal{Z}, \quad \text{for some } \kappa > 0.
    \end{align*}
\end{assumption}

\begin{assumption}
    \textnormal{(Consistency)} \label{assumption_consistency}\\ The estimators of the nuisance functions are sup-norm consistent:
    \begin{align*}
        \sup_{x \in \mathcal{X}, z \in \mathcal{Z}} &\vert \hat \mu_d(z,x) - \mu_d(z,x) \vert \xrightarrow{p} 0 \\
        \sup_{x \in \mathcal{X}, z \in \mathcal{Z}} &\vert \hat \pi_d(z,x) - \pi_d(z,x) \vert \xrightarrow{p} 0 \\
        \sup_{w \in \mathcal{W}} &\vert \hat \lambda_z(w) - \lambda_z(w) \vert \xrightarrow{p} 0 \\
        \sup_{w \in \mathcal{W}} &\vert \tilde g_z(w) - g_z(w) \vert \xrightarrow{p} 0
    \end{align*}
\end{assumption}

\begin{assumption}
    \textnormal{(Risk decay)} \label{assumption_risk_decay}\\ The products of the estimation errors for the outcome and propensity models decay as
    \begin{align*}
        \E\left[(\hat \mu_d(Z_i,X_i) - \mu_d(Z_i,X_i) )^2\right] \E\left[(\hat \pi_d(Z_i,X_i) - \pi_d(Z_i,X_i) )^2\right] = o_p\left(\frac{1}{N}\right) \\
        \E\left[(\tilde g_z(W_i) - g_z(W_i) )^2\right] \E\left[(\hat \lambda_z(W_i) - \lambda_z(W_i) )^2\right] = o_p\left(\frac{1}{N}\right)
    \end{align*}
    Note that the expectation refers to $X_i$ and $Z_i$ or $W_i$ taken $\hat \mu_d, \hat \pi_d, \tilde g_z$ and $\hat \lambda_z$ as given. If both nuisance parameters are estimated at the parametric ($\sqrt{N}$-consistent) rate, then the product of the errors would be bounded by $O_p\left(\frac{1}{N^2}\right)$. Hence, it is sufficient for the estimators of the nuisance parameters to be $N^{1/4}$-consistent.
\end{assumption}
\newpage
\begin{assumption}
    \textnormal{(Boundness of conditional variances)} \label{assumption_boundness}\\
    The conditional variances of the outcomes and score functions are bounded:
    \begin{align*}
      & \sup_{w \in \mathcal{W}} \Var(\delta(H_i) | Z_i = z, W_i = w) < \epsilon_{z1} < \infty \\
      &\sup_{w \in \mathcal{W}} \Var(\hat \delta(H_i) | Z_i = z, W_i = w) < \epsilon_{z0} < \infty \\
      & \sup_{x \in \mathcal{X}, z \in \mathcal{Z}} \Var(Y_i | D_i = d, Z_i = z, X_i = x) < \epsilon_{d} < \infty
    \end{align*}
    for some constants $\epsilon_{z1}, \epsilon_{z0}, \epsilon_{d} > 0$.
\end{assumption}
\begin{assumption}
    \textnormal{(Stability)} \label{assumption_stability}\\
    The second step regression estimator $\hat \E[\dots | \dots]$ has to be stable in the sense of Definition \ref{definition_stability} in Appendix \ref{Asymptotics_BGATE} (Definition 1 in \cite{Kennedy:2023}).
\end{assumption}
\defcitealias{Chernozhukov:2018}{Chernozhukov et al., 2018}  
Assumptions \ref{assumption_overlap} to \ref{assumption_boundness} made are standard in the DML literature \citepalias{Chernozhukov:2018}. The only difference is that these assumptions are applied for the first and the second estimation step. Assumption \ref{assumption_risk_decay} is needed to ensure that the product of the nuisance function errors converges faster than or equal to $\sqrt{N}$. $L_2$ convergence rates of various machine learning methods adhere to these properties when sparsity conditions are met. For example, \cite{Belloni:2013} shows that under approximate sparsity, meaning that the sorted absolute values of the coefficients decay quickly, the error of the Lasso estimator is of order $O\left(\sqrt{\frac{s \log(\max(p,N))}{N}}\right)$ with $p$ being the regressors and $s$ the unknown number of true coefficients in the oracle model. Similar rates also depending on the sparsity level, have been shown for shallow regression trees or honest and arbitrarily deep regression forests \citep{Wager:2016, Syrgkanis:2020}, boosting in sparse linear models \citep{Luo:2016} or a class of deep neural nets \citep{Farrell:2021}. For detailed information on how sparsity conditions depend on the parameters of the predictors, we refer to the cited references. Assumption \ref{assumption_stability} is needed because, in the second estimation step, we regress an already estimated quantity $\hat \delta$ on $Z_i$ and $W_i$. Stability can be perceived as a type of stochastic equicontinuity for a nonparametric regression. \cite{Kennedy:2023} proves that linear smoothers, such as linear regression, random forests or nearest neighbour matching, are stable.

Given these assumptions, we can derive the main theoretical result:

\begin{theorem}\label{theorem_asymptotics}
    Under Assumptions \ref{assumption_overlap} to \ref{assumption_stability}, the proposed estimation strategy for the $\Delta$BGATE obeys $\sqrt{N}(\hat \theta^{\Delta B} - \theta^{\Delta B}) \xrightarrow{d} N(0, V^\star)$
with $V^\star = \E[\phi^{\Delta B}(H_i; \theta^{\Delta B}, \eta)^2]$.
\end{theorem}

Theorem \ref{theorem_asymptotics} states that the estimator is $\sqrt N$-consistent and asymptotically normal. See Appendix \ref{Asymptotics_BGATE} for the proof.

\subsection{Automatic Debiased Machine Learning} \label{2_autoDML}

The previous section shows that DML relies on a double-robust score function that depends on the propensity score. As extreme propensity scores can be problematic and distort the results \citep[e.g.][]{Lechner:2024, Busso:2014, Frolich:2004}, we propose a second estimation strategy that does not rely on the propensity score. We use the Auto-DML framework \citep{Chernozhukov:2022a, Chernozhukov:2022b} that relies on using Riesz representers instead of propensity scores. Hence, we proceed in three steps, similarly to the DML estimation strategy outlined above. The usual double robust score function using the Riesz representer is estimated in the first step. In the second step, the score function is regressed on the two indicator variables defined by the different values of the moderator variable $Z_i$ and the covariates we want to balance $W_i$. Last, we take the difference between the two groups defined by the moderator and average over the distribution of $W_i$ in the population.

The estimated double robust score function is given by the following expression:
\begin{align*}
 \hat \phi^{\Delta B}_{Riesz}(h; \theta^{\Delta B}, \hat \eta) = & \hat g_1(w) -
 \hat g_0(w) + \hat\alpha_{z}(w)(\hat \delta(h) - \hat g_z(w)) - \theta^{\Delta B}_{Riesz} 
\end{align*}
 with
\begin{align*}
\delta(h) &= \mu_1(z,x) - \mu_0(z,x) + \alpha_{d}(z,x)(y - \mu_d(z,x)) \\
\mu_d(z,x) &= \E[Y_i | D_i = d, Z_i = z, X_i = x], \quad g_z(w) = \E[\delta(H_i)| Z_i = z, W_i = w]. 
\end{align*}

Again, the score function $\hat \delta(h)$ has to be estimated, the product of the nuisance function errors for $\hat \delta(h)$ must converge faster than or equal to $\sqrt N$, and cross-fitting with K-folds ($K > 1$) must be used. Similarly, in the second estimation step, the product of the nuisance function errors has to converge with $\sqrt N$ and cross-fitting with J-folds ($J > 1$) has to be used.

The variance of $\hat \theta^{\Delta B}_{Riesz}$ is given by
\begin{align*} 
 \Var(\hat \theta^{\Delta B}_{Riesz}) &= \Var(\phi^{\Delta B}_{Riesz}(H_i; \theta^{\Delta B}_{Riesz}, \eta)) = \E[ \phi^{\Delta B}_{Riesz}(H_i; \theta^{\Delta B}_{Riesz}, \eta)^2]
\end{align*}
and is estimated by $\widehat{\Var}(\hat \theta^{\Delta B}_{Riesz}) = \frac{1}{N}\sum_{j = 1}^J \sum_{i \in S_j}[ \hat \phi^{\Delta B}_{Riesz}(H_i; \theta^{\Delta B}_{Riesz}, \hat\eta)]^2$. Algorithm \ref{alg:BGATE-Learner_auto} in Appendix \ref{Estimation_procedures} depicts the implementation of the Auto-DML estimator. The proof of the Auto-DML estimator for the BGATE is beyond the scope of this paper.

\subsection{Reweighting Approach} \label{2_resampling}

Last, we suggest an estimation strategy independent of the specific method a researcher wants to use to estimate GATEs. The idea is to change the data so that the distribution of the balancing variables is the same across the groups defined by the moderator variable. After having balanced the data, the BGATE can be estimated using any method for estimating GATEs as the GATE on the balanced data equals the BGATE. 

An example of a reweighting strategy is as follows: In the first step, each observation is duplicated such that we have an identical observation for each group defined by the moderator variable. In the second step, a nearest-neighbour matching procedure is performed. For each observation in the expanded dataset, the covariate values for the balancing variables $W_i$ are compared with those of all other observations that share the same treatment assignment. Using the Mahalanobis distance metric, each observation is matched to the "nearest" observation from the original dataset with similar covariate values. This ensures that the matched pairs are comparable in terms of observed covariates. When $W_i$ is a vector of several variables, we adjust for their covariance structure using the inverse covariance matrix (i.e., using the so-called Mahalanobis distance for matching). The covariate values of the matched observations are then assigned to the duplicated observation in the reference dataset. This substitution allows each observation to represent a counterfactual scenario where it maintains similar covariate characteristics but with the opposite treatment assignment. The exact procedure for reweighting the data is depicted in Algorithm \ref{alg:BGATE-Learner_resampling} in Appendix \ref{Estimation_procedures}.

After reweighting the data, any estimator for $\Delta$GATE can be employed to estimate the $\Delta$BGATE (e.g., the DML-based version outlined as Algorithm \ref{alg:DR-Learner_GATE} in Online Appendix \ref{Estimation_procedures}). 
However, the variance estimator must be adjusted to account for the different weights each observation receives. The derivation of the variance is shown in Online Appendix \ref{Variance_estimation_resampling}. While the formal proof of the asymptotic properties of the reweighting strategy is beyond the scope of this paper, it performs well in the simulations.



\section{Simulation Study}\label{Simulation_study}

\subsection{Data Generating Process (DGP)} \label{DGP_BGATE}

We start with simulating a p-dimensional covariate matrix $X_{i,p}$ with p=10. The first two covariates are drawn from a uniform distribution $X_{i,0}, X_{i,1} \sim \mathcal{U}[0,1]$ and the remaining covariates from a normal distribution $X_{i,2} \dots, X_{i,p-1} \sim \mathcal{N}\left(0.5, \sqrt{1/12}\right)$. All covariates have a mean of 0.5 and a standard deviation of $\sqrt{1/12}$. The moderator variable $Z_i$ is drawn from a Bernoulli distribution with probability $P(Z_i = 1|X_{i,0}, X_{i,1}) = (0.1 + 0.8  \beta(X_{i,0} \times X_{i,1};2,4))$. $\beta(X_{i,0} \times X_{i,1};2,4)$ denotes the cdf of a beta distribution with the shape parameters $a = 2$ and $b = 4$. The propensity score is created similarly as in \cite*{Kuenzel:2019} and \cite*{Wager:2018}. The treatment variable $D_i$ is drawn from a Bernoulli distribution with probability $P(D_i = 1|X_{i,0}, X_{i,1}, X_{i,2}, X_{i,5}, Z_i) = \left(0.2 + 0.6 \beta( \frac{X_{i,0} + X_{i,1} + X_{i,2} + X_{i,5} + Z_i}{5}; 2, 4)\right)$.

The response functions under treatment and non-treatment, and the two states of the moderator variable are specified. The highly non-linear non-treatment response function is specified similarly as in \cite{Nie:2021} and is given by
\begin{align*}
\mu_{0}(X_i) = \sin(\pi \times X_{i,0} \times X_{i,1}) + (X_{i,2} - 0.5)^2 + 0.1X_{i,3} + 0.3X_{i,5}
\end{align*}
The response functions under treatment depend on $Z_i$ and are defined as:
\begin{align*}
 \mu_{1}(1,X_i) &= \mu_{0}(X_i) + \sin(4.9X_{i,0}) + \sin(2X_{i,1}) + 0.7X_{i,2}^4 + 0.4X_{i,5} + 0.2 \\
\mu_{1}(0,X_i) &=\mu_{0}(X_i) + \sin(1.4X_{i,0}) + \sin(6X_{i,1}) + 0.6X_{i,2}^2 + 0.3X_{i,5}.
\end{align*}

They are chosen such that the $\Delta$BGATE is different from the $\Delta$GATE. Last, we simulate the potential outcomes as $Y_i^{d}(z) = \mu_{d}(z,X_i)  + e_{i,d,z}$ $\forall z \in \{0,1\}$ with noise $e_{i,d,z}  \sim \mathcal{N}(0,1)$. Summing up, the data consists of an observable quadruple $(y_{i,r}, d_{i,r}, z_{i,r}, x_{i,r} )$. 

\subsection{Effects of Interest, Simulation Design and Estimators}

The parameters of interest are two $\Delta$BGATEs and a $\Delta$GATE, namely:
\begin{align*}
 \theta^{\Delta GATE} &= \E[ \E[Y_i^1 - Y_i^0 | Z_i = 1] - \E[Y_i^1 - Y_i^0 | Z_i = 0]] \\
 \theta^{\Delta BGATE}_{X_0} &= \E[\E[Y_i^1 - Y_i^0 | Z_i = 1, X_{i,0}] - \E[Y_i^1 - Y_i^0 | Z_i = 0, X_{i,0}]] \\
\theta^{\Delta BGATE}_{X_2} &= \E[\E[Y_i^1 - Y_i^0 | Z_i = 1, X_{i,2}] - \E[Y_i^1 - Y_i^0 | Z_i = 0, X_{i,2}]]
\end{align*}
$X_{i,0}$ is unbalanced across the two moderator groups, whereas $X_{i,2}$ is balanced. Hence, $\theta^{\Delta B}_{X_0}$ is different from $\theta^{\Delta G}$, whereas $\theta^{\Delta B}_{X_2} $ is equal to $\theta^{\Delta G}$. We generate $R = 2,000$ samples of size $N = 1,000$, $R = 1,000$ samples of size $N = 2,500$, $R = 500$ samples of size $N = 5,000$, and $R = 250$ samples of size $N = 10,000$. Since the variance is doubled when the sample size is halved, we make the number of replications proportional to the sample size (to reduce the computational costs of the simulations). The true values are estimated on a sample with $N = 1,000,000$. The parameters of interest are estimated by the three different estimation strategies outlined above, i.e. DML, Auto-DML, and the reweighting strategy. 

The algorithm used for the DML estimation strategy is depicted in Algorithm \ref{alg:BGATE-Learner_ap} in Appendix \ref{Estimation_procedures}. We use two folds in both estimation steps ($K = 2$, $J = 2$). All nuisance functions in the DML estimation strategy are estimated using random forests (number of trees: 1000). As shown by \cite*{Bach:2024}, tuning the learners used to estimate the nuisance parameters is crucial. Table \ref{Optimal tuning parameters} in Appendix \ref{RF tuning} shows the hyperparameters used. 

The algorithm used for the Auto-DML estimation strategy is depicted in Algorithm \ref{alg:BGATE-Learner_auto} in Appendix \ref{Estimation_procedures}, and we use again two folds in the first and two folds in the second estimation step ($K = 2$, $J = 2$). The algorithm is implemented by using a neural net as proposed in \cite{Chernozhukov:2022c}. More details about implementing the specific neural net can be found in Appendix \ref{autoDML RieszNet}.

Last, the effects are estimated by first reweighting the data so that the distribution of the balancing variables is the same across the groups defined by the moderator variable. The algorithm used for reweighting is depicted in Algorithm \ref{alg:BGATE-Learner_resampling} in Appendix \ref{Estimation_procedures}. After having balanced the data, we use the DML estimation strategy to estimate the $\Delta$GATE with the adjusted variance estimator as explained in Subsection \ref{2_resampling} (Algorithm \ref{alg:DR-Learner_GATE} in Appendix \ref{Estimation_procedures}).

\subsection{Results}

Table \ref{tab:results_simulation} shows the simulation results for the different sample sizes, parameters of interest, and estimators. Results with additional performance measures can be found in Appendix \ref{Results BGATE non-linear}.

Comparing the different estimators shows that DML has the smallest root mean squared error (rmse) in almost all cases, followed by Auto-DML. As expected, the reweighting strategy often leads to a slightly higher standard deviation (std), which results in a higher rmse. The coverage for Auto-DML is sometimes worse, as the bias of the effect or the standard error is too large. It is possible that the performance of the Auto-DML can be improved by tuning the neural networks for this particular estimation task. Such an additional investigation is, however, left to future research. Additionally, the sample size for using a neural net should be large enough, which might not be the case for a sample size of 1,250. This might be a reason why the coverage for Auto-DML is poor for the estimation of the $\Delta$GATE (78.3\%). Concluding, the DML estimator performs best in the simulations. However, this finding may not generalize to cases where DML is known to have performance issues, such as when propensity scores become extreme. In such cases the suggested Auto-DML estimator or the reweighting method might be more appropriate.

\begin{table}[h!]
    \centering
    \caption{Simulation results}
\label{tab:results_simulation}
    \begin{adjustbox}{max width=0.72\textwidth}
        \begin{threeparttable}
        \begin{tabular}{llrrrr|rrrr} \toprule
        &&\multicolumn{4}{c}{$N = 1,250$} &\multicolumn{4}{c}{$N = 2,500$} \\ \midrule
        effect&estimator  & bias & std & rmse &  cov.  & bias & std & rmse &  cov. \\  \midrule
        $\theta^{\Delta B}_{X_0}$ & DML & 0.007& 0.174 & 0.174& 0.937&  -0.023 & 0.119& 0.121&0.943\\
        $\theta^{\Delta B}_{X_0}$ & Auto-DML & 0.017 & 0.174& 0.175 & 0.905 & -0.001 & 0.123 & 0.123 &  0.922 \\
        $\theta^{\Delta B}_{X_0}$ & reweighting & -0.055 & 0.177& 0.186 &  0.955 & -0.038 &  0.125 &  0.131 & 0.957 \\ \midrule
        $\theta^{\Delta B}_{X_2}$ & DML & -0.005& 0.154& 0.154&0.943 & -0.007&0.105&0.105&0.944\\
        $\theta^{\Delta B}_{X_2}$ & Auto-DML & 0.009& 0.155& 0.155& 0.924&0.011&0.105&0.106&0.941\\
        $\theta^{\Delta B}_{X_2}$ & reweighting &-0.018 & 0.166 &0.167 &0.966 &-0.013&0.116&0.117&0.965\\  \midrule
        $\theta^{\Delta G}$ & DML & -0.008&0.150&0.150&0.944&-0.007&0.105&0.106&0.946\\
        $\theta^{\Delta G}$ & Auto-DML & 0.004&0.157&0.157&0.783&0.009&0.108&0.108&0.910\\
        \midrule
        &&\multicolumn{4}{c}{$N = 5,000$} &\multicolumn{4}{c}{$N = 10,000$}\\     \midrule
        effect&estimator  & bias & std & rmse &  cov.  & bias & std & rmse &  cov. \\  \midrule
        $\theta^{\Delta B}_{X_0}$ & DML &-0.023&0.076&0.080& 0.948&-0.018&0.056& 0.059& 0.916\\
        $\theta^{\Delta B}_{X_0}$ & Auto-DML &-0.035&0.082&0.090&0.902&-0.031&0.060&0.067& 0.864\\
        $\theta^{\Delta B}_{X_0}$ & reweighting &-0.031&0.088& 0.094&0.956&-0.025&0.064& 0.069& 0.940\\ \midrule
        $\theta^{\Delta B}_{X_2}$ & DML &-0.008&0.069&0.069&0.954& -0.003&0.051&0.051& 0.940\\
        $\theta^{\Delta B}_{X_2}$ & Auto-DML &0.006&0.070&0.070&0.950&0.000& 0.053& 0.053& 0.928\\
        $\theta^{\Delta B}_{X_2}$ & reweighting &-0.008&0.077&0.077&0.982&-0.002&0.056&0.056&0.964 \\  \midrule
        $\theta^{\Delta G}$ & DML &-0.002&0.070&0.070&0.958&-0.001&0.052&0.052& 0.928\\
        $\theta^{\Delta G}$ & Auto-DML &0.005&0.072&0.072&0.944&0.004&0.053&0.053& 0.952\\
        \midrule
    \end{tabular}
    \begin{tablenotes}[flushleft]
        \small
     \item \textit{Note:} This table shows results for different sample sizes, different effects and different estimators. The results for additional performance measures can be found in Appendix \ref{Results BGATE non-linear}. The $\Delta$GATE does not require reweighting.
    \end{tablenotes}
\end{threeparttable}
\end{adjustbox}
\end{table}

Comparing the results across different sample sizes, the std and rmse approximately halve by increasing the sample size by four, which suggests a $\sqrt{N}$ convergence of the estimator already for the comparatively small sample size used in the simulations. This pattern can be observed for all estimators. 

\begin{figure}[h!]
    \centering
    \begin{minipage}{\textwidth}
        \caption{Simulation results: $\hat \theta^{\Delta B}_{X_0}$ versus $\hat \theta^{\Delta G}$}
    \label{fig:BGATE0_vs_GATE}
    \includegraphics[width = \textwidth]{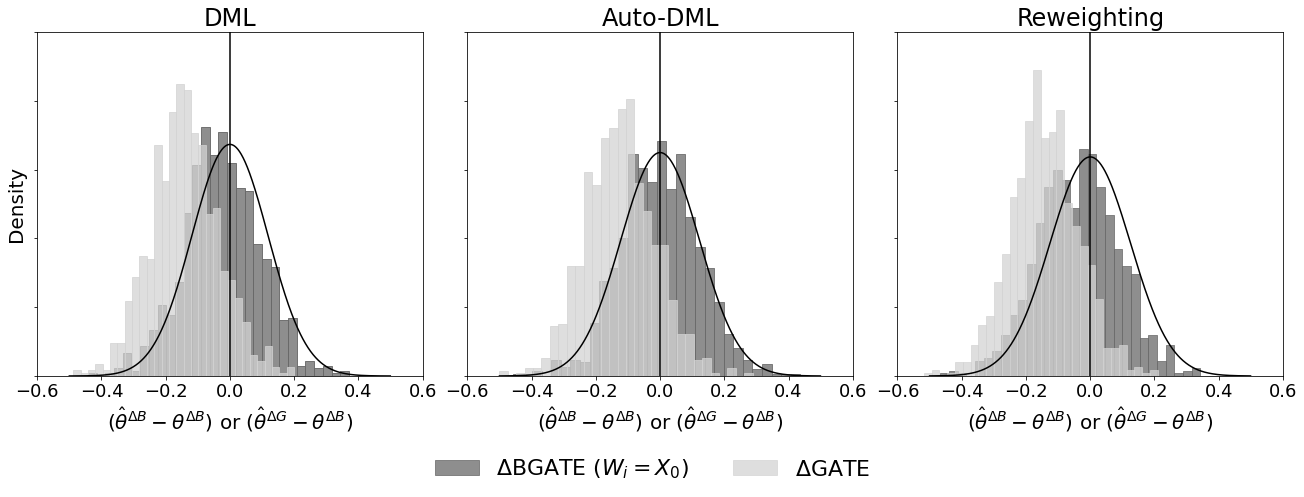}
    \scriptsize \textit{Note:} The figure depicts the distribution of the errors of $\hat{\theta}^{\Delta B}_{X_0}$ vs. $\hat{\theta}^{\Delta G}$ and the normal distribution with the standared deviation of $\hat{\theta}^{\Delta B}$. Column (1) shows results for DML, Column (2) for Auto-DML and Column (3) for the reweighting method. The results are for $N = 2,500$ and $R = 1,000$. The overlapping distribution of both effects results in a color that combines elements of both shades, creating a blended appearance. 
    \end{minipage}
\end{figure}

Next, we compare the different parameters of interest. Figure \ref{fig:BGATE0_vs_GATE} depicts the distributions of the errors for $\hat \theta^{\Delta B}_{X_0}$ and $ \hat \theta^{\Delta G}$ if the effect of interest is $ \theta^{\Delta B}_{X_0}$. Estimating $\hat \theta^{\Delta G}$ leads to a different result, as the variable $X_{i,0}$ is not balanced across the two groups of $Z_i$.

\begin{figure}[h!]
    \centering
    \begin{minipage}{\textwidth}
        \caption{Simulation results: $\hat \theta^{\Delta B}_{X_2}$ versus $\hat \theta^{\Delta G}$}
    \label{fig:BGATE2_vs_GATE}
    \includegraphics[width = \textwidth]{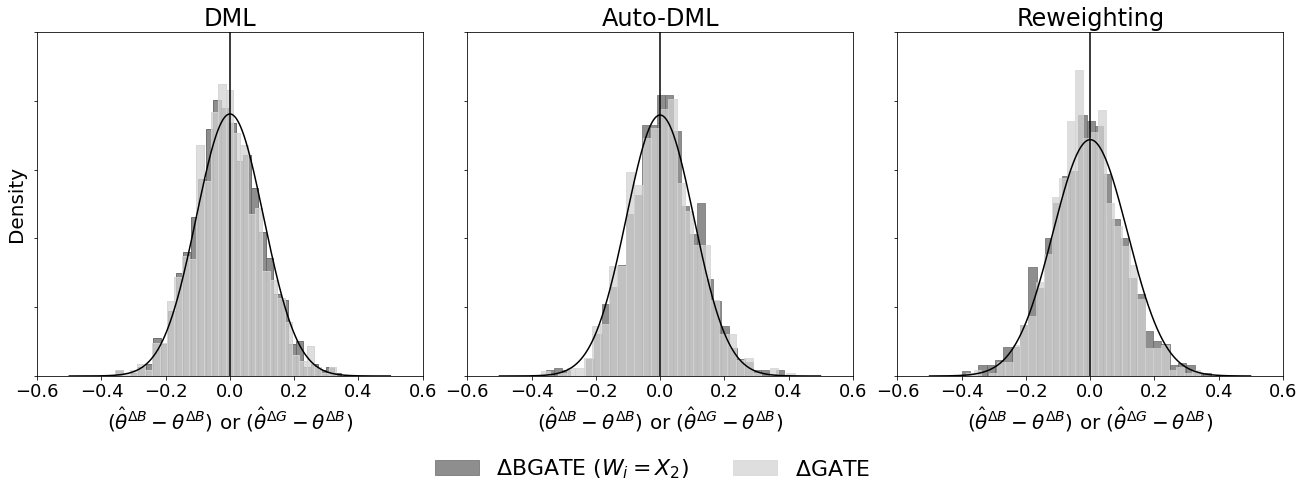}
    \scriptsize \textit{Note:} The figure depicts the distribution of the errors of $\hat{\theta}^{\Delta B}_{X_2}$ vs. $\hat{\theta}^{\Delta G}$ and the normal distribution with the standared deviation of $\hat{\theta}^{\Delta B}$. Column (1) shows results for DML, Column (2) for Auto-DML and Column (3) for the reweighting method. The results are for $N = 2,500$ and $R = 1,000$. The overlapping distribution of both effects results in a color that combines elements of both shades, creating a blended appearance.
    \end{minipage}
\end{figure}

In contrast, $X_{i,2}$ is already balanced across the two groups of $Z_i$. Figure \ref{fig:BGATE2_vs_GATE} depicts the distributions of the errors for $\hat \theta^{\Delta B}_{X_2}$ and $\hat \theta^{\Delta G}$ with the effect of interest being $\theta^{\Delta B}_{X_2}$. As expected, estimating $\hat \theta^{\Delta G}$ leads to the same result as estimating $\hat \theta^{\Delta B}_{X_2}$ because $X_{i,2}$ is already balanced. Hence, if the covariate(s) is (are) not balanced, it is important to differentiate between the two effects.

\newpage
\section{Empirical Example}\label{Empirical_application} 

An empirical example is taken from the literature that evaluates the effects of active labour market policies (ALMP) on unemployed individuals. \cite*{Card:2018} summarize the estimates of more than 200 studies on the effects of ALMPs in a meta analysis. Generally, they find higher positive effects for females, long-term unemployed, and low-income individuals. Several recent studies use causal machine learning to analyze the heterogeneous impacts of such ALMPs. For example, \cite{Cockx:2023} examine Belgian unemployed individuals, finding positive medium-term program effects, especially for recent immigrants with low local language proficiency. \cite{Burlat:2024} reports that positive impacts of technical training in Eastern France vary by education. \cite*{Knaus:2022} use Swiss administrative data from 2003 to evaluate ALMPs with various causal machine learning methods. They find effect heterogeneity in the first six months after the start of the job search programs. The heterogeneity relates to labor market characteristics and nationality. Individuals with disadvantaged labor market characteristics, like low-income or low-labor market attachment, benefit more from the programs. Similarly, foreigners benefit more. These heterogeneous effects can be explained by the indirect costs of the programs (due to not searching intensively for a job during a program and therefore needing more time to find a job), which are lower for more disadvantaged and foreign individuals.

\defcitealias{Knaus:2020}{Lechner, Knaus, Huber, Frölich, Behncke, Mellace \& Strittmatter (2020)}  
\subsection{Data}
To explore the method in an ALMP setting, we use the publicly available dataset from \citetalias{Knaus:2020}. Using this adminiatrative data, we study the effect of job search programs on the employment status of Swiss unemployed in 2003. \cite{Knaus:2022} provide an extensive description of the data. 
The dataset provides information on individuals' participation in a job search program, where $(d = 1)$ indicates participation and $(d = 0)$ indicates non-participation. Two outcome variables are considered: a short-term outcome, 
$(Y_i^{short})$, which represents the number of months an individual was employed during the first six months following the start of the program, and a medium-term outcome, $(Y_i^{long})$, which represents the number of months employed during the last six months available in the dataset (months 28 to 33 after the program's start). The effect is claimed to be identified in an unconfoundedness setting. Using the same specification as \cite{Knaus:2022}, we include several covariates on the individuals' socioeconomic background and the labor market history $(X_i)$.

A concern regarding the treatment definition arises because caseworkers can assign individuals to the program anytime during their unemployment spell. Therefore, individuals with more favourable labor market characteristics might be overrepresented in the control group, as they already found a job by the time a caseworker would have assigned them to the program. To overcome this issue, we follow \cite{Knaus:2022} and predict (pseudo) treatment start dates for each individual in the control group. To ensure consistency in treatment definitions between participants and non-participants, we restrict the analysis to individuals who remain unemployed at their (pseudo) treatment start dates. The final sample consists of 84,582 unemployed individuals. We use this dataset to illustrate the proposed method and check whether the heterogeneities found by \cite{Knaus:2022} are due to the variables identified by the authors or whether other underlying variables possibly confound them. 

\subsection{Summary Statistics}

For the sake of brevity, we only consider effect heterogeneity concerning nationality and employability. The employability variable reflects the caseworkers' assessment of whether the unemployed individual is easy or difficult to reemploy. The first part of Table \ref{emp_descriptives_moderator_short} shows selected covariate means for Swiss and non-Swiss individuals among the treated and non-treated individuals. The second part of the table shows the same for easy and hard to employ individuals among the treated and non-treated individuals. This helps to better understand which variables might account for the variation in treatment effects of the two groups. More descriptives can be found in Appendix \ref{data descriptives}.

\begin{table}[h!]
    \centering
    \caption{Empirical analysis: Descriptive statistics for the treatment and moderator variables}
\label{emp_descriptives_moderator_short}
    \begin{adjustbox}{max width=0.9\textwidth}
    \begin{threeparttable}
\begin{tabular}{lrrrr|rrrr}
\toprule
&\multicolumn{4}{c}{\textbf{Swiss or not}} &\multicolumn{4}{c}{\textbf{Employability}}\\ \midrule
&\multicolumn{2}{c}{\textbf{Treated}} &\multicolumn{2}{c}{\textbf{Non-treated}} &\multicolumn{2}{c}{\textbf{Treated}} &\multicolumn{2}{c}{\textbf{Non-treated}}\\ [1ex]
&\textbf{No} & \textbf{ Yes} &\textbf{No} & \textbf{Yes} &\textbf{Easy} & \textbf{ Hard} &\textbf{Easy} & \textbf{Hard}\\ [1ex]
Variable &Mean &Mean & Mean& Mean &Mean &Mean & Mean& Mean\\ \midrule
Age & 35.62 & 38.09 & 36.42 & 36.69 & 36.98 & 39.23 & 36.27 & 38.48\\
Female & 0.40 & 0.47 & 0.41 & 0.45 & 0.45 & 0.43 & 0.44 & 0.45\\
Married & 0.67 & 0.36 & 0.70 & 0.35 & 0.46 & 0.53 & 0.48 & 0.55\\
Mother tongue not Swiss language & 0.65 & 0.10 & 0.67 & 0.10 & 0.28 & 0.36 & 0.31 & 0.43\\
Past annual income & 41703 & 47916 & 38057 & 43941 & 46459 & 41129 & 42802 & 34523\\
Previous job: manager & 0.05 & 0.09 & 0.04 & 0.09 & 0.08 & 0.05 & 0.08 & 0.04\\
Previous job: primary sector & 0.07 & 0.05 & 0.13 & 0.08 & 0.06 & 0.07 & 0.10 & 0.09\\
Previous job: secondary sector & 0.18 & 0.15 & 0.14 & 0.13 & 0.16 & 0.15 & 0.13 & 0.13\\
Previous job: tertiary sector & 0.54 & 0.67 & 0.48 & 0.65 & 0.63 & 0.62 & 0.58 & 0.57\\
Previous job: missing sector & 0.21 & 0.13 & 0.26 & 0.15 & 0.15 & 0.16 & 0.19 & 0.21\\
Previous job: self-employed & 0.00 & 0.00 & 0.01 & 0.01 & 0.00 & 0.01 & 0.01 & 0.01\\
Previous job: skilled worker & 0.47 & 0.72 & 0.45 & 0.70 & 0.65 & 0.53 & 0.62 & 0.47\\
Previous job: unskilled worker & 0.47 & 0.16 & 0.49 & 0.17 & 0.24 & 0.39 & 0.27 & 0.46\\
Qualification: some degree & 0.35 & 0.73 & 0.31 & 0.72 & 0.62 & 0.47 & 0.59 & 0.39\\
Qualification: semiskilled & 0.18 & 0.13 & 0.21 & 0.13 & 0.14 & 0.18 & 0.15 & 0.20\\
Qualification: unskilled & 0.40 & 0.13 & 0.40 & 0.12 & 0.21 & 0.31 & 0.21 & 0.36\\
Qualification: skilled without degree & 0.07 & 0.02 & 0.08 & 0.03 & 0.03 & 0.04 & 0.05 & 0.05\\
No of unemployment spells last 2 years & 0.54 & 0.35 & 0.80 & 0.53 & 0.37 & 0.72 & 0.58 & 0.99\\
 \midrule
Number of observations & 4423 & 8575& 28169 &  43415 & 11396 & 1602& 61411 &  10173 \\ \bottomrule
\end{tabular}
\begin{tablenotes}[flushleft]
    \footnotesize
 \item \textit{Note:} This table shows the mean of some covariates included in the analysis. Column (1), (2), (5) and (6) show it for treated individuals, column (3), (4), (7) and (8) for non-treated individuals. Column (1) and (3) show it for non-Swiss individuals, column (2) and (4) for Swiss individuals and column (5) and (7) for easily employable individuals, column (6) and (8) for hard employable individuals.
\end{tablenotes}
\end{threeparttable}
\end{adjustbox}
\end{table}

There are some differences in covariates related to the previous labor market history, namely in past income, previous job and qualifications and the number of unemployment spells in the last two years. Furthermore, some sociodemographic characteristics, such as being married, also differ. Finally, as expected, there are more foreign than Swiss individuals with a mother tongue other than German, French, Italian or Raeto-Romansh. Similarly, hard to employ individuals are more likely to not have a Swiss mother tongue. Other variables, such as age or gender, are already well balanced, so balancing these covariates should not change the effect much. Nevertheless, we include age and gender to avoid that they will be no longer balanced after balancing some (previously unbalanced) other covariates (since they might correlate with the newly balanced covariates).

\subsection{Empirical Results}

For the empirical application we focus on the DML estimator, as it performs best in the simulation study and we have asymptotic properties for it. 
The first two columns of Table \ref{results_empirical} show the different effects considered in the analysis. These effects include the $\Delta$GATE, a $\Delta$BGATE with already balanced covariates, such as age and gender, a $\Delta$BGATE with additionally adding marital status, an extended $\Delta$BGATE balancing additionally unbalanced covariates like past annual income, previous job, and qualification variables. Then, we further add mother tongue to the variables to be balanced. Finally, the analysis considers a $\Delta$BGATE that balances all covariates included in the study.  

Columns three to six of Table \ref{results_empirical} depict the results for the different effects in the first six months, a period that is commonly called ``lock-in'' period. As a reference point, the average treatment effect is $\hat \theta_{lock-in} = -0.785$ ($0.021$). $\hat \theta^{\Delta G}$ shows that the difference in treatment effects between Swiss and non-Swiss individuals is statistically and economically significant. Hence, it seems that the program works better for foreigners. As pointed out above, the interpretation is not straightforward because the two groups have unbalanced covariates. After explicitely balancing already balanced sociodemographic characteristics like age, gender, and also marital status the coefficient remains relatively stable. When balancing the labor market history, including covariates like past income and previous job details, there is a notable reduction in the coefficient. This directly translates  into an increase in the compositional effect (column 6) that shows the part of $\hat \theta^{\Delta G}$ that comes from differences in the distribution of covariates. As anticipated, additionally, balancing mother tongue results in an even lower difference between Swiss and non-Swiss individuals of only $0.088$, which is not statistically significant at the 5\% significance level. Balancing all covariates included in the analysis does not reduce the difference further. Hence, it becomes evident that mother tongue disparities and differences in the labor market history significantly contribute to the variance in treatment effects between Swiss and non-Swiss individuals.

\begin{table}[h!]
    \centering
    \caption{Empirical analysis: Results}
\label{results_empirical}
    \begin{adjustbox}{max width=\textwidth}
        \begin{threeparttable}
        \begin{tabular}{clrrrr|rrrr} \\\toprule
       & & \multicolumn{4}{c}{\textbf{Lock-in effect}} & \multicolumn{4}{c}{\textbf{Mid-term effect}}\\
       && \multicolumn{3}{c}{$\Delta$BGATE} & comp. effect &\multicolumn{3}{c}{$\Delta$BGATE} & comp. effect\\  \midrule
        \multicolumn{2}{l}{Effect} & Coef & SE & P-value & Coef  & Coef & SE & P-value & Coef\\  \midrule
        \multicolumn{10}{c}{\textbf{Non-Swiss vs. Swiss individuals}}\\ \midrule
        $\theta^{\Delta G}$  & none & 0.192 & 0.042 & 0.000 & 0.000 & 0.059 & 0.064 & 0.350 & 0.000\\
$\theta_1^{\Delta B}$ & age, female & 0.211 & 0.047 & 0.000 & -0.019 & 0.059 & 0.068 & 0.381 & 0.000\\
$\theta_2^{\Delta B}$ & + married & 0.216 & 0.053 & 0.000 & -0.024 & 0.072 & 0.072 & 0.320 & -0.013\\
$\theta_3^{\Delta B}$ & + past income and unemployment, & 0.103 & 0.051 & 0.044 & 0.089 & 0.098 & 0.072 & 0.173 & -0.039\\
& previous job, qualification&&&&&&&&\\
$\theta_4^{\Delta B}$ & + mother tongue not Swiss language & 0.088 & 0.050 & 0.078 & 0.104 & 0.103 & 0.081 & 0.205 & -0.044\\
$\theta_5^{\Delta B}$ & all covariates included in the analysis& 0.075 & 0.048 & 0.118 & 0.117 & 0.068 & 0.076 & 0.369 & -0.009\\
 \midrule
        \multicolumn{10}{c}{\textbf{Easy to employ vs. hard to employ individuals}}\\ \midrule
        $\theta^{\Delta G}$ & none & -0.371 & 0.055 & 0.000&  0.000 & -0.257 & 0.088 & 0.004 &  0.000 \\
$\theta_1^{\Delta B}$ & age, female & -0.323 & 0.075 & 0.000 & -0.048 & -0.274 & 0.128 & 0.033 & 0.017\\
$\theta_2^{\Delta B}$ & + married & -0.320 & 0.066 & 0.000 & -0.051 & -0.648 & 0.423 & 0.126 & 0.391\\
$\theta_3^{\Delta B}$ & + past income and unemployment, & -0.184 & 0.067 & 0.006 & -0.187 & -0.214 & 0.132 & 0.104 & -0.043\\
& previous job, qualification &&&&&&&&\\
$\theta_4^{\Delta B}$ & + mother tongue not Swiss language & -0.189 & 0.064 & 0.003 & -0.182 & -0.304 & 0.175 & 0.082 & 0.047\\
$\theta_5^{\Delta B}$ & all covariates included in the analysis & -0.034 & 0.154 & 0.827 & -0.337 & -0.337 & 0.159 & 0.034 & 0.080\\
 \midrule
    \end{tabular}
    \begin{tablenotes}[flushleft]
        \small
     \item \textit{Note:} This table shows the effects of interest and results of the empirical example ($N=84,582$). Column (3) to (6) show the effects for the short term, while column (7) to (10) show the effects for the medium term. The compositional effect (comp. effect) is the part of the $\Delta$GATE that comes from differences in the distribution of covariates. The first part of the table shows the results for Swiss and non-Swiss individuals, the second part for easy and hard to employ individuals.
    \end{tablenotes}
\end{threeparttable}
\end{adjustbox}
\end{table}

Similarly, we find that the program has a significantly different effect on easily employable individuals compared to hard to employ individuals (columns 7 to 10). It works better for hard to employ individuals and the $\hat \theta^{\Delta G}$ is $-0.371$ ($0.055$). Again, balancing sociodemographic characteristics does only slightly change the result. Balancing the labor market history reduces the difference in treatment effects to $-0.184$ ($0.067$), which is halving the difference. Interestingly, the difference does not reduce to zero after balancing the labor market history. Hence, this employability variable created by the caseworkers does not seem to be based only on the labor market history. Balancing mother tongue does not change the result further, however, balancing for all remaining covariates included in the analysis does reduce the difference to basically zero.

The average medium-term effect is $\hat \theta_{medium} = 0.012$ ($0.033$), which is very small and not significantly different from zero. We also do not find any relevant differences between Swiss and non-Swiss individuals in the medium-term. However, there is a difference between easily and hard to employ individuals. It seems that even in the medium-term, the program works better for hard to employ individuals. This difference becomes even more pronounced when balancing other covariates. This directly relates to the findings of the meta-analysis of \cite{Card:2018}, where the job search program appears more effective for such individuals.

Concluding, these results highlight the importance of carefully interpreting group average treatment effects, as the difference in treatment effects between some groups diminish or vanish after balancing the distribution of other covariates.

\section{Conclusion}\label{Conclusion}

This paper presents a novel approach for analyzing and interpreting treatment heterogeneity in an unconfoundedness setting. We introduce a parameter called $\Delta$BGATE for measuring the difference in treatment effects between different groups while accounting for variations in covariates. This paper proposes an estimator based on DML for discrete treatments and moderators, demonstrating its consistency and asymptotic normality under standard conditions. Additionally, we outline two alternative estimation strategies. The first one is the so-called Auto-DML, which is a DML-type estimator with the additional advantage that the already well-documented non-robustness of DML estimators to extreme estimated propensity scores can be avoided. The second alternative estimator is a reweighting method that allows the use of any estimator that is consistent for the GATE applied to the reweighted data. A simulation study shows the practicability of these estimation strategies. An empirical example illustrates the proposed estimand and underlines that seeming causal heterogeneity may be caused by an underlying different distribution of other covariates. The proposed new parameter allows a more informative interpretation of heterogeneity and, thus, a better understanding of the differential impact of decisions. Future research could extend the estimation approach to continuous treatments and moderators. This paper shows identification in an unconfoundedness setting. It would be interesting to extend it to an instrumental variable setting for the treatment, the moderator, or both. More extensive research on how to tune the RieszNet and providing a thorough analysis of the asymptotic properties of Auto-DML and the proposed reweighting estimator are fruitful areas for further research. More extensive simulation studies would lead to a more comprehensive picture of the finite sample properties of the proposed estimators.

\newpage

\bibliographystyle{apacite}
\bibliography{library}

\newpage
\begin{appendices}
\renewcommand\thetable{\thesection.\arabic{table}} 
\counterwithin{table}{section}
\renewcommand\thefigure{\thesection.\arabic{figure}} 
\counterwithin{figure}{section}
\section{Appendix: Causal Interpretation of Differences in Treatment Effects} \label{Extension_Causal_Moderation}

\subsection{Effect of Interest}

This appendix considers the case when the analyst wants to interpret the difference in treatment effects between two groups causally. Then a different but closely related parameter is needed, namely the \textit{causal balanced group average treatment effect ($\Delta$CBGATE)}. 

To define the $\Delta$CBGATE we first introduce potential outcomes that additionally depend on the moderator $Z_i$, i.e. $(Y_i^{0,0}, Y_i^{1,0}, Y_i^{0,1}, Y_i^{1,1})$. The new parameter of interest is defined as follows:
\begin{equation*}
   \theta^{\Delta C} = \E\left[ \left(Y_i^{1,1} - Y_i^{0,1} \right) - \left(Y_i^{1,0} - Y_i^{0,0} \right) \right] = \E\left[ \tau(X_i,1) - \tau(X_i,0) \right]
\end{equation*}

The $\Delta$CBGATE represents the difference between the two groups, balancing the distribution of all the other covariates. If the identifying assumptions in the next section hold, then it follows that the $\Delta$CBGATE is fully balanced and other covariates do not influence this difference. The conditions to obtain a causally interpretable $\Delta$CBGATE may be challenging in many applications. In a setting with a randomized treatment and a randomized moderator variable, the $\Delta$CBGATE is identified without further assumptions. 

\subsection{Identifiying Assumptions}

In addition to the usual identifying assumptions stated in Section \ref{2_identifying_assumptions}, we apply the unconfoundedness setting to the moderator variable to be able to interpret the $\Delta$CBGATE causally. Hence, the following assumptions have to be changed or added in comparison to the assumptions for the $\Delta$BGATE in Section \ref{2_identifying_assumptions}.

\begin{assumption}\label{assumption_CIA_CBGATE} \textnormal{(Additional identifying assumptions)}
    \begin{enumerate}[label=(\alph*)]
            \item CIA for moderator: $(Y_i^{1,1}, Y_i^{0,1}, Y_i^{1,0}, Y_i^{0,0}) \perp Z_i | X_i = x \quad \forall x \in \mathcal{X}$
            \item CS for moderator: $0 < P(Z_i = z | X_i = x) < 1, \quad \forall z \in \{0,1\}, \forall x \in \mathcal{X}$
            \item Exogenity of confounders and moderator: $Z_i^0 = Z_i^1, \quad X_i^{0,0} = X_i^{0,1}$
            \item Stable Unit Treatment Value Assumption (SUTVA): $Y_i = \sum_{z = 0}^{1}\sum_{d = 0}^{1} I(d,z) Y_i^{d,z}$
    \end{enumerate}
\end{assumption}

In some applications, Assumption 7. (a) may be fulfilled by conditioning only on a subset of $X_i$. One example is if the distribution of some $X_i$'s is already balanced across the two groups of $Z_i$. New about the exogeneity assumption is that the moderator must not influence the confounders in a way related to the outcome variable. This assumption is non-standard and might be hard to fulfil in some applications. It often happens that a moderator variable such as gender influences other covariates, violating this assumption.

\begin{lemma} \label{MGATE_lemma}
    Under Assumption \ref{assumption_CIA_CBGATE} the parameter $\theta^{\Delta C}= \E\left[\left(Y_i^{1,1} - Y_i^{0,1} \right) - \left(Y_i^{1,0} - Y_i^{0,0} \right)\right]$ is identified as $\E\left[\mu_{1}(1,X_i) - \mu_{0}(1,X_i) - \mu_{1}(0,X_i) + \mu_{0}(0,X_i)\right]$ with $\mu_{d}(z,x) = \E[Y_i | D_i = d, Z_i = z, X_i = x]$.
\end{lemma}
    
For the proof of Lemma \ref{MGATE_lemma}, see Section \ref{_MGATE}. The $\Delta CBGATE$ is the same effect as a fully balanced $\Delta BGATE$ for which Assumption \ref{assumption_CIA_CBGATE} holds.

\subsection{Estimation and Inference}

To obtain an efficient and flexible estimator, we use DML. The estimated Neyman-orthogonal score based on the efficient influence function for the $\Delta$CBGATE is given by: 
\defcitealias{Chernozhukov:2018}{Chernozhukov et al., 2018} 
\begin{align*}
    \hat \phi^{\Delta C}(h; \theta^{\Delta C},  \hat \eta) &=  \hat \mu_{1}(1,x) -   \hat \mu_{0}(1,x) -  \hat \mu_{1}(0,x) +  \hat \mu_{0}(0,x) \\ &+ \frac{dz(y -  \hat \mu_{1}(1,x))}{ \hat\omega_{1,1}(x)} - \frac{(1-d)z(y -  \hat \mu_{0}(1,x))}{ \hat \omega_{0,1}(x)} \\ &-\frac{d(1-z)(y -  \hat \mu_{1}(0,x))}{ \hat \omega_{1,0}(x)} + \frac{(1-d)(1-z)(y -  \hat \mu_{0}(0,x))}{ \hat \omega_{0,0}(x)} -  \theta^{\Delta C}
\end{align*}
with $ \hat \omega_{d,z}(x) =  \hat P(D_i = d, Z_i = z | X_i = x)$, $  \hat \mu_{d}(z,x) =  \hat \E[Y_i | D_i = d, Z_i = z, X_i = x]$ and the nuisance parameters $ \hat \eta = (  \hat \mu_{d}(z,x),  \hat \omega_{d,z}(x))$. Using this double robust moment condition, the $\Delta$CBGATE is estimated at a convergence rate of $\sqrt{N}$, even if the nuisance parameters are estimated with slower rates \citep*{Smucler:2019}. However, they must converge such that the product of the convergence rates of the outcome regression score and the propensity score estimator is faster than or equal to $\sqrt{N}$. Many common machine learning algorithms are known to converge at a rate faster or equal to $N^{1/4}$ under certain conditions but slower than $\sqrt{N}$ \citepalias{Chernozhukov:2018}. Furthermore, as before, we use cross-fitting with K-folds. An estimator based on these elements is $\sqrt{N}$-consistent, asymptotically normal, and asymptotically efficient \citep{Kennedy:2022}. The proof can be found in Appendix \ref{Appendix_CBGATE_proof}.

The variance of $ \hat \theta^{\Delta C}$ can be defined as
\begin{align*} 
    \Var(\hat \theta^{\Delta C})  &= \Var( \phi^{\Delta C}(H_i; \theta^{\Delta C},  \eta)) \\&=\E[ \phi^{\Delta C}(H_i; \theta^{\Delta C}, \eta)^2] - \underbrace{\E[ \phi^{\Delta C}(H_i; \theta^{\Delta C}, \eta)]^2}_{=0} \\ &= \E[\phi^{\Delta C}(H_i; \theta^{\Delta C}, \eta)^2]
\end{align*}
and can be estimated as follows $\widehat{\Var}(\hat \theta^{\Delta C}) = \frac{1}{N}\sum_{k = 1}^K \sum_{i \in S_k}[\hat \phi^{\Delta C}(H_i; \theta^{\Delta C},  \hat\eta)]^2$.

\subsection{Implementation}
Concerning the practical implementation, any machine learner with the required convergence rates can be used to estimate the nuisance parameters. Note that since $P(D_i=d, Z_i=z|X_i = x)$ can be rewritten as $P(D_i=d |X_i=x, Z_i = z)\cdot P(Z_i=z |X_i = x)$, there are two basic versions of the estimator. One version is based on directly estimating $P(D_i=d, Z_i=z|X_i = x)$. In contrast, the second version is based on estimating $P(D_i=d |X_i = x, Z_i = z)$ and $P(Z_i=z |X_i = x)$ separately in the full sample and subsequently obtaining the estimate of $P(D_i=d, Z_i=z|X_i = x)$ as the product of these two estimates. For the proof that the second version of the estimator is also $\sqrt{N}$-consistent and asymptotically normal, see Appendix \ref{Asymptotics}. Same as for the $\Delta$BGATE, we normalize the weights (e.g.,$\frac{dz}{\omega_{1,1}(x)}$) to ensure that they do not have much more weight than the outcome regression (see Algorithm \ref{alg:normalization} in Appendix \ref{Estimation_procedures}). The proposed Algorithm \ref{alg:DDR-Learner} for the $\Delta$CBGATE is also summarized in Appendix \ref{Estimation_procedures}. In a previous version of this paper, a small simulation study was conducted to show the finite sample properties of the estimator (https://arxiv.org/abs/2401.08290).
\newpage

\section{Appendix: Theory} \label{Appendix_A}
\subsection{Notation}\label{Notation}

The model considered here is more general than in the main body of the paper; let the treatment be $d \in \{0,1, \dots, m\}$ and the moderator be $z \in \{0,1,2, \dots, v \}$ variables which are discrete.
To identify the effects, we rely on the usual potential outcomes by treatment $(Y_i^{0}, Y_i^{1}, \dots, Y_i^{m})$. Furthermore, we have an indicator function $I(d) = \mathds{1}(D_i = d)$, which takes the value one if $D_i = d$ and zero otherwise, and an indicator function $I(z) = \mathds{1}(Z_i = z)$ which takes the value one if $Z_i = z$ and zero otherwise. The analysis examines pairwise comparisons between two treatments, denoted as $m$ and $l$, and two moderator groups represented as $u$ and $v$.

\subsection{Decomposition of the $\Delta$GATE} \label{Decomposition_GATE}

In this subsection we show how the $\Delta$GATE can be decomposed into a direct effect and two compositional effects. 
\begin{align*}
    \Delta GATE = &\E[Y_i^1 - Y_i^0 | Z_i = 1] - \E[Y_i^1 - Y_i^0 | Z_i = 0] \\ =& \frac{P(Z_i = 0)}{P(Z_i = 1)}\E[\E[Y_i^1 - Y_i^0 |W_i, Z_i = 1]] - \E[\E[Y_i^1 - Y_i^0 | W_i, Z_i = 1] | Z_i = 0] \\&- \frac{P(Z_i = 1)}{P(Z_i = 0)}\E[\E[Y_i^1 - Y_i^0 |W_i, Z_i = 0]] - \E[\E[Y_i^1 - Y_i^0 | W_i, Z_i = 0] | Z_i = 1] \\
    = &\underbrace{\E[\E[Y_i^1 - Y_i^0 | W_i, Z_i = 1] - \E[Y_i^1 - Y_i^0 | W_i, Z_i = 0] ]}_{\Delta BGATE} \\ &+ \frac{\E[\E[Y_i^1 - Y_i^0|W_i, Z_i = 1]] - \E[\E[Y_i^1 - Y_i^0|W_i, Z_i = 1]]P(Z_i = 1)}{P(Z_i = 1)} \\ &- \frac{\E[\E[Y_i^1 - Y_i^0|W_i, Z_i = 0]] - \E[\E[Y_i^1 - Y_i^0|W_i, Z_i = 0]]P(Z_i = 0)}{P(Z_i = 0)} \\ &- \frac{\E[\E[Y_i^1 - Y_i^0 |W_i, Z_i = 1]|Z_i = 0]P(Z_i = 0)}{P(Z_i = 1)} \\ &+ \frac{\E[\E[Y_i^1 - Y_i^0 |W_i, Z_i = 0]| Z_i = 1]P(Z_i = 1)}{P(Z_i = 0)} \\
    = &\underbrace{\E[\E[Y_i^1 - Y_i^0 | W_i, Z_i = 1] - \E[Y_i^1 - Y_i^0 | W_i, Z_i = 0] ]}_{\Delta BGATE} \\ &+ \frac{\E[\E[Y_i^1 - Y_i^0 | W_i, Z_i = 1]](1 - P(Z_i = 1)) - \E[\E[Y_i^1 - Y_i^0 |W_i, Z_i = 1]|Z_i = 0]P(Z_i = 0)}{P(Z_i = 1)}   \\  & - \frac{\E[\E[Y_i^1 - Y_i^0 | W_i, Z_i = 0]](1 - P(Z_i = 0)) - \E[\E[Y_i^1 - Y_i^0 |W_i, Z_i = 0]|Z_i = 1]P(Z_i = 1)}{P(Z_i = 0)}
\end{align*}
    \begin{align*}
    = &\underbrace{\E[\E[Y_i^1- Y_i^0 | W_i, Z_i = 1] - \E[Y_i^1- Y_i^0 | W_i, Z_i = 0]]}_{\text{direct effect: $\Delta$BGATE}} \\&+ \underbrace{\frac{P(Z_i = 0)}{P(Z_i = 1)}\E[\E[Y_i^1- Y_i^0 |W_i, Z_i = 1]] - \E[\E[Y_i^1- Y_i^0 |W_i, Z_i = 1]| Z_i = 0]}_{\text{compositional effect (1)}} \\&- \underbrace{\frac{P(Z_i = 1)}{P(Z_i = 0)}\E[\E[Y_i^1- Y_i^0 |W_i, Z_i = 0]] - \E[\E[Y_i^1- Y_i^0 |W_i, Z_i = 0]| Z_i = 1]}_{\text{compositional effect (2)}}
\end{align*}

\subsection{$\Delta$BGATE} \label{Appendix_A_identification}
\subsubsection{Indentification Based on the Outcome Regression} \label{_BGATE}

In this subsection, the identification of the BGATE with Assumption \ref{assumption_CIA} stated in Section \ref{2_identifying_assumptions} is shown. Due to the linearity in expectations assumption, the identification for the $\Delta$BGATE directly follows. Please recall that $\mu_d(z,x) = \E[Y_i | D_i = d, Z_i=z, X_i=x]$.
\begin{align*}
    \E\Bigl[\E[&Y_i^l - Y_i^m \mid  Z_i = z, W_i] \Bigr] \\
     = \E \Bigl[&\E \left[ \E \left[ Y_i^l - Y_i^m \mid X_i, Z_i \right] \mid  Z_i = z, W_i  \right] \Bigr] \\
    = \E \Bigl[&\E \left[ \E [ Y_i^l \mid X_i, Z_i] - \E [ Y_i^m \mid X_i, Z_i] \mid  Z_i = z, W_i \right] \Bigr]\\
    = \E \Bigl[&\E \left[ \E [ Y_i^l \mid  D_i = l, X_i, Z_i] - \E [ Y_i^m \mid D_i = m, X_i, Z_i] \mid  Z_i = z, W_i \right] \Bigr]\\
    = \E \Bigl[&\E \left[ \E [ Y_i \mid  D_i = l, X_i, Z_i] - \E [ Y_i \mid D_i = m, X_i, Z_i] \mid  Z_i = z, W_i \right] \Bigr] \\
    = \E \Bigl[&\E \left[ \mu_l(Z_i,X_i) - \mu_m(Z_i,X_i)\mid Z_i = z, W_i \right] \Bigr]
\end{align*}

The first equality is derived from the law of iterated expectations, while the second equality follows from the linearity in expectations. The third equality is based on Assumption \ref{assumption_CIA}, and the fourth is derived from the law of total expectation.

\subsubsection{Identification Based on the Doubly Robust Score} \label{DR_Identificaction_BGATE}
The BGATE can also be identified with the doubly robust score function (see Section \ref{2_estimation} in the main body of the text). Again, due to the linearity in expectations, this directly implies that the $\Delta$BGATE can be identified. Please recall that $\pi_d(z,x) = P(D_i = d| Z_i = z, X_i = x)$ and $\lambda_z(w) = P(Z_i = z| W_i = w)$. For better readability $\mu_l(z,x) - \mu_m(z,x) + \frac{I(l)(y - \mu_l(z,x))}{\pi_l(z,x)} - \frac{I(m)(y - \mu_m(z,x))}{\pi_m(z,x)}$ is denoted as $\delta_{l,m}(h)$ and $\E[\delta_{l,m}(H_i) | Z_i = z, W_i = w]$ as $g_{l,m,z}(w)$.
\begin{align*}
\E[&\E[Y_i^{l} - Y_i^{m} | Z_i = z, W_i]] \\
&= \E\Biggl[\E\Biggl[g_{l,m,z}(W_i) + \frac{I(z)(\delta_{l,m}(H_i) - g_{l,m,z}(W_i))}{\lambda_z(W_i)} \bigg | W_i  \Biggr] \Biggr] \\ 
&= \E\Biggl[g_{l,m,z}(W_i) + \E\Biggl[ \frac{I(z)(\delta_{l,m}(H_i) - g_{l,m,z}(W_i))}{\lambda_z(W_i)} \bigg | W_i \Biggr] \Biggr] \\ 
&= \E\Biggl[g_{l,m,z}(W_i)+ \E\Biggl[ \frac{I(z)(\delta_{l,m}(H_i) - g_{l,m,z}(W_i))}{\lambda_z(W_i)} \bigg | W_i , Z = z \Biggr] \lambda_z(W_i) \Biggr] \\
&= \E\biggl[g_{l,m,z}(W_i) + \E[\delta_{l,m}(H_i) - g_{l,m,z}(W_i)| Z_i = z, W_i ]\biggr] = \E[g_{l,m,z}(W_i)]
\end{align*}
The first equality follows from the law of iterated expectations, and the third from the law of total expectation. Hence, the BGATE can be identified if the nuisance functions are correctly specified. Due to the double robustness property, the BGATE is also identified if only the outcome regression or the propensity score is correctly specified.

\textit{Correctly specified propensity score $\lambda_z(w)$ and wrongly specified outcome regression $\bar g_{l,m,z}(w)$:}
\begin{align*}
    \E[&\E[Y_i^{l} - Y_i^{m} | Z_i = z, W_i]] \\ 
    &= \E\Biggl[\E\Biggl[\bar g_{l,m,z}(W_i) + \frac{I(z)(\delta_{l,m}(H_i) - \bar g_{l,m,z}(W_i))}{\lambda_z(W_i)} \bigg | W_i  \Biggr] \Biggr] \\ 
    &= \E\Biggl[\bar g_{l,m,z}(W_i) + \E\Biggl[ \frac{I(z)(\delta_{l,m}(H_i) - \bar g_{l,m,z}(W_i))}{\lambda_z(W_i)} \bigg | W_i \Biggr] \Biggr] \\ 
    &= \E\Biggl[\bar g_{l,m,z}(W_i) + \E\Biggl[ \frac{I(z)(\delta_{l,m}(H_i) - \bar g_{l,m,z}(W_i))}{\lambda_z(W_i)} \bigg | W_i , Z_i = z \Biggr] \lambda_z(W_i) \Biggr] \\
    &= \E\biggl[\bar g_{l,m,z}(W_i) + \E[\delta_{l,m}(H_i) - \bar g_{l,m,z}(W_i)| Z_i = z, W_i ]\biggr]\\
    &= \E\biggl[\bar g_{l,m,z}(W_i) +  g_{l,m,z}(W_i)- \bar g_{l,m,z}(W_i)\biggr] = \E[ g_{l,m,z}(W_i)]
\end{align*}

\textit{Correctly specified outcome regression $g_{l,m,z}(w)$ and wrongly specified propensity score $\bar \lambda_z(w)$:}
\begin{align*}
    \E[&\E[Y_i^{l} - Y_i^{m} | Z_i = z, W_i]] \\ 
    &= \E\Biggl[\E\Biggl[g_{l,m,z}(W_i) + \frac{I(z)(\delta_{l,m}(H_i) - g_{l,m,z}(W_i))}{\bar \lambda_z(W_i)} \bigg | W_i  \Biggr] \Biggr] \\ 
    &= \E\Biggl[g_{l,m,z}(W_i) + \E\Biggl[ \frac{I(z)(\delta_{l,m}(H_i) - g_{l,m,z}(W_i))}{\bar \lambda_z(W_i)} \bigg | W_i \Biggr] \Biggr] \\ 
    &= \E\Biggl[g_{l,m,z}(W_i) + \E\Biggl[ \frac{I(z)(\delta_{l,m}(H_i) - g_{l,m,z}(W_i))}{\bar \lambda_z(W_i)} \bigg | W_i , Z_i = z \Biggr] \lambda_z(W_i) \Biggr] \\
    &= \E\Biggl[g_{l,m,z}(W_i) + \frac{\lambda_z(W_i)}{\bar \lambda_z(W_i)}\left(g_{l,m,z}(W_i) - \E\left[ \delta_{l,m}(H_i)\mid W_i, Z_i = z \right]\right)  \Biggr] = \E[g_{l,m,z}(W_i)]
\end{align*}

Hence, the BGATE is identified if the outcome regression or the propensity score is correctly specified.

\subsubsection{Asymptotic Properties} \label{Asymptotics_BGATE}

The following proof is for $\theta_{l,m,u,v}^B$. However, due to the linearity in expectations, it directly follows that the proof is also valid for $\theta_{l,m,u,v}^{\Delta B}$. In a first step, let us define the following terms for easier readability:
\begin{align*}
    \pi_d(z,x) &= P(D_i=d | Z_i = z, X_i = x) \\
    \mu_d(z,x) &= \E[Y_i | D_i = d, Z_i=z, X_i=x] \\
    \delta_{l,m}(h) &= \mu_l(z,x) - \mu_m(z,x) + \frac{I(l)(y - \mu_l(z,x))}{\pi_l(z,x)} - \frac{I(m)(y - \mu_m(z,x))}{\pi_m(z,x)} \\
    \lambda_z(w) &= P(Z_i=z | W_i=w) \\
    g_{l,m,z}(w) &= \E[\delta_{l,m}(H_i) | Z_i = z, W_i = w] \\
    \tau_{l,m,z}(\delta_{l,m}(h),w) &= g_{l,m,z}(w) + \frac{I(z) (\delta_{l,m}(h)- g_{l,m,z}(w))}{\lambda_z(w)} \\
    \theta_{l,m,z}^B &= \E[\tau_{l,m,z}(\delta_{l,m}(H_i), W_i)]
\end{align*} 
The goal is to show that
\begin{align*}
    &\sqrt{N}(\hat{\theta}_{l,m,z}^B - \theta_{l,m,z}^{B \star}) \xrightarrow{d} N(0, V^\star)\\
    V^\star & = \E\left[ \left(\tau_{l,m,z}(\delta_{l,m}(H_i), W_i) - \theta_{l,m,z}^{B}\right)^2\right]
\end{align*}
with $\theta_{l,m,z}^{B \star}$ being an oracle estimator of $\theta_{l,m,z}^B$ if all nuisance functions would be known. Then $\theta_{l,m,z}^{B \star}$ is an i.i.d. average, hence:
\begin{align*}
    \sqrt{N}(\theta_{l,m,z}^{B\star} - \theta_{l,m,z}^B) \xrightarrow{d} N(0, V^\star)  \quad \text{with} \quad
    V^\star = \E\left[ \left(\tau_{l,m,z}(\delta_{l,m}(H_i), W_i) - \theta_{l,m,z}^{B}\right)^2\right]
\end{align*}

\begin{theorem}
    Let $\mathcal{I}_1$ and $\mathcal{I}_2$ be two half samples such that $\vert \mathcal{I}_1 \vert = \vert \mathcal{I}_1 \vert = \frac{N}{2} $. Furthermore, let $\mathcal{I}_{11}$, $\mathcal{I}_{12}$, $\mathcal{I}_{21}$ and $\mathcal{I}_{22}$ be four quarter samples such that $\vert \mathcal{I}_{11} \vert = \vert \mathcal{I}_{12} \vert = \vert \mathcal{I}_{21} \vert = \vert \mathcal{I}_{22} \vert = \frac{\vert\mathcal{I}_1\vert}{2} = \frac{\vert\mathcal{I}_2\vert}{2}  = \frac{N}{4}$ The second split is independent conditional on the first split. Define the estimator as follows:
    \begin{align*}
        \hat \theta_{l,m,z}^B &= \frac{\vert \mathcal{I}_{11} \vert}{N} \hat \theta^{B,\mathcal{I}_{11}}_{l,m,z} + \frac{\vert \mathcal{I}_{12} \vert}{N} \hat \theta^{B,\mathcal{I}_{12}}_{l,m,z} + \frac{\vert \mathcal{I}_{21} \vert}{N} \hat \theta^{B,\mathcal{I}_{21}}_{l,m,z} + \frac{\vert \mathcal{I}_{22} \vert}{N} \hat \theta^{B, \mathcal{I}_{22}}_{l,m,z} \\
        \hat \theta_{l,m,z}^{B, \mathcal{I}_{11}} &= \frac{1}{\vert \mathcal{I}_{11} \vert} \sum_{\mathcal{I}_{11}} \hat \tau_{l,m,z}^{\mathcal{I}_{12}}(\hat \delta_{l,m}^{\mathcal{I}_{2}}(H_i), W_i) \\
        \hat \theta_{l,m,z}^{B, \mathcal{I}_{12}} &= \frac{1}{\vert \mathcal{I}_{12} \vert} \sum_{\mathcal{I}_{12}} \hat \tau_{l,m,z}^{\mathcal{I}_{11}}(\hat \delta_{l,m}^{\mathcal{I}_{2}}(H_i), W_i) \\
        \hat \theta_{l,m,z}^{B, \mathcal{I}_{21}} &= \frac{1}{\vert \mathcal{I}_{21} \vert} \sum_{\mathcal{I}_{21}} \hat \tau_{l,m,z}^{\mathcal{I}_{22}} (\hat \delta_{l,m}^{\mathcal{I}_{1}}(H_i), W_i)\\
        \hat \theta_{l,m,z}^{B, \mathcal{I}_{22}} &= \frac{1}{\vert \mathcal{I}_{22} \vert} \sum_{\mathcal{I}_{22}} \hat \tau_{l,m,z}^{\mathcal{I}_{21}}(\hat \delta_{l,m}^{\mathcal{I}_{1}}(H_i), W_i)
    \end{align*}
Then, if Assumptions 5 to 9 hold, it follows that 
\begin{align*}
    \sqrt{N}(\hat \theta_{l,m,z}^B - \theta_{l,m,z}^B) \xrightarrow{d} N(0, V^\star) \quad \text{with} \quad V^\star = \E\left[ \left(\tau_{l,m,z}(\delta_{l,m}(H_i), W_i) - \theta_{l,m,z}^{B}\right)^2\right]
\end{align*}

\end{theorem}
\textit{Proof}.
\begin{align*}
    \sqrt{N} &(\hat \theta_{l,m,z}^B - \theta_{l,m,z}^B)= \sqrt{N}(\hat \theta_{l,m,z}^B - \theta_{l,m,z}^{B\star} + \theta_{l,m,z}^{B\star} - \theta_{l,m,z}^B) \\
    &= \sqrt{N}(\hat \theta_{l,m,z}^B - \theta_{l,m,z}^{B\star}) + \sqrt{N}(\theta_{l,m,z}^{B\star} - \theta_{l,m,z}^B) \\
    &= \sqrt{N}\left(\frac{\vert \mathcal{I}_{11} \vert}{N} \hat \theta_{l,m,z}^{B, \mathcal{I}_{11}} + \frac{\vert \mathcal{I}_{12} \vert}{N} \hat \theta_{l,m,z}^{B, \mathcal{I}_{12}} + \frac{\vert \mathcal{I}_{21} \vert}{N} \hat \theta_{l,m,z}^{B, \mathcal{I}_{21}} + \frac{\vert \mathcal{I}_{22} \vert}{N} \hat \theta_{l,m,z}^{B, \mathcal{I}_{22}} - \theta_{l,m,z}^{B\star}\right) + \sqrt{N}(\theta_{l,m,z}^{B\star} - \theta_{l,m,z}^B) \\
    &= \sqrt{N} \biggl(\frac{\vert \mathcal{I}_{11} \vert}{N} \hat \theta_{l,m,z}^{B, \mathcal{I}_{11}} + \frac{\vert \mathcal{I}_{12} \vert}{N} \hat \theta_{l,m,z}^{B, \mathcal{I}_{12}} + \frac{\vert \mathcal{I}_{21} \vert}{N} \hat \theta_{l,m,z}^{B, \mathcal{I}_{21}} + \frac{\vert \mathcal{I}_{22} \vert}{N} \hat \theta_{l,m,z}^{B, \mathcal{I}_{22}} \\
    & \quad -\frac{\vert \mathcal{I}_{11} \vert}{N} \theta_{l,m,z}^{B \star, \mathcal{I}_{11}} - \frac{\vert \mathcal{I}_{12} \vert}{N} \theta_{l,m,z}^{B \star, \mathcal{I}_{12}} - \frac{\vert \mathcal{I}_{21} \vert}{N} \theta_{l,m,z}^{B \star, \mathcal{I}_{21}} - \frac{\vert \mathcal{I}_{22} \vert}{N}  \theta_{l,m,z}^{B \star, \mathcal{I}_{22}}\biggr) + \sqrt{N} (\theta_{l,m,z}^{B\star} - \theta_{l,m,z}^B) \\
    &= \sqrt{N} \biggl(\frac{\vert \mathcal{I}_{11} \vert}{N} \hat \theta_{l,m,z}^{B, \mathcal{I}_{11}} -\frac{\vert \mathcal{I}_{11} \vert}{N} \theta_{l,m,z}^{B \star, \mathcal{I}_{11}} \biggr) + 
    \sqrt{N} \biggl( \frac{\vert \mathcal{I}_{12} \vert}{N} \hat \theta_{l,m,z}^{B, \mathcal{I}_{12}} - \frac{\vert \mathcal{I}_{12} \vert}{N}  \theta_{l,m,z}^{B \star, \mathcal{I}_{12}} \biggr) \\ & \quad + \sqrt{N} \biggl(\frac{\vert \mathcal{I}_{21} \vert}{N} \hat \theta_{l,m,z}^{B, \mathcal{I}_{21}} - \frac{\vert \mathcal{I}_{21} \vert}{N}  \theta_{l,m,z}^{B \star, \mathcal{I}_{21}} \biggr) +
    + \sqrt{N} \biggl( \frac{\vert \mathcal{I}_{22} \vert}{N} \hat \theta_{l,m,z}^{B, \mathcal{I}_{22}}  - \frac{\vert \mathcal{I}_{22} \vert}{N} \theta_{l,m,z}^{B \star, \mathcal{I}_{22}}\biggr) \\ & \quad + \sqrt{N} (\theta_{l,m,z}^{B\star} - \theta_{l,m,z}^B)
\end{align*}
We must show that the first four terms converge to zero in probability. It is enough to show this for the first term only, as the same steps can also be directly applied to the remaining three terms. The first summation can be decomposed as follows:
\begin{align*}
     \hat \theta_{l,m,z}^{B, \mathcal{I}_{11}} - \theta_{l,m,z}^{B \star, \mathcal{I}_{11}} &= \frac{1}{|\mathcal{I}_{11}|} \sum_{i \in I_{11}} \Biggl(\hat \tau_{l,m,z}(\hat \delta_{l,m}(H_i)^{\mathcal{I}_2},W_i) - \tau^\star_{l,m,z}( \delta_{l,m}(H_i),W_i) \Biggr)\\
     =\frac{1}{|\mathcal{I}_{11}|} \sum_{i \in \mathcal{I}_{11}} \Biggl(&\hat \E[\hat \delta_{l,m}(H_i)^{\mathcal{I}_{2}} | Z_i = z, W_i]^{\mathcal{I}_{12}} + \frac{I(z) (\hat \delta_{l,m}(H_i)^{\mathcal{I}_{2}} - \hat \E[\hat \delta_{l,m}(H_i)^{\mathcal{I}_{2}}| Z_i = z, W_i]^{\mathcal{I}_{12}})}{\hat \lambda_z(W_i)^{\mathcal{I}_{12}}}  \\
     &- \E[\delta_{l,m}(H_i) | Z_i = z, W_i] - \frac{I(z) (\delta_{l,m}(H_i) -\E[\delta_{l,m}(H_i) | Z_i = z, W_i])}{\lambda_z(W_i)}\Biggr) \\
     = \frac{1}{|\mathcal{I}_{11}|} \sum_{i \in \mathcal{I}_{11}} \Biggl(&\hat \E[\hat \delta_{l,m}(H_i)^{\mathcal{I}_{2}} | Z_i = z, W_i]^{\mathcal{I}_{12}} + \frac{I(z) (\hat \delta_{l,m}(H_i)^{\mathcal{I}_{2}} - \hat \E[\hat \delta_{l,m}(H_i)^{\mathcal{I}_{2}}  | Z_i = z, W_i]^{\mathcal{I}_{12}})}{\hat \lambda_z(W_i)^{\mathcal{I}_{12}}}  \\
     &- \hat \E[\delta_{l,m}(H_i) | Z_i = z, W_i]^{\mathcal{I}_{12}} - \frac{I(z) (\delta_{l,m}(H_i) -\hat \E[\delta_{l,m}(H_i) | Z_i =z, W_i]^{\mathcal{I}_{12}})}{\hat \lambda_z(W_i)^{\mathcal{I}_{12}}} \\
     &+ \hat \E[\delta_{l,m}(H_i) | Z_i = z, W_i]^{\mathcal{I}_{12}} + \frac{I(z) (\delta_{l,m}(H_i) -\hat \E[\delta_{l,m}(H_i) | Z_i = z, W_i]^{\mathcal{I}_{12}})}{\hat \lambda_z(W_i)^{\mathcal{I}_{12}}} \\
     &- \E[\delta_{l,m}(H_i) | Z_i = z, W_i] - \frac{I(z) (\delta_{l,m}(H_i) -\E[\delta_{l,m}(H_i) | Z_i = z, W_i])}{\lambda_z(W_i)}\Biggr) 
\end{align*}

From now on, the terms are denoted as follows: $\hat g_{l,m,z}(w) = \hat \E[\hat \delta_{l,m}(H_i) | Z_i = z, W_i = w] $,  $ \tilde g_{l,m,z}(w) = \hat \E[\delta_{l,m}(H_i)| Z_i = z, W_i = w]$ and $g_{l,m,z}(w) = \E[\delta_{l,m}(H_i)| Z_i = z, W_i = w]$.
Next, analyze:
\begin{align*}
  \frac{1}{|\mathcal{I}_{11}|} \sum_{i \in \mathcal{I}_{11}} \Biggl(&\hat{\tau}_{l,m,z}(\hat \delta_{l,m}(H_i)^{\mathcal{I}_2},W_i) - \tilde{\tau}_{l,m,z}( \delta_{l,m}(H_i),W_i) + \tilde{\tau}_{l,m,z}(\delta_{l,m}(H_i),W_i) - \tau_{l,m,z}^{\star}(\delta_{l,m}(H_i),W_i)\Biggr) \\
    &= \underbrace{\frac{1}{|\mathcal{I}_{11}|} \sum_{i \in \mathcal{I}_{11}} \Biggl(\hat g_{l,m,z}(W_i)^{\mathcal{I}_{12}} + \frac{I(z) (\hat \delta_{l,m}(H_i)^{\mathcal{I}_{2}} -  \hat g_{l,m,z}(W_i)^{\mathcal{I}_{12}})}{\hat \lambda_z(W_i)^{\mathcal{I}_{12}}}}_{\text{Part A}}
    \\ & \quad - \underbrace{\tilde g_{l,m,z}(W_i)^{\mathcal{I}_{12}} - \frac{I(z) (\delta_{l,m}(H_i) -\tilde g_{l,m,z}(W_i)^{\mathcal{I}_{12}})}{\hat \lambda_z(W_i)^{\mathcal{I}_{12}}}\Biggr)}_{\text{Part A}}  \\
    & \quad + \underbrace{\frac{1}{|\mathcal{I}_{11}|} \sum_{i \in \mathcal{I}_{11}}  \Biggl(\tilde g_{l,m,z}(W_i)^{\mathcal{I}_{12}} + \frac{I(z) (\delta_{l,m}(H_i)
-\tilde g_{l,m,z}(W_i)^{\mathcal{I}_{12}})}{\hat \lambda_z(W_i)^{\mathcal{I}_{12}}}\Biggr) }_{\text{Part B}}  \\ & \quad - \underbrace{\Biggl(g_{l,m,z}(W_i) - \frac{I(z) (\delta_{l,m}(H_i) -g_{l,m,z}(W_i))}{\lambda_z(W_i)}\Biggr) }_{\text{Part B}} 
\end{align*}

\textbf{Part A} \\
We start with Part A and show that $\hat \theta^B_{l,m,z}$ converges in probability to $\tilde \theta^B_{l,m,z}$ fast enough. It is possible to rewrite it in the following way:
\begin{align*}
    &\frac{1}{|\mathcal{I}_{11}|} \sum_{i \in \mathcal{I}_{11}} \Biggl(\hat g_{l,m,z}(W_i)^{\mathcal{I}_{12}} + \frac{I(z) (\hat \delta_{l,m}(H_i)^{I_2} -\hat g_{l,m,z}(W_i)^{\mathcal{I}_{12}})}{\hat \lambda_z(W_i)^{\mathcal{I}_{2}}}
    \\ &- \tilde g_{l,m,z}(W_i)^{\mathcal{I}_{12}} - \frac{I(z) (\delta_{l,m}(H_i) - \tilde g_{l,m,z}(W_i)^{\mathcal{I}_{12}})}{\hat \lambda_z(W_i)^{\mathcal{I}_{12}}}\Biggr) \\
    &=\underbrace{\frac{1}{|\mathcal{I}_{11}|} \sum_{i \in \mathcal{I}_{11}} \Biggl((\hat g_{l,m,z}(W_i)^{\mathcal{I}_{12}}  - \tilde g_{l,m,z}(W_i)^{\mathcal{I}_{12}}) \left(1 - \frac{I(z)}{\lambda_z(W_i)} \right) \Biggr)}_{\text{Part A.1}}\\
    &+ \underbrace{\frac{1}{|\mathcal{I}_{11}|} \sum_{i \in \mathcal{I}_{11}} \Biggl(I(z)(\delta_{l,m}(H_i) - \tilde g_{l,m,z}(W_i)^{\mathcal{I}_{12}}) \left(\frac{1}{\lambda_z(W_i)} -  \frac{1}{\hat \lambda_z(W_i)^{\mathcal{I}_{12}}}\right) \Biggr)}_{\text{Part A.2}}\\ 
    &+ \underbrace{\frac{1}{|\mathcal{I}_{11}|} \sum_{i \in \mathcal{I}_{11}} \Biggl(I(z)(\hat \delta_{l,m}(H_i)^{\mathcal{I}_{2}} - \hat g_{l,m,z}(W_i)^{\mathcal{I}_{12}}) \left( \frac{1}{\hat \lambda_z(W_i)^{\mathcal{I}_{12}}} -  \frac{1}{\lambda_z(W_i)}\right) \Biggr)}_{\text{Part A.3}}\\
    &+ \underbrace{\frac{1}{|\mathcal{I}_{11}|} \sum_{i \in \mathcal{I}_{11}} \Biggl(\frac{I(z)}{\lambda_z(W_i)} (\hat \delta_{l,m}(H_i)^{\mathcal{I}_{2}} - \delta_{l,m}(H_i))\Biggr)}_{\text{Part A.4}}
\end{align*}

Using the $L_2$-norm and the fact that after conditioning on $\mathcal{I}_{12}$ and $\mathcal{I}_{2}$ the summands become mean-zero and independent, we can show that the term $A.1$ converges in probability to zero:
\begin{align*}
    &\E\left[\left(\frac{1}{|\mathcal{I}_{11}|} \sum_{i \in \mathcal{I}_{11}} \left(\hat g_{l,m,z}(W_i)^{\mathcal{I}_{12}}  - \tilde g_{l,m,z}(W_i)^{\mathcal{I}_{12}} \right) \left(1 - \frac{I(z)}{\lambda_z(W_i)} \right) \right)^2\right]\\
    &= \E\left[\E\left[\left(\frac{1}{|\mathcal{I}_{11}|} \sum_{i \in \mathcal{I}_{11}} \left(\hat g_{l,m,z}(W_i)^{\mathcal{I}_{12}}  - \tilde g_{l,m,z}(W_i)^{\mathcal{I}_{12}}\right) \left(1 - \frac{I(z)}{\lambda_z(W_i)} \right) \right)^2 \bigg | \mathcal{I}_{12}, \mathcal{I}_{2}\right]\right]\\
    &= \E\left[\Var\left[\frac{1}{|\mathcal{I}_{11}|} \sum_{i \in \mathcal{I}_{11}} \left(\hat g_{l,m,z}(W_i)^{\mathcal{I}_{12}} - \tilde g_{l,m,z}(W_i)^{\mathcal{I}_{12}}\right) \left(1 - \frac{I(z)}{\lambda_z(W_i)} \right) \bigg | \mathcal{I}_{12},\mathcal{I}_{2}\right]\right] \\
    &= \frac{1}{|\mathcal{I}_{11}|}\E\left[\Var\left[\left(\hat g_{l,m,z}(W_i)^{\mathcal{I}_{12}} - \tilde g_{l,m,z}(W_i)^{\mathcal{I}_{12}}\right) \left(1 - \frac{I(z)}{\lambda_z(W_i)} \right) \bigg | \mathcal{I}_{12}, \mathcal{I}_{2}\right]\right] \\
    &= \frac{1}{|\mathcal{I}_{11}|}\E\left[\E\left[\left(\hat g_{l,m,z}(W_i)^{\mathcal{I}_{12}} - \tilde g_{l,m,z}(W_i)^{\mathcal{I}_{12}}\right)^2 \left(\frac{1}{\lambda_z(W_i)}-1 \right) \bigg | \mathcal{I}_{12},\mathcal{I}_{2}\right]\right] \\
    & \leq \frac{1}{\kappa |\mathcal{I}_{11}|}\E\left[\left(\hat g_{l,m,z}(W_i)^{\mathcal{I}_{12}} - \tilde g_{l,m,z}(W_i)^{\mathcal{I}_{12}}\right)^2 \right] =  \frac{o_p(1)}{N}
\end{align*}
The second equality follows because after conditioning on $\mathcal{I}_{12}$ and $\mathcal{I}_{2}$ the summands become mean-zero and independent. The last inequality follows from Assumption \ref{assumption_boundness}. For the last equality, we need the $L_2$-norm of $\hat g_{l,m,z}(w) - \tilde g_{l,m,z}(w)$ to converge in probability and the fact that $\vert \mathcal{I}_{11} \vert = \frac{N}{4}$. \cite{Kennedy:2023} establishes the fact that
\begin{align*}
    \hat g_{l,m,z}(w) &- \tilde g_{l,m,z}(w) \\ & = \hat\E[ \E[\hat \delta_{l,m}(H_i) - \delta_{l,m}(H_i)|Z_i = z, W_i = w, \mathcal{I}_{2}]|Z_i = z, W_i = w] + o_p(R_z^\star (w))
\end{align*}
with
\begin{align*}
    R_z^\star (w)^2 = \E\Biggl[\Biggl(&\tilde g_{l,m,z}(W_i) + \frac{I(z) (\delta_{l,m}(H_i) -\tilde g_{l,m,z}(W_i))}{\hat \lambda_z(W_i)}
    \\ - &g_{l,m,z}(W_i) - \frac{I(z) (\delta_{l,m}(H_i) -g_{l,m,z}(W_i))}{\lambda_z(W_i)}\Biggr)^2\Biggr]
\end{align*}
as long as the regression estimator $\hat\E[\dots| \dots]$ is stable (Definition 1 in \cite{Kennedy:2023}, Proposition B1 in \cite{Rambachan:2022}, Definition \ref{definition_stability} below) with respect to distance $a$ and $a(\hat \delta, \delta) \xrightarrow{p} 0$. 
\begin{definition}{\text{(Stability)}} \label{definition_stability} \\
    Suppose that the test $I_1$ and training sample $I_2$ are independent. Let:
    \begin{enumerate}
        \item $\hat \delta (h)$ be an estimate of a function $\delta(h)$ using the training data $\mathcal{I}_2$
        \item $\hat b(w) = \E[\hat \delta_{l,m}(H_i) - \delta_{l,m}(H_i)| Z_i = z, W_i, \mathcal{I}_{2}]$ the conditional bias of the estimator $\hat \delta$
        \item $\hat \E[\delta(H_i)| Z_i = z, W_i = w] = \tilde g_w(x)$ denote a generic regression estimator that regresses outcomes on covariates in the test sample $\mathcal{I}_1$
    \end{enumerate}
    Then the regression estimator $\hat \E$ is stable with respect to distance metric $a$ at $Z_i = z$ and $W_i = w$ if:
    \begin{align*}
        \frac{\hat g_z(w)- \tilde g_z(w) - \hat \E[\hat b(W_i) | Z_i = z, W_i = w]}{\sqrt{\E\left[\tilde g_z(w) - g_z(w)\right]^2}} \xrightarrow{p} 0
    \end{align*}
whenever $a(\hat \delta, \delta) \xrightarrow{p} 0$
\end{definition}

Stability can be perceived as a type of stochastic equicontinuity for a nonparametric regression. They prove that linear smoothers, such as linear regressions, series regressions, nearest neighbour matching and random forests satisfy this stability condition. We will show in Part B that the oracle estimator $R_z^\star (w)$ is $o_p(1/\sqrt{N})$.

Furthermore, \cite{Kennedy:2023} shows in Theorem 2 that the bias term $\hat b(w)$ of an estimator regressing a doubly-robust pseudo-outcome ($\delta_{l,m}(H_i)$) on convariates $W_i$ can be expressed as:
\begin{equation*}
    \hat b(w) = \sum_{d=0}^1 \frac{(\hat \pi_d(z,x) - \pi_d(z,x))(\hat \mu_d(z,x) - \mu_d(z,x))}{d\hat  \pi_d(z,x)  + (1-d)(1- \hat \pi_d(z,x))}
\end{equation*}
Therefore, as long as the estimated nuisance parameters $\hat \pi_d(z,x)$ and $\hat \mu_d(z,x)$ converge in probability to the true nuisance parameters, which is given by Assumption \ref{assumption_consistency}, the bias term will converge in probability to zero. In conclusion, the term A1 is $o_p(1/ \sqrt{N})$.

The second term $A.2$ can again be analyzed as follows:
\begin{align*}
&\E\left[\left(\frac{1}{|\mathcal{I}_{11}|} \sum_{i \in \mathcal{I}_{11}} \left(I(z)(\delta_{l,m}(H_i) - \tilde g_{l,m,z}(W_i)^{\mathcal{I}_{12}}) \left(\frac{1}{\lambda_z(W_i)} - \frac{1}{\hat \lambda_z(W_i)^{\mathcal{I}_{12}}}\right)\right)\right)^2\right] \\
&= \E\left[\E\left[\left(\frac{1}{|\mathcal{I}_{11}|} \sum_{i \in \mathcal{I}_{11}} \left(I(z)(\delta_{l,m}(H_i) - \tilde g_{l,m,z}(W_i)^{\mathcal{I}_{12}}) \left(\frac{1}{\lambda_z(W_i)} - \frac{1}{\hat \lambda_z(W_i)^{\mathcal{I}_{12}}}\right) \right)\right)^2 \bigg | \mathcal{I}_{12}, \mathcal{I}_{2} \right]\right] \\
&= \E\left[\Var\left[\frac{1}{|\mathcal{I}_{11}|} \sum_{i \in \mathcal{I}_{11}} \left(I(z)(\delta_{l,m}(H_i) - \tilde g_{l,m,z}(W_i)^{\mathcal{I}_{12}}) \left(\frac{1}{\lambda_z(W_i)} - \frac{1}{\hat \lambda_z(W_i)^{\mathcal{I}_{12}}}\right) \right) \vert \mathcal{I}_{12}, \mathcal{I}_{2}\right] \right] \\
&= \frac{1}{|\mathcal{I}_{11}|}\E\left[\Var\left[I(z)(\delta_{l,m}(H_i) - \tilde g_{l,m,z}(W_i)^{\mathcal{I}_{12}}) \left(\frac{1}{\lambda_z(W_i)} - \frac{1}{\hat \lambda_z(W_i)^{\mathcal{I}_{12}}}\right) \bigg | \mathcal{I}_{12}, \mathcal{I}_{2}\right] \right] \\
&= \frac{1}{|\mathcal{I}_{11}|}\E\left[\E\left[I(z)\left(\delta_{l,m}(H_i) - \tilde g_{l,m,z}(W_i)^{\mathcal{I}_{12}}\right)^2 \left( \frac{1}{\lambda_z(W_i)} -  \frac{1}{\hat \lambda_z(W_i)^{\mathcal{I}_{12}}}\right)^2 \bigg | \mathcal{I}_{12}, \mathcal{I}_{2}\right] \right] \\
&\leq \frac{1}{|\mathcal{I}_{11}|}(1-\kappa)\E\left[\left(\delta_{l,m}(H_i) - \tilde g_{l,m,z}(W_i)^{\mathcal{I}_{12}}\right)^2 \left( \frac{1}{\lambda_z(W_i)} - \frac{1}{\hat \lambda_z(W_i)^{\mathcal{I}_{12}}}\right)^2 \right] \\
&\leq \frac{1}{|\mathcal{I}_{11}|}\epsilon_{z1}(1-\kappa)\E\left[\left( \frac{1}{\lambda_z(W_i)} - \frac{1}{\hat \lambda_z(W_i)^{\mathcal{I}_{12}}}\right)^2 \right]  = \frac{o_p(1)}{N}
\end{align*}
The third equality follows because the summands are mean-zero and independent. The last equality is true due to Assumption \ref{assumption_consistency} and \ref{assumption_boundness}, the fact that the MSE for the inverse weights decays at the same rate as the MSE for the propensities and the fact that $|\mathcal{I}_{11}| = N/4$. Hence, the term $A.2$ is $o_p(1/\sqrt{N})$.

Similarly, this can be shown for the term $A.3$:
\begin{align*}
    & \E \left[\left(\frac{1}{|\mathcal{I}_{11}|} \sum_{i \in \mathcal{I}_{11}} \left(I(z)(\hat \delta_{l,m}(H_i)^{\mathcal{I}_{2}} - \hat g_{l,m,z}(W_i)^{\mathcal{I}_{12}}) \left( \frac{1}{\hat \lambda_z(W_i)^{I_{12}}} -  \frac{1}{ \lambda_z(W_i)}\right) \right)\right)^2\right]\\
    &=\E \left[\E \left[\left(\frac{1}{|\mathcal{I}_{11}|} \sum_{i \in \mathcal{I}_{11}} \left(I(z)(\hat \delta_{l,m}(H_i)^{\mathcal{I}_{2}} - \hat g_{l,m,z}(W_i)^{\mathcal{I}_{12}}) \left( \frac{1}{\hat \lambda_z(W_i)^{\mathcal{I}_{12}}} - \frac{1}{\lambda_z(W_i)}\right) \right)\right)^2 \bigg | \mathcal{I}_{12}, \mathcal{I}_{2}\right]\right]\\
    &=\E \left[\Var \left[\frac{1}{|\mathcal{I}_{11}|} \sum_{i \in \mathcal{I}_{11}} \left(I(z)(\hat \delta_{l,m}(H_i)^{\mathcal{I}_{2}} - \hat g_{l,m,z}(W_i)^{\mathcal{I}_{12}}) \left( \frac{1}{\hat \lambda_z(W_i)^{\mathcal{I}_{12}}} - \frac{1}{\lambda_z(W_i)}\right) \right) \bigg | \mathcal{I}_{12}, \mathcal{I}_{2}\right]\right] \\
    &=\frac{1}{|\mathcal{I}_{11}|}\E \left[\Var \left[I(z)(\hat \delta_{l,m}(H_i)^{\mathcal{I}_{2}} - \hat g_{l,m,z}(W_i)^{\mathcal{I}_{12}}) \left(\frac{1}{\hat \lambda_z(W_i)^{\mathcal{I}_{12}}} - \frac{1}{\lambda_z(W_i)}\right) \bigg | \mathcal{I}_{12}, \mathcal{I}_{2} \right]\right] \\
    &= \frac{1}{|\mathcal{I}_{11}|}\E \left[\E\left[I(z)\left(\hat \delta_{l,m}(H_i)^{I_2} - \hat g_{l,m,z}(W_i)^{\mathcal{I}_{12}}\right)^2 \left( \frac{1}{\hat \lambda_z(W_i)^{\mathcal{I}_{12}}} - \frac{1}{\lambda_z(W_i)}\right)^2 \bigg | \mathcal{I}_{12}, \mathcal{I}_{2}\right]\right] \\
    &\leq \frac{1}{|\mathcal{I}_{11}|}(1-\kappa)\E \left[\left(\hat \delta_{l,m}(H_i)^{\mathcal{I}_{2}} - \hat g_{l,m,z}(W_i)^{\mathcal{I}_{12}}\right)^2 \left( \frac{1}{\hat \lambda_z(W_i)^{\mathcal{I}_{12}}} - \frac{1}{\lambda_z(W_i)}\right)^2 \right] \\
    &\leq \frac{1}{|\mathcal{I}_{11}|}\epsilon_{z0}(1-\kappa)\E \left[\left( \frac{1}{\hat \lambda_z(W_i)^{\mathcal{I}_{12}}} - \frac{1}{\lambda_z(W_i)}\right)^2 \right] = \frac{o_p(1)}{N}
\end{align*}
Again, the mean-zero and independence property of the summands is needed. The last equality is true due to Assumption \ref{assumption_consistency} and \ref{assumption_boundness}, the fact that the MSE for the inverse weights decays at the same rate as the MSE for the propensities and the fact that $|\mathcal{I}_{11}| = N/4$. Hence, the term $A.3$ is $o_p(1/\sqrt{N})$.

Because $\delta_{l,m}(h)$ is a doubly robust score, a similar approach as for the other parts can be used again for the term $A.4$. Due to the linearity in expectations assumption, only the first part of the score function can be considered. The second part follows analogously. The term $A.4$ can be rewritten as follows:

\begin{align*}
    &\frac{1}{|\mathcal{I}_{11}|} \sum_{i \in \mathcal{I}_{11}} \Biggl(\frac{I(z)}{\lambda_z(W_i)} (\hat \delta_{l,m}(H_i)^{\mathcal{I}_{2}} - \delta_{l,m}(H_i))\Biggr) \\
    &= \frac{1}{|\mathcal{I}_{11}|} \sum_{i \in \mathcal{I}_{11}} \frac{I(z)}{\lambda_z(W_i)}\Biggl( \hat \mu_d(Z_i,X_i)^{\mathcal{I}_{2}} + \frac{I(d)(Y_i- \hat \mu_d(Z_i,X_i)^{\mathcal{I}_{2}})}{\hat \pi_d(Z_i,X_i)^{\mathcal{I}_{2}}} -  \mu_d(Z_i,X_i) - \frac{I(d)(Y_i- \mu_d(Z_i,X_i))}{\pi_d(Z_i,X_i)} \Biggr) \\
    &= \underbrace{\frac{1}{|\mathcal{I}_{11}|} \sum_{i \in \mathcal{I}_{11}} \Biggl( \frac{I(z)}{\lambda_z(W_i)}\Biggl( \hat \mu_d (Z_i,X_i)^{\mathcal{I}_{2}} - \mu_d(Z_i,X_i) \Biggr)\left(1 - \frac{I(d)}{\pi_d(Z_i,X_i)}\right)\Biggr)}_{\text{Part A.4.1}} \\
    &+ \underbrace{\frac{1}{|\mathcal{I}_{11}|} \sum_{i \in \mathcal{I}_{11}} \Biggl(\frac{I(z)}{\lambda_z(W_i)}\Biggl(I(d)(Y_i- \mu_d(Z_i,X_i))\Biggr)\left(\frac{1}{\hat \pi_d(Z_i,X_i)^{\mathcal{I}_{2}}} - \frac{1}{\pi_d(Z_i,X_i)}\right) \Biggr)}_{\text{Part A.4.2}} \\
    &+ \underbrace{\frac{1}{|\mathcal{I}_{11}|} \sum_{i \in \mathcal{I}_{11}} \Biggl(\frac{I(z)}{\lambda_z(W_i)}I(d)\Biggl(\hat \mu_d(Z_i,X_i)^{\mathcal{I}_{2}} - \mu_d(Z_i,X_i)\Biggr)\left(\frac{1}{\hat \pi_d(Z_i,X_i)^{\mathcal{I}_{2}}} - \frac{1}{\pi_d(Z_i,X_i)}\right) \Biggr)}_{\text{Part A.4.3}}
\end{align*}

After conditioning on $I_{2}$, the summands used to build the term are mean-zero and independent. The squared $L_2$-norm of $A.4.1$ looks as follows:
\begin{align*}
    &\E\left[ \left(\frac{1}{|\mathcal{I}_{11}|} \sum_{i \in \mathcal{I}_{11}} \Biggl( \frac{I(z)}{\lambda_z(W_i)}\Biggl(\hat \mu_d(Z_i,X_i)^{\mathcal{I}_{2}} - \mu_d(Z_i,X_i) \Biggr)\left(1 - \frac{I(d)}{\pi_d(Z_i,X_i)}\right)\Biggr)\right)^2\right] \\
   &= \E\left[ \E\left[  \left(\frac{1}{|\mathcal{I}_{11}|} \sum_{i \in \mathcal{I}_{11}} \Biggl( \frac{I(z)}{\lambda_z(W_i)}\Biggl( \hat \mu_d(Z_i,X_i)^{\mathcal{I}_{2}} -  \mu_d(Z_i,X_i) \Biggr)\left(1 - \frac{I(d)}{\pi_d(Z_i,X_i)}\right)\Biggr)\right)^2 \bigg | \mathcal{I}_{2} \right]\right] \\
   &=\E\left[\Var \left[ \left(\frac{1}{|\mathcal{I}_{11}|} \sum_{i \in \mathcal{I}_{11}} \Biggl( \frac{I(z)}{\lambda_z(W_i)}\Biggl(\hat \mu_d(Z_i,X_i)^{\mathcal{I}_{2}} -  \mu_d(Z_i,X_i) \Biggr)\left(1 - \frac{I(d)}{\pi_d(Z_i,X_i)}\right)\Biggr)\right) \bigg | \mathcal{I}_{2} \right]\right] \\
   &=\frac{1}{|\mathcal{I}_{11}|}\E\left[\Var \left[\frac{I(z)}{\lambda_z(W_i)}\Biggl( \hat \mu_d(Z_i,X_i)^{\mathcal{I}_{2}} - \mu_d(Z_i,X_i) \Biggr)\left(1 - \frac{I(d)}{\pi_d(Z_i,X_i)}\right) \bigg | \mathcal{I}_{2} \right]\right] \\
   &=\frac{1}{|\mathcal{I}_{11}|}\E\left[\E \left[\frac{I(z)}{\lambda_z(W_i)^2} \Biggl( \hat \mu_d(Z_i,X_i)^{\mathcal{I}_{2}} - \mu_d(Z_i,X_i) \Biggr)^2\left(1 - \frac{I(d)}{\pi_d(Z_i,X_i)}\right)^2 \bigg | \mathcal{I}_{2} \right]\right] \\
   &\leq \frac{1}{\kappa^3|\mathcal{I}_{11}|} \E \left[\Biggl( \hat \mu_d(Z_i,X_i)^{\mathcal{I}_{2}} - \mu_d(Z_i,X_i) \Biggr)^2 \right] = \frac{o_p(1)}{N}
\end{align*}

The third line follows because the summands are mean-zero and independent. The last line follows from Assumption \ref{assumption_overlap} and \ref{assumption_consistency} and the fact that $|{\mathcal{I}_{11}}| = N/4$. Hence, Part $A.4.1$ is $o_p(1/\sqrt{N})$.

Similarly, using the squared $L_2$-norm of $A.4.2$:
\begin{align*}
    &\E\left[ \left(\frac{1}{|\mathcal{I}_{11}|} \sum_{i \in \mathcal{I}_{11}} \Biggl(\frac{I(z)}{\lambda_z(W_i)}\Biggl(I(d)(Y_i- \mu_d(Z_i,X_i))\Biggr)\left(\frac{1}{\hat \pi_d(Z_i,X_i)^{\mathcal{I}_{2}}} - \frac{1}{\pi_d(Z_i,X_i)}\right) \Biggr)\right)^2\right] \\
    &= \E\left[ \E\left[\left(\frac{1}{|\mathcal{I}_{11}|} \sum_{i \in \mathcal{I}_{11}} \Biggl(\frac{I(z)}{\lambda_z(W_i)}\Biggl(I(d)(Y_i- \mu_d(Z_i,X_i))\Biggr)\left(\frac{1}{\hat \pi_d(Z_i,X_i)^{\mathcal{I}_{2}}} - \frac{1}{\pi_d(Z_i,X_i)}\right) \Biggr)\right)^2 \bigg | \mathcal{I}_{2}\right]\right] \\
    &= \E\left[ \Var \left[\frac{1}{|\mathcal{I}_{11}|} \sum_{i \in \mathcal{I}_{11}} \Biggl(\frac{I(z)}{\lambda_z(W_i)}\Biggl(I(d)(Y_i- \mu_d(Z_i,X_i))\Biggr)\left(\frac{1}{\hat \pi_d(Z_i,X_i)^{\mathcal{I}_{2}}} - \frac{1}{\pi_d(Z_i,X_i)}\right) \Biggr)\bigg | \mathcal{I}_{2} \right]\right] \\
    &= \frac{1}{|\mathcal{I}_{11}|} \E\left[ \Var \left[ \frac{I(z)}{\lambda_z(W_i)}\Biggl(I(d)(Y_i- \mu_d(Z_i,X_i))\Biggr)\left(\frac{1}{\hat \pi_d(Z_i,X_i)^{\mathcal{I}_{2}}} - \frac{1}{\pi_d(Z_i,X_i)}\right) \bigg | \mathcal{I}_{2} \right]\right] \\
    &= \frac{1}{|\mathcal{I}_{11}|} \E\left[\E \left[ \frac{I(z)}{\lambda_z(W_i)^2}\Biggl( I(d)(Y_i- \mu_d(Z_i,X_i))\Biggr)^2\left(\frac{1}{\hat \pi_d(Z_i,X_i)^{\mathcal{I}_{2}}} - \frac{1}{\pi_d(Z_i,X_i)}\right)^2 \bigg | \mathcal{I}_{2} \right]\right] \\
    &\leq \frac{1}{\kappa^2|\mathcal{I}_{11}|}(1-\kappa) \E\left[\E \left[\Biggl((Y_i- \mu_d(Z_i,X_i))\Biggr)^2\left(\frac{1}{\hat \pi_d(Z_i,X_i)^{\mathcal{I}_{2}}} - \frac{1}{\pi_d(Z_i,X_i)}\right)^2 \bigg | \mathcal{I}_{2} \right]\right] \\
    &\leq \frac{1}{\kappa^2|\mathcal{I}_{11}|}(1-\kappa)\epsilon_1 \E\left[\left(\frac{1}{\hat \pi_d(Z_i,X_i)^{\mathcal{I}_{2}}} - \frac{1}{\pi_d(Z_i,X_i)}\right)^2 \right] = \frac{o_p(1)}{N}
\end{align*}

The third line follows from the fact that the summands are mean-zero and independent. The last two inequalitites follow from Assumption \ref{assumption_overlap}, \ref{assumption_consistency} and \ref{assumption_boundness} and the fact that $|\mathcal{I}_{11}| = N/4$. 

Last, using the $L_1$-norm of $A.4.3$:
\begin{align*}
    \E\Biggl[\bigg| &\frac{1}{|\mathcal{I}_{11}|} \sum_{i \in \mathcal{I}_{11}} \Biggl(\frac{I(z)}{\lambda_z(W_i)}I(d)\Biggl(\hat \mu_d(Z_i,X_i)^{\mathcal{I}_{2}} -  \mu_d(Z_i,X_i)\Biggr)\left(\frac{1}{\hat \pi_d(Z_i,X_i)^{\mathcal{I}_{2}}} - \frac{1}{\pi_d(Z_i,X_i)}\right) \Biggr) \bigg| \Biggr] \\
    &\leq \frac{1}{\kappa}\E\Biggl[\bigg|\frac{1}{|\mathcal{I}_{11}|} \sum_{i \in \mathcal{I}_{11}} \Biggl(\hat \mu_d(Z_i,X_i)^{\mathcal{I}_{2}} -  \mu_d(Z_i,X_i)\Biggr)\left(\frac{1}{\hat \pi_d(Z_i,X_i)^{\mathcal{I}_{2}}} - \frac{1}{\pi_d(Z_i,X_i)}\right) \bigg| \Biggr] \\
    &\leq \frac{1}{\kappa} \E\Biggl[\frac{1}{|\mathcal{I}_{11}|} \sum_{i \in \mathcal{I}_{11}} \Biggl( \bigg| \hat \mu_d(Z_i,X_i)^{\mathcal{I}_{2}} - \mu_d(Z_i,X_i)\bigg|  \bigg| \frac{1}{\hat \pi_d(Z_i,X_i)^{\mathcal{I}_{2}}} - \frac{1}{\pi_d(Z_i,X_i)} \bigg|\Biggr) \Biggr] \\
    &= \frac{1}{\kappa} \E\Biggl[\bigg| \hat \mu_d(Z_i,X_i)^{\mathcal{I}_{2}} - \mu_d(Z_i,X_i) \bigg| \bigg|\frac{1}{\hat \pi_d(Z_i,X_i)^{\mathcal{I}_{2}}} - \frac{1}{\pi_d(Z_i,X_i)}\bigg| \Biggr] \\
    &\leq \frac{1}{\kappa} \E\Biggl[\left( \hat \mu_d(Z_i,X_i)^{\mathcal{I}_{2}} -  \mu_d(Z_i,X_i) \right)^2\Biggr]^{1/2} \E\Biggl[\left(\frac{1}{\hat \pi_d(Z_i,X_i)^{\mathcal{I}_{2}}} - \frac{1}{\pi_d(Z_i,X_i)}\right)^2 \Biggr]^{1/2} = \frac{o_p(1)}{\sqrt{N}}
   \end{align*}
The first inequality follows from Assumption \ref{assumption_overlap}, the second inequality follows from Cauchy-Schwarz, the last line from Assumption \ref{assumption_risk_decay} and the last equality from Assumption \ref{assumption_consistency} and the fact that $|\mathcal{I}_{11}| = N/4$. Hence, term $A.4.3$ is $o_p(1/\sqrt{N})$.

Summing up, we have shown that Part $A$ is $o_p(1/ \sqrt{N})$ as long as the product of the estimation errors decays faster than $1/\sqrt{N}$, which is given by Assumption \ref{assumption_risk_decay}.
\begin{align*}
    \frac{1}{|\mathcal{I}_{11}|} \sum_{i \in \mathcal{I}_{11}} \hat{\tau}_{l,m,z}(\hat \delta_{l,m}(H_i)^{\mathcal{I}_2},W_i)^{\mathcal{I}_{11}} - \frac{1}{|\mathcal{I}_{11}|} \sum_{i \in \mathcal{I}_{11}}  \tilde \tau_{l,m,z}( \delta_{l,m}(H_i), W_i)^{\mathcal{I}_{11}}= o_p\left(\frac{1}{\sqrt{N}}\right)	
\end{align*}
Hence, it follows that 
\begin{align*}
    \frac{|\mathcal{I}_{11}|}{N} \hat{\theta}_{l,m,z}^{B, \mathcal{I}_{11}} - \frac{|\mathcal{I}_{11}|}{N} \tilde{\theta}_{l,m,z}^{B, \mathcal{I}_{11}}= o_p\left(\frac{1}{\sqrt{N}}\right)	
\end{align*}

\textbf{Part B} \\
In the next step, the term $B$ is considered. It is the same proof as the usual proof for the average treatment effect \citep{Wager:2020} since we assume that the true pseudo-outcome $\delta_{l,m}(h)$ is known. The term can be rewritten as follows:
\begin{align*}
    &\frac{1}{|\mathcal{I}_{11}|} \sum_{i \in \mathcal{I}_{11}} \Biggl(\tilde g_{l,m,z}(W_i)^{\mathcal{I}_{12}} + \frac{I(z) (\delta_{l,m}(H_i) -\tilde g_{l,m,z}(W_i)^{\mathcal{I}_{12}})}{\hat \lambda_z(W_i)^{\mathcal{I}_{12}}}
    - g_{l,m,z}(W_i) - \frac{I(z) (\delta_{l,m}(H_i) -g_{l,m,z}(W_i))}{\lambda_z(W_i)}\Biggr) \\
    &= \underbrace{\frac{1}{|\mathcal{I}_{11}|} \sum_{i \in \mathcal{I}_{11}} \Biggl( (\tilde g_{l,m,z}(W_i)^{\mathcal{I}_{12}} - g_{l,m,z}(W_i))  \left(1- \frac{I(z)}{\lambda_z(W_i)}\right)\Biggr)}_{\text{Part B.1}} \\
    &\quad + \underbrace{\frac{1}{|\mathcal{I}_{11}|} \sum_{i \in \mathcal{I}_{11}} \Biggl(I(z) (\delta_{l,m}(H_i) - g_{l,m,z}(W_i)) \left(\frac{1}{\hat \lambda_z(W_i)^{\mathcal{I}_{12}}} - \frac{1}{\lambda_z(W_i)}\right) \Biggr)}_{\text{Part B.2}}  \\
    &\quad + \underbrace{\frac{1}{|\mathcal{I}_{11}|} \sum_{i \in \mathcal{I}_{11}} \Biggl(I(z) (\tilde g_{l,m,z}(W_i)^{\mathcal{I}_{12}} - g_{l,m,z}(W_i)) \left(\frac{1}{\lambda_z(W_i)} - \frac{1}{\hat \lambda_z(W_i)^{\mathcal{I}_{12}}}\right)\Biggr)}_{\text{Part B.3}} 
\end{align*}
Still following \cite{Wager:2020}, we can show that all three terms converge to zero in probability. For the term $B.1$, after conditioning on $\mathcal{I}_{12}$, the summands used to build the term are mean-zero and independent. Using the squared $L_2$-norm of $B.1$:

\begin{align*}
    &\E\Biggl[\Biggl(\frac{1}{|\mathcal{I}_{11}|} \sum_{i \in \mathcal{I}_{11}} (\tilde g_{l,m,z}(W_i)^{\mathcal{I}_{12}} - g_{l,m,z}(W_i)) \left(1- \frac{I(z)}{\lambda_z(W_i)}\right)\Biggr)^2 \Biggr] \\
    &= \E\Biggl[\E\Biggl[\Biggl(\frac{1}{|\mathcal{I}_{11}|} \sum_{i \in \mathcal{I}_{11}} (\tilde g_{l,m,z}(W_i)^{\mathcal{I}_{12}} - g_{l,m,z}(W_i)) \left(1- \frac{I(z)}{\lambda_z(W_i)}\right)\Biggr)^2 \bigg | \mathcal{I}_{12} \Biggr] \Biggr] \\
    &= \E\Biggl[\Var\Biggl[ \frac{1}{|\mathcal{I}_{11}|} \sum_{i \in \mathcal{I}_{11}} \Biggl((\tilde g_{l,m,z}(W_i)^{\mathcal{I}_{12}} - g_{l,m,z}(W_i))\left(1- \frac{I(z)}{g_{l,m,z}(W_i)}\right)   \Biggr)\bigg | \mathcal{I}_{12} \Biggr] \Biggr] \\
    &= \frac{1}{|\mathcal{I}_{11}|}\E\Biggl[\Var\Biggl[\left(\tilde g_{l,m,z}(W_i)^{\mathcal{I}_{12}} - g_{l,m,z}(W_i)\right)  \left(1- \frac{I(z)}{\lambda_z(W_i)}\right)\bigg | \mathcal{I}_{12} \Biggr] \Biggr] \\
    &= \frac{1}{|\mathcal{I}_{11}|}\E\Biggl[\E\Biggl[\left(\tilde g_{l,m,z}(W_i)^{\mathcal{I}_{12}} - g_{l,m,z}(W_i)\right)^2  \left(\frac{1}{\lambda_z(W_i)} - 1\right) \bigg | \mathcal{I}_{12} \Biggr] \Biggr] \\
    & \leq \frac{1}{\kappa |\mathcal{I}_{11}|} \E\Biggl[(\tilde g_{l,m,z}(W_i)^{\mathcal{I}_{12}} - g_{l,m,z}(W_i))^2  \Biggr] = \frac{o_p(1)}{N}
\end{align*}
The second equality follows because the summands are mean-zero and independent. The last line follows from Assumption \ref{assumption_overlap} and \ref{assumption_consistency} and the fact that $|\mathcal{I}_{11}| = N/4$. Hence, the term $B.1$ is $o_p(1/\sqrt{N})$.

Similarly, using the squared $L_2$-norm of $B.2$:
\begin{align*}
    &\E\Biggl[\Biggl( \frac{1}{|\mathcal{I}_{11}|} \sum_{i \in \mathcal{I}_{11}} \Biggl(I(z)(\delta_{l,m}(H_i) - g_{l,m,z}(W_i)) \left(\frac{1}{\hat \lambda_z(W_i)^{\mathcal{I}_{12}}} - \frac{1}{\lambda_z(W_i)}\right) \Biggr)\Biggr)^2 \Biggr] \\
    &=\E\Biggl[\E\Biggl[ \Biggl( \frac{1}{|\mathcal{I}_{11}|} \sum_{i \in \mathcal{I}_{11}} \Biggl(I(z) (\delta_{l,m}(H_i) - g_{l,m,z}(W_i)) \left(\frac{1}{\hat \lambda_z(W_i)^{\mathcal{I}_{12}}} - \frac{1}{\lambda_z(W_i)}\right)\Biggr) \Biggr)^2 \bigg | \mathcal{I}_{12} \Biggr]\Biggr] \\
    &=\E\Biggl[\Var\Biggl[\frac{1}{|\mathcal{I}_{11}|} \sum_{i \in \mathcal{I}_{11}} \Biggl(I(z) (\delta_{l,m}(H_i) - g_{l,m,z}(W_i)) \left(\frac{1}{\hat \lambda_z(W_i)^{\mathcal{I}_{12}}} - \frac{1}{\lambda_z(W_i)}\right) \Biggr) \bigg | \mathcal{I}_{12} \Biggr]\Biggr] \\
    &=\frac{1}{|\mathcal{I}_{11}|}\E\Biggl[\Var\Biggl[I(z) \left(\delta_{l,m}(H_i) - g_{l,m,z}(W_i)\right) \left(\frac{1}{\hat \lambda_z(W_i)^{\mathcal{I}_{12}}} - \frac{1}{\lambda_z(W_i)}\right) \bigg | \mathcal{I}_{12} \Biggr]\Biggr] \\
    &=\frac{1}{|\mathcal{I}_{11}|}\E\Biggl[\E\Biggl[I(z)(\delta_{l,m}(H_i) - g_{l,m,z}(W_i))^2 \left(\frac{1}{\hat \lambda_z(W_i)^{\mathcal{I}_{12}}} - \frac{1}{\lambda_z(W_i)}\right)^2 \bigg | \mathcal{I}_{12} \Biggr]\Biggr] \\
    &=\frac{1}{|\mathcal{I}_{11}|}\E\Biggl[I(z)\left(\delta_{l,m}(H_i) - g_{l,m,z}(W_i)\right)^2 \left(\frac{1}{\hat \lambda_z(W_i)^{\mathcal{I}_{12}}} - \frac{1}{\lambda_z(W_i)}\right)^2 \Biggr] \\
    &\leq \frac{1}{|\mathcal{I}_{11}|}(1-\kappa)\E\Biggl[\left(\delta_{l,m}(H_i) - g_{l,m,z}(W_i)\right)^2 \left(\frac{1}{\hat \lambda_z(W_i)^{\mathcal{I}_{12}}} - \frac{1}{\lambda_z(W_i)}\right)^2 \Biggr] \\
    &\leq \frac{1}{|\mathcal{I}_{11}|}(1-\kappa)\epsilon_{z1}\E\Biggl[\left(\frac{1}{\hat \lambda_z(W_i)^{\mathcal{I}_{12}}} - \frac{1}{\lambda_z(W_i)}\right)^2 \Biggr] = \frac{o_p(1)}{N}
\end{align*}
Again, the second equality follows because the summands are mean-zero and independent. The last two inequalities follow from Assumption \ref{assumption_consistency} and \ref{assumption_boundness}, the fact that the MSE for the inverse weights decays at the same rate as the MSE for the propensities and the fact that $|\mathcal{I}_{11}| = N/4$. Hence, the term $B.2$ is $o_p(1/ \sqrt{N})$.

Last, using the $L_1$-norm of $B.3$:
\begin{align*}
 \E\Biggl[\bigg| &\frac{1}{|\mathcal{I}_{11}|} \sum_{i \in \mathcal{I}_{11}} \Biggl( I(z)(\tilde g_{l,m,z}(W_i)^{\mathcal{I}_{12}} - g_{l,m,z}(W_i)) \left(\frac{1}{\lambda_z(W_i)} - \frac{1}{\hat \lambda_z(W_i)^{\mathcal{I}_{12}}}\right)\Biggr) \bigg| \Biggr] \\
 & \leq \E\Biggl[\frac{1}{|\mathcal{I}_{11}|} \sum_{i \in \mathcal{I}_{11}} I(z)\bigg| \tilde g_{l,m,z}(W_i)^{\mathcal{I}_{12}} - g_{l,m,z}(W_i)\bigg| \bigg|\frac{1}{\lambda_z(W_i)} - \frac{1}{\hat \lambda_z(W_i)^{\mathcal{I}_{12}}} \bigg|  \Biggr] \\
 &= \frac{1}{|\mathcal{I}_{11}|} \sum_{i \in \mathcal{I}_{11}} \E\Biggl[ I(z)\bigg| \tilde g_{l,m,z}(W_i)^{\mathcal{I}_{12}} - g_{l,m,z}(W_i)\bigg| \bigg|\frac{1}{\lambda_z(W_i)} - \frac{1}{\hat \lambda_z(W_i)^{\mathcal{I}_{12}}} \bigg|  \Biggr] \\
 &= \E\Biggl[ I(z)\bigg| \tilde g_{l,m,z}(W_i)^{\mathcal{I}_{12}} - g_{l,m,z}(W_i)\bigg| \bigg|\frac{1}{\lambda_z(W_i)} - \frac{1}{\hat \lambda_z(W_i)^{\mathcal{I}_{12}}} \bigg|  \Biggr] \\
 & \leq \E\Biggl[ I(z) (\tilde g_{l,m,z}(W_i)^{\mathcal{I}_{12}} - g_{l,m,z}(W_i) )^2 \Biggr]^{1/2} \E\Biggl[I(z) \left( \frac{1}{\lambda_z(W_i)} - \frac{1}{\hat \lambda_z(W_i)^{\mathcal{I}_{12}}} \right)^2 \Biggr]^{1/2}  \\
 & \leq \E\Biggl[(\tilde g_{l,m,z}(W_i)^{\mathcal{I}_{12}} - g_{l,m,z}(W_i) )^2 \Biggr]^{1/2} \E\Biggl[\left( \frac{1}{\lambda_z(W_i)} - \frac{1}{\hat \lambda_z(W_i)^{\mathcal{I}_{12}}} \right)^2 \Biggr]^{1/2} = \frac{o_p(1)}{\sqrt{N}}
\end{align*}
The first inequality follows from Cauchy-Schwarz, the fourth line from Assumption \ref{assumption_risk_decay} and the last equality from Assumption \ref{assumption_consistency} and the fact that $|\mathcal{I}_{11}| = N/4$. Hence, term $B.2$ is $o_p(1/\sqrt{N})$.

Hence, we have shown that Part $B$ is $o_p(1/ \sqrt{N})$ as long as the product of the estimation errors decays faster than $1/\sqrt{N}$, which is given by Assumption \ref{assumption_risk_decay}. Summing up, in Part B, we have shown that:
\begin{align*}
    \frac{1}{|\mathcal{I}_{11}|} \sum_{i \in \mathcal{I}_{11}} \tilde{\tau}_{l,m,z}(\delta(H_i),W_i)^{\mathcal{I}_{11}} - \frac{1}{|\mathcal{I}_{11}|} \sum_{i \in \mathcal{I}_{11}}  \tau_{l,m,z}^{\star}(\delta(H_i), W_i)^{\mathcal{I}_{11}}= o_p\left(\frac{1}{\sqrt{N}}\right)	
\end{align*}
Hence, it follows that 
\begin{align*}
    \frac{|\mathcal{I}_{11}|}{N}  \tilde{\theta}_{l,m,z}^{B, \mathcal{I}_{11}} - \frac{|\mathcal{I}_{11}|}{N} \theta_{l,m,z}^{B \star, \mathcal{I}_{11}}= o_p\left(\frac{1}{\sqrt{N}}\right)	
\end{align*}

and therefore,
\begin{align*}
    \frac{|\mathcal{I}_{11}|}{N} \hat{\theta}^{B, \mathcal{I}_{11}} - \frac{|\mathcal{I}_{11}|}{N}  \theta^{B \star, \mathcal{I}_{11}}= o_p\left(\frac{1}{\sqrt{N}}\right)	
\end{align*}

Putting all the parts together, we conclude that:
\begin{align*}
    \sqrt{N}(\hat \theta^B - \theta^B) &\\
    &= \underbrace{\sqrt{N} \biggl(\frac{\vert \mathcal{I}_{11} \vert}{N} \hat \theta^{B, \mathcal{I}_{11}} -\frac{\vert \mathcal{I}_{11} \vert}{N} \theta^{B \star, \mathcal{I}_{11}} \biggr)}_{=o_p(1)} + 
    \underbrace{\sqrt{N} \biggl( \frac{\vert \mathcal{I}_{12} \vert}{N} \hat \theta^{B, \mathcal{I}_{12}} - \frac{\vert \mathcal{I}_{12} \vert}{N} \theta^{B \star, \mathcal{I}_{12}} \biggr)}_{=o_p(1)} \\
    &+ \underbrace{\sqrt{N} \biggl(\frac{\vert \mathcal{I}_{21} \vert}{N} \hat \theta^{B, \mathcal{I}_{21}} - \frac{\vert \mathcal{I}_{21} \vert}{N}  \theta^{B \star \mathcal{I}_{21}} \biggr)}_{=o_p(1)}  +
    + \underbrace{\sqrt{N} \biggl( \frac{\vert \mathcal{I}_{22} \vert}{N} \hat \theta^{B, \mathcal{I}_{22}}  - \frac{\vert \mathcal{I}_{22} \vert}{N} \theta^{B \star \mathcal{I}_{22}}\biggr)}_{=o_p(1)}  \\ &+ \underbrace{\sqrt{N} (\theta^{B\star} - \theta^B)}_{\xrightarrow{d} N(0, V^\star)}
\end{align*}

Hence, the estimator $\hat \theta^B$ is $\sqrt{N}$-consistent and asymptotically normal.

\subsection{$\Delta$CBGATE} \label{Appendix_CBGATE_proof}
\subsubsection{Identification Based on the Outcome Regression} \label{_MGATE}

This subsection identifies the $\Delta$CBGATE with the assumptions from Section \ref{2_identifying_assumptions}. Please recall that $\mu_{d}(z,x) = \E[Y_i | D_i = d, Z_i = z, X_i = x]$. The aim is to show that we can estimate the estimand of interest  $\E \left[ \left( Y_i^{m,v} - Y_i^{l,v} \right) - \left( Y_i^{m,u}  - Y_i^{l,u}  \right) \right]$.
\begin{align*}
    \E [( Y_i^{m,v} &- Y_i^{l,v} ) - (Y_i^{m,u}  - Y_i^{l,u} ) ] \\
     = \E \Bigl[&\E \left[ \left( Y_i^{m,v} - Y_i^{l,v} \right) - \left( Y_i^{m,u}  - Y_i^{l,u}  \right) \mid X_i  \right]\Bigr] \\
    =\E \Bigl[&\E \left[ Y_i^{m,v} \mid X_i \right] - \E \left[ Y_i^{l,v} \mid X_i \right]  \nonumber - \E \left[ Y_i^{m,u} \mid X_i \right] + \E \left[ Y_i^{l,u} \mid X_i \right] \Bigr]\\
    =\E \Bigl[&\E \left[ Y_i^{m,v} \mid Z_i = v, X_i \right] - \E \left[ Y_i^{l,v} \mid Z_i = v, X_i \right] - \E \left[ Y_i^{m,u} \mid Z_i = u, X_i \right] + \E \left[ Y_i^{l,u} \mid Z_i = u, X_i \right] \Bigr]\\
    = \E \Bigl[& \E \left[ Y_i^{m,v} \mid D_i = m, Z_i = v,  X_i \right] - \E \left[ Y_i^{l,v} \mid D_i = l, Z_i = v,X_i \right] \nonumber \\  - &\E \left[ Y_i^{m,u} \mid D_i = m, Z_i = u,X_i \right] + \E \left[ Y_i^{l,u} \mid D_i = l, Z_i = u, X_i \right] \Bigr]\\
    = \E\Bigl[&\E[Y_i | D_i = m, Z_i = v, X_i] - \E[Y_i | D_i = l, Z_i = v, X_i] \\-& \E[Y_i | D_i = m, Z_i = u, X_i]  + \E[Y_i | D_i = l, Z_i = u, X_i] \Bigr] \\
    = \E[&\mu_{m}(v,X_i) - \mu_{l}(v,X_i) - \mu_{m}(u,X_i) + \mu_{l}(u,X_i)]
\end{align*}

The first equality follows from the law of iterated expectations, the second from the linearity of expectations and the third from Assumption \ref{assumption_CIA_CBGATE} (a). The fourth equality follows from Assumption \ref{assumption_CIA} (a), and the fifth equality follows the law of total expectation.

\subsubsection{Identification Based on the Doubly Robust Score} \label{DR_Identificaction}

The parameter of interest can also be identified with the doubly robust score function. Please recall that $\omega_{d,z}(x) = P(D_i = d, Z_i = z | X_i = x)$ and $I(d,z) = \mathds{1}(D_i = d \wedge Z_i = z)$. It is enough to show that the average potential outcome $\E[Y_i^{d,z}]$ is identified due to the linearity of expectation property:
\begin{align*}
    \E[Y_i^{d,z}] &= \E\left[\E\left[ \mu_{d}(z,X_i) + \frac{I(d,z)(Y_i - \mu_{d}(z,X_i))}{\omega_{d,z}(X_i)} \vert X_i \right] \right] \\ 
  &= \E \Biggl[\E\Biggl[\mu_{d}(z,X_i) +  \frac{I(d,z)(Y_i - \mu_{d}(z,X_i))}{\omega_{d,z}(X_i)}| D_i = d, Z_i=z, X_i \Biggr]\omega_{d,z}(X_i) \Biggr]\\
     &= \E[\E[\mu_{d}(z,X_i) + Y_i - \mu_{d}(z,X_i)| D_i = d, Z_i=z, X_i]] \\
     &= \E[\E[Y_i| D_i = d, Z_i=z, X_i]] \\
     &= \E[\mu_{d}(z,X_i)]
\end{align*}
The first equality is due to the linearity of expectations, and the second is due to the law of total expectations.

\subsubsection{Neyman Orthogonality} \label{Neyman ortho}

To be able to use machine learning algorithms to estimate the nuisance functions, the score function has to be Neyman orthogonal. This section follows closely \cite{Knaus:2022b}. We use the following APO (average potential outcome) building block to build the double-double robust estimator:
\begin{align*}
    \Gamma_{d,z}(h) = \mu_{d}(z,x) + \frac{I(d,z)(y-\mu_{d}(z,x))}{\omega_{d,z}(x)} 
\end{align*}

The estimate of interest $\theta^{\Delta C}$ = $\Delta$CBGATE can be built from the four different APO's, and this looks as follows:
\begin{equation*}
    \theta^{\Delta C} = \E[\Gamma_{m,v}(H_i) - \Gamma_{l,v}(H_i) - \Gamma_{m,u}(H_i) + \Gamma_{l,u}(H_i)]
\end{equation*}

Let us show that the APO is Neyman-orthogonal. The score looks the following
\begin{align*}
    \E\left[\underbrace{\mu_{d}(z,X_i) + \frac{I(d,z)(Y_i-\mu_{d}(z,X_i))}{\omega_{d,z}(X_i) } - \psi_{d,z}}_{\phi(H_i; \psi_{d,z}, \mu_{d}(z,X_i), \omega_{d,z}(X_i) )} \right] = 0
\end{align*}
with $\psi_{d,z} = \E[\Gamma_{d,z}(H_i)]$.
A score $\phi(h; \psi_{d,z}, \mu_{d}(z,x), \omega_{d,z}(x))$ is Neyman-orthogonal if its Gateaux derivative w.r.t. to the nuisance parameters is in expectation zero at the true nuisance parameters. This means the following:
\begin{equation*}
    \partial_r\E[\phi(H_i; \psi_{d,z}, \mu_{d}(z,X_i) + r(\tilde{\mu}_{d}(z,X_i) - \mu_{d}(z,X_i)), \omega_{d,z}(X_i) + r(\tilde{\omega}_{d,z}(X_i) - \omega_{d,z}(X_i))\mid X_i ]\mid_{r = 0} = 0.
\end{equation*}

To show that the APO is Neyman-orthogonal, we first have to add the perturbation to the nuisance parameters of the score:
\begin{align*}
    \phi(&h; \psi_{d,z}, \mu + r(\tilde{\mu}_{d}(z,x) - \mu_{d}(z,x)), \omega_{d,z}(x) + r(\tilde{\omega}_{d,z}(x)- \omega_{d,z}(x))) \\
    &= (\mu_{d}(z,x) + r(\tilde{\mu}_{d}(z,x) - \mu_{d}(z,x))) + \frac{I(d,z)y - I(d,z)(\mu_{d}(z,x) + r(\tilde{\mu}_{d}(z,x) - \mu_{d}(z,x)))}{\omega_{d,z}(x) + r(\tilde{\omega}_{d,z}(x) - \omega_{d,z}(x))} \\ & \quad - \psi_{d,z}
\end{align*}

In a second step, the conditional expectation is taken
\begin{align*}
    \E&[ \phi(H_i; \psi_{d,z}, \mu_{d}(z,X_i) + r(\tilde{\mu}_{d}(z,X_i) - \mu_{d}(z,X_i)), \omega_{d,z}(X_i) + r(\tilde{\omega}_{d,z}(X_i) - \omega_{d,z}(X_i))) \mid X_i=x ] \\
    &= \E\Biggl[ (\mu_{d}(z,X_i) + r(\tilde{\mu}_{d}(z,X_i) - \mu_{d}(z,X_i))) \\ &\quad + \frac{I(d,z)Y_i - I(d,z)(\mu_{d}(z,X_i) + r(\tilde{\mu}_{d}(z,X_i) - \mu_{d}(z,X_i)))}{\omega_{d,z}(X_i) + r(\tilde{\omega}_{d,z}(X_i) - \omega_{d,z}(X_i))} - \psi_{d,z} \bigg | X_i=x \Biggr] \\
    &= (\mu_{d}(z,X_i) + r(\tilde{\mu}_{d}(z,X_i) - \mu_{d}(z,X_i))) + \E\Biggl[ \frac{I(d,z)Y_i}{\omega_{d,z}(X_i) + r(\tilde{\omega}_{d,z}(X_i) - \omega_{d,z}(X_i))} \bigg | X_i=x \Biggr] \\ & \quad- \E\Biggl[ \frac{I(d,z)(\mu_{d}(z,X_i) + r(\tilde{\mu}_{d}(z,X_i) - \mu_{d}(z,X_i)))}{\omega_{d,z}(X_i) + r(\tilde{\omega}_{d,z}(X_i) - \omega_{d,z}(X_i))} \bigg | X_i =x\Biggr] - \psi_{d,z}\\
    &= (\mu_{d}(z,x) + r(\tilde{\mu}_{d}(z,x) - \mu_{d}(z,x))) + \frac{\mu_{d}(z,x) \omega_{d,z}(x)}{\omega_{d,z}(x) + r(\tilde{\omega}_{d,z}(x) - \omega_{d,z}(x))} \\ & \quad - \frac{\omega_{d,z}(x) (\mu_{d}(z,x) + r(\tilde{\mu}_{d}(z,x) - \mu_{d}(z,x)))}{\omega_{d,z}(x) + r(\tilde{\omega}_{d,z}(x) - \omega_{d,z}(x))} - \psi_{d,z}
\end{align*}

The third equality follows from:
\begin{align*}
    \E[I(d,z)Y_i | X_i =x] &= \E[I(d,z)\sum_d \sum_z I(d,z)Y_i^{d,z} | X_i = x ] \\ &= \E[I(d,z)Y_i^{d,z} | X_i = x]  \\ &= \mu_{d}(z,x) \omega_{d,z}(x) 
\end{align*}

In a third step, the derivative with respect to r is taken:
\begin{align*}
    \partial_r \E[&\phi(H_i; \psi_{d,z}, \mu_{d}(z,X_i) + r(\tilde{\mu}_{d}(z,X_i) - \mu_{d}(z,X_i)), \omega_{d,z}(X_i) + r(\tilde{\omega}_{d,z}(X_i) - \omega_{d,z}(X_i))) \mid X_i = x ] \\
    &= (\tilde{\mu}_{d}(z,x) - \mu_{d}(z,x)) - \frac{\mu_{d}(z,x) \omega_{d,z}(x) (\tilde{\omega}_{d,z}(x) - \omega_{d,z}(x))}{(\omega_{d,z}(x) + r(\tilde{\omega}_{d,z}(x) - \omega_{d,z}(x)))^2} 
    \\&- \frac{\omega_{d,z}(x)(\tilde{\mu}_{d}(z,x) - \mu_{d}(z,x))(\omega_{d,z}(x) + r(\tilde{\omega}_{d,z}(x) - \omega_{d,z}(x)))}{(\omega_{d,z}(x) + r(\tilde{\omega}_{d,z}(x) - \omega_{d,z}(x)))^2}
    \\&- \frac{\omega_{d,z}(x)(\mu_{d}(z,x) - r(\tilde{\mu}_{d}(z,x) - \mu_{d}(z,x)))(\tilde{\omega}_{d,z}(x) - \omega_{d,z}(x)) }{(\omega_{d,z}(x) + r(\tilde{\omega}_{d,z}(x) - \omega_{d,z}(x)))^2}
\end{align*}

Finally, evaluate at the true nuisance values, i.e. set r = 0:
\begin{align*}
    \partial_r \E[&\phi(H_i; \psi_{d,z}, \mu_{d}(z,X_i) + r(\tilde{\mu}_{d}(z,X_i) - \mu_{d}(z,X_i)), \omega_{d,z}(X_i) + r(\tilde{\omega}_{d,z}(X_i) - \omega_{d,z}(X_i))) \mid X_i = x ] \mid_{r = 0}\\
    &= (\tilde{\mu}_{d}(z,x) - \mu_{d}(z,x)) - \frac{\mu_{d}(z,x) \omega_{d,z}(x) (\tilde{\omega}_{d,z}(x) - \omega_{d,z}(x))}{\omega_{d,z}(x)^2} \\ &- \frac{\omega_{d,z}(x)(\tilde{\mu}_{d}(z,x) - \mu_{d}(z,x)) \omega_{d,z}(x) - \omega_{d,z}(x) \mu_{d}(z,x) (\tilde{\omega}_{d,z}(x) - \omega_{d,z}(x)) }{\omega_{d,z}(x)^2} \\
    &= (\tilde{\mu}_{d}(z,x) - \mu_{d}(z,x)) - \frac{\mu_{d}(z,x) \omega_{d,z}(x) (\tilde{\omega}_{d,z}(x) - \omega_{d,z}(x))}{\omega_{d,z}(x)^2} \\ &- \frac{\omega_{d,z}(x)^2(\tilde{\mu}_{d}(z,x) - \mu_{d}(z,x))}{\omega_{d,z}(x)^2} + \frac{\omega_{d,z}(x) \mu_{d}(z,x) (\tilde{\omega}_{d,z}(x) - \omega_{d,z}(x)) }{\omega_{d,z}(x)^2} \\
    &= 0
\end{align*}
\newpage
\subsubsection{Asymptotic Properties of the Estimator with Two Propensity Scores} \label{Asymptotics}

The asymptotic properties of the $\Delta$CBGATE estimator with two propensity scores are investigated in this subsection. The following assumptions are imposed:

\begin{assumption}
    \textnormal{(Overlap)}\label{assumption_overlap_CBGATE}\\ 
    The propensity scores $\lambda_z(w)$ and $\pi_d(z,x)$ are bounded away from 0 and 1:
    \begin{align*}
        \kappa < \lambda_z(x), \pi_d(z,x), \hat \lambda_z(x), \hat \pi_d(z,x) < 1- \kappa \quad \forall x \in \mathcal{X}, z \in \mathcal{Z}, 
    \end{align*}
  for some $\kappa > 0$.  
\end{assumption}

\begin{assumption}
    \textnormal{(Consistency)} \label{assumption_consistency_CBGATE}\\ The estimators of the nuisance functions are sup-norm consistent:
    \begin{align*}
        \sup_{x \in \mathcal{X}, z \in \mathcal{Z}} \vert \hat \mu_d(z,x) - \mu_d(z,x) \vert \xrightarrow{p} 0, \quad 
        \sup_{x \in \mathcal{X}, z \in \mathcal{Z}} \vert \hat \pi_d(z,x) - \pi_d(z,x) \vert \xrightarrow{p} 0, \quad
        \sup_{x \in \mathcal{X}} \vert \hat \lambda_z(x) - \lambda_z(x) \vert \xrightarrow{p} 0 
    \end{align*}
\end{assumption}

\begin{assumption}
    \textnormal{(Risk decay)} \label{assumption_risk_decay_CBGATE}\\ The products of the estimation errors for the outcome and propensity models decays as
    \begin{align*}
        \E\left[(\hat \mu_d(Z_i,X_i) - \mu_d(Z_i,X_i) )^2\right] \E\left[(\hat \pi_d(Z_i,X_i) - \pi_d(Z_i,X_i) )^2\right] = o_p\left(\frac{1}{N}\right) \\
        \E\left[(\hat \mu_d(Z_i,X_i) - \mu_d(Z_i,X_i) )^2\right] \E\left[(\hat \lambda_z(X_i) - \lambda_z(X_i) )^2\right] = o_p\left(\frac{1}{N}\right)
    \end{align*}
    If both nuisance parameters are estimated with the parametric ($\sqrt{N}$-consistent) rate, then the product of the errors would be bounded by $O_p\left(\frac{1}{N^2}\right)$. Hence, it is sufficient for the estimators of the nuisance parameters to be $N^{1/4}$-consistent.
\end{assumption}

\begin{assumption}
    \textnormal{(Boundness of conditional variances)} \label{assumption_boundness_CBGATE}\\
    The conditional variances of the outcome is bounded:
    \begin{align*}
        \sup_{x \in \mathcal{X}, z \in \mathcal{Z}} \Var(Y_i| D_i  = d, Z_i = z, X_i = x) < \epsilon_{d} < \infty
    \end{align*}
\end{assumption}
\defcitealias{Chernozhukov:2018}{Chernozhukov et al., 2018}  
The assumptions made are standard in the DML literature \citepalias{Chernozhukov:2018}. Given these assumptions, the following Theorem can be derived:

\begin{theorem}\label{theorem_asymptotics_CBGATE}
    Under Assumptions \ref{assumption_overlap_CBGATE} to \ref{assumption_boundness_CBGATE}, the proposed estimation strategy for the $\Delta$BGATE obeys $\sqrt{N}(\hat \theta_{l,m,u,v}^{\Delta C} - \theta_{l,m,u,v}^{\Delta C}) \xrightarrow{d} N(0, V^\star)$ with $V^\star = \E[\phi^{\Delta C}(H_i; \theta^{\Delta C}, \hat\eta)^2]$.
\end{theorem}

It follows from Theorem \ref{theorem_asymptotics_CBGATE} that the estimator is $\sqrt N$-consistent and asymptotically normal. The proof looks as follows:

In a first step, define the following terms for easier readability:
\begin{align*}
    \pi_d(z,x) &= P(D_i=d | Z_i = z, X_i = x) \\
    \mu_d(z,x) &= \E[Y_i | D_i = d, Z_i = z, X_i = x] \\
    \lambda_z(x) &= P(Z_i=z | X_i = x) \\
    \zeta_{d,z} &= \E\left[\mu_d(z,x) + \frac{I(d,z) (y -\mu_d(z,x))}{\pi_d(z,x) \lambda_z(x)}\right] \\
    \Gamma_{d,z}(h) &= \mu_d(z,x) + \frac{I(d,z) (y -\mu_d(z,x))}{\pi_d(z,x) \lambda_z(x)} \\
    \theta_{l,m,u,v}^{\Delta C} &= \E[\Gamma_{l,u}(H_i) - \Gamma_{m,u}(H_i) -\Gamma_{l,v}(H_i) + \Gamma_{m,v}(H_i)]
\end{align*} 
Due to the linearity in expectations it is enough to focus on $\zeta_{d,z}$, hence we want to show that
\begin{align*}
    \sqrt{N}(\hat \zeta_{d,z} - \zeta^{\star}_{d,z}) \xrightarrow{d} N(0, V^\star)\quad \text{with} \quad
    V^\star = \E\left[ \left(\hat \Gamma_{d,z}(H_i) - \zeta_{d,z}\right)^2\right]
\end{align*}
with $\zeta^{\star}_{d,z}$ being an oracle estimator of $\zeta_{d,z}$ if all nuisance functions would be known. Then $\zeta^{\star}_{d,z}$ is an i.i.d. average, hence:
\begin{align*}
    \sqrt{N}(\zeta^{\star}_{d,z} - \zeta_{d,z}) \xrightarrow{d} N(0, V^\star) \quad \text{with} \quad
    V^\star = \E\left[\left(\hat \Gamma_{d,z}(H_i) - \zeta_{d,z} \right)^2\right]
\end{align*}

\begin{theorem}
    Let $\mathcal{I}_1$ and $\mathcal{I}_2$ be two half samples such that $\vert \mathcal{I}_1 \vert = \vert \mathcal{I}_1 \vert = \frac{N}{2} $. Define the estimator as follows:
    \begin{align*}
        \hat \zeta_{d,z} &= \frac{\vert \mathcal{I}_{1} \vert}{N} \hat \zeta^{\mathcal{I}_{1}}_{d,z} + \frac{\vert \mathcal{I}_{2} \vert}{N} \hat \zeta^{\mathcal{I}_{2}}_{d,z} \\
        \hat \zeta_{d,z}^{\mathcal{I}_{1}} &= \frac{1}{\vert \mathcal{I}_{1} \vert} \sum_{\mathcal{I}_{1}} \hat \Gamma_{d,z}^{\mathcal{I}_{2}}(H_i) \\
        \hat \zeta_{d,z}^{\mathcal{I}_{2}} &= \frac{1}{\vert \mathcal{I}_{2} \vert} \sum_{\mathcal{I}_{2}} \hat \Gamma_{d,z}^{\mathcal{I}_{1}}(H_i)
    \end{align*}
Then, if Assumptions 5 to 8 it hold, it follows that $\sqrt{N}(\hat \zeta_{d,z} - \zeta_{d,z}) \xrightarrow{d} N(0, V^\star)$ with $ V^\star = \E\left[ \left( \hat \Gamma_{d,z}(H_i) - \zeta_{d,z}\right)^2\right]$.

\end{theorem}
\textit{Proof}.
\begin{align*}
    \sqrt{N}(\hat \zeta_{d,z} &- \zeta_{d,z}) \\ &= \sqrt{N}(\hat \zeta_{d,z} - \zeta_{d,z}^{\star} + \zeta_{d,z}^{\star} - \zeta_{d,z}) \\
    &= \sqrt{N}(\hat \zeta_{d,z} - \zeta_{d,z}^{\star}) + \sqrt{N}(\zeta_{d,z}^{\star} - \zeta_{d,z}) \\
    &= \sqrt{N}\left(\frac{\vert \mathcal{I}_{1} \vert}{N} \hat \zeta_{d,z}^{\mathcal{I}_{1}} + \frac{\vert \mathcal{I}_{2} \vert}{N} \hat \zeta_{d,z}^{\mathcal{I}_{2}} - \zeta_{d,z}^{\star}\right) + \sqrt{N}(\zeta_{d,z}^{\star} - \zeta_{d,z}) \\
    &= \sqrt{N} \biggl(\frac{\vert \mathcal{I}_{1} \vert}{N} \hat \zeta_{d,z}^{ \mathcal{I}_{1}} + \frac{\vert \mathcal{I}_{2} \vert}{N} \hat \zeta_{d,z}^{ \mathcal{I}_{2}} -\frac{\vert \mathcal{I}_{1} \vert}{N} \zeta_{d,z}^{\star, \mathcal{I}_{1}} - \frac{\vert \mathcal{I}_{2} \vert}{N} \zeta_{d,z}^{\star, \mathcal{I}_{2}} \biggr) + \sqrt{N} (\zeta_{d,z}^{\star} - \zeta_{d,z}) \\
    &= \sqrt{N} \biggl(\frac{\vert \mathcal{I}_{1} \vert}{N} \hat \zeta_{d,z}^{ \mathcal{I}_{1}} -\frac{\vert \mathcal{I}_{1} \vert}{N} \zeta_{d,z}^{\star, \mathcal{I}_{1}} \biggr) + 
    \sqrt{N} \biggl( \frac{\vert \mathcal{I}_{2} \vert}{N} \hat \zeta_{d,z}^{\mathcal{I}_{2}} - \frac{\vert \mathcal{I}_{2} \vert}{N}\zeta_{d,z}^{\star, \mathcal{I}_{2}} \biggr) + \sqrt{N} (\zeta_{d,z}^{\star} - \zeta_{d,z})
\end{align*}
The goal is to show that the first two terms converge to zero in probability. It is enough to show this for the first term only, as the same steps can also be directly applied to the remaining second term.

Hence,
\begin{align*}
     &\hat \zeta_{d,z}^{\mathcal{I}_{1}} - \zeta_{d,z}^{\star, \mathcal{I}_{1}}
    \\ &= \frac{1}{\vert \mathcal{I}_{1} \vert}\sum_{i \in \mathcal{I}_{1}} \Biggl(\hat \mu_d(Z_i,X_i)^{\mathcal{I}_{2}} - \mu_d(Z_i,X_i) + \frac{I(d,z) (Y_i - \hat \mu_d(Z_i,X_i)^{\mathcal{I}_{2}})}{\hat \pi_d(Z_i,X_i)^{\mathcal{I}_{2}} \hat \lambda_z(X_i)^{\mathcal{I}_{2}}} - \frac{I(d,z) (Y_i - \mu_d(Z_i,X_i))}{\pi_d(Z_i,X_i) \lambda_z(X_i)} \Biggr) \\
    &= \underbrace{\frac{1}{\vert \mathcal{I}_{1} \vert}\sum_{i \in \mathcal{I}_{1}} \Biggl( \biggl(\hat \mu_d(Z_i,X_i)^{\mathcal{I}_{2}} - \mu_d(Z_i,X_i)\biggr) \biggl(1- \frac{I(d,z)}{\pi_d(Z_i,X_i) \lambda_z(X_i)} \biggr)\Biggr)}_{\text{Part 1}} \\
    & \quad + \underbrace{\frac{1}{\vert \mathcal{I}_{1} \vert}\sum_{i \in \mathcal{I}_{1}} \Biggl(I(d,z)(Y_i-\mu_d(Z_i,X_i)) \frac{1}{\hat \pi_d (Z_i,X_i)^{\mathcal{I}_{2}} \hat \lambda_z(X_i)^{\mathcal{I}_{2}} \pi_d(Z_i,X_i)}(\pi_d(Z_i,X_i) -  \hat \pi_d(Z_i,X_i)^{\mathcal{I}_{2}})\Biggr)}_{\text{Part 2}} \\
    &\quad +\underbrace{\frac{1}{\vert \mathcal{I}_{1} \vert}\sum_{i \in \mathcal{I}_{1}} \Biggl( I(d,z)(Y_i-\mu_d(Z_i,X_i)) \frac{1}{ \pi_d (Z_i,X_i) \hat \lambda_z(X_i)^{\mathcal{I}_{2}} \lambda_z(X_i)}(\lambda_z(X_i) -  \hat \lambda_z(X_i)^{\mathcal{I}_{2}})\Biggr)}_{\text{Part 3}} \\
    & \quad + \underbrace{\frac{1}{\vert \mathcal{I}_{1} \vert}\sum_{i \in \mathcal{I}_{1}} \Biggl( I(d,z)(\hat \mu_d(Z_i,X_i)^{\mathcal{I}_{2}}  -\mu_d(Z_i,X_i)) \left(\frac{1}{ \pi_d (Z_i,X_i) \lambda_z(X_i)} - \frac{1}{ \hat \pi_d (Z_i,X_i)^{\mathcal{I}_{2}} \hat \lambda_z(X_i)^{\mathcal{I}_{2}}} \right)\Biggr)}_{\text{Part 4}}
\end{align*}

The proof is based on \citet{Wager:2020}. All four terms converge to zero in probability. For Part $1$, after conditioning on $\mathcal{I}_{2}$, the summands used to build the term are mean-zero and independent. Using the squared $L_2$-norm of Part $1$:
\begin{align*}
    &\E\Biggl[\Biggl(\frac{1}{\vert \mathcal{I}_{1} \vert}\sum_{i \in \mathcal{I}_{1}} \Biggl( \biggl(\hat \mu_d(Z_i,X_i)^{\mathcal{I}_{2}} - \mu_d(Z_i,X_i)\biggr) \biggl(1- \frac{I(d,z)}{\pi_d(Z_i,X_i) \lambda_z(X_i)} \biggr)\Biggr)\Biggr)^2 \Biggr] \\
    &= \E\Biggl[\E\Biggl[\Biggl(\frac{1}{\vert \mathcal{I}_{1} \vert}\sum_{i \in \mathcal{I}_{1}} \Biggl( \biggl(\hat \mu_d(Z_i,X_i)^{\mathcal{I}_{2}} - \mu_d(Z_i,X_i)\biggr) \biggl(1- \frac{I(d,z)}{\pi_d(Z_i,X_i) \lambda_z(X_i)} \biggr)\Biggr)\Biggr)^2 \bigg | \mathcal{I}_{2} \Biggr] \Biggr] \\
    &= \E\Biggl[\Var\Biggl[\frac{1}{|\mathcal{I}_{1}|} \sum_{i \in \mathcal{I}_{1}} \Biggl(\hat \mu_d(Z_i,X_i)^{\mathcal{I}_{2}} - \mu_d(Z_i,X_i)\biggr) \biggl(1- \frac{I(d,z)}{\pi_d(Z_i,X_i) \lambda_z(X_i)}  \Biggr)\bigg | \mathcal{I}_{2} \Biggr] \Biggr] \\
    &= \frac{1}{|\mathcal{I}_{1}|}\E\Biggl[\Var\Biggl[\Biggl(\hat \mu_d(Z_i,X_i)^{\mathcal{I}_{2}} - \mu_d(Z_i,X_i)\biggr) \biggl(1- \frac{I(d,z)}{\pi_d(Z_i,X_i) \lambda_z(X_i)}\Biggr)\bigg | \mathcal{I}_{2} \Biggr] \Biggr] \\
    &= \frac{1}{|\mathcal{I}_{1}|}\E\Biggl[\E\Biggl[\Biggl(\hat \mu_d(Z_i,X_i)^{\mathcal{I}_{2}} - \mu_d(Z_i,X_i)\biggr)^2 \biggl(\frac{1}{\pi_d(Z_i,X_i) \lambda_z(X_i)} -1   \Biggr)\bigg | \mathcal{I}_{2} \Biggr] \Biggr] \\
    & \leq \frac{1}{\kappa^2 |\mathcal{I}_{1}|} \E\Biggl[\Biggl(\hat \mu_d(Z_i,X_i)^{\mathcal{I}_{2}} - \mu_d(Z_i,X_i)\biggr)^2  \Biggr] = \frac{o_p(1)}{N}
\end{align*}
The second equality follows because the summands are mean-zero and independent. The last line follows from Assumption \ref{assumption_overlap_CBGATE} and \ref{assumption_consistency_CBGATE} and the fact that $|\mathcal{I}_{1}| = N/2$. Hence, Part $1$ is $o_p(1/\sqrt{N})$.

Similarly, using the squared $L_2$-norm of Part $2$:
\begin{align*}
    &\E\Biggl[\Biggl( \frac{1}{\vert \mathcal{I}_{1} \vert}\sum_{i \in \mathcal{I}_{1}} \Biggl(  \frac{I(d,z)(Y_i-\mu_d(Z_i,X_i))}{\hat \pi_d (Z_i,X_i)^{\mathcal{I}_{2}} \hat \lambda_z(X_i)^{\mathcal{I}_{2}} \pi_d(Z_i,X_i)}(\pi_d(Z_i,X_i) -  \hat \pi_d(Z_i,X_i)^{\mathcal{I}_{2}})\Biggr)\Biggr)^2 \Biggr] \\
    &=\E\Biggl[\E\Biggl[ \Biggl( \frac{1}{\vert \mathcal{I}_{1} \vert}\sum_{i \in \mathcal{I}_{1}} \Biggl(  \frac{I(d,z)(Y_i-\mu_d(Z_i,X_i))}{\hat \pi_d (Z_i,X_i)^{\mathcal{I}_{2}} \hat \lambda_z(X_i)^{\mathcal{I}_{2}} \pi_d(Z_i,X_i)}(\pi_d(Z_i,X_i) -  \hat \pi_d(Z_i,X_i)^{\mathcal{I}_{2}})\Biggr) \Biggr)^2 \bigg | I_2 \Biggr]\Biggr] \\
    &=\E\Biggl[\Var\Biggl[\frac{1}{|\mathcal{I}_{1}|} \sum_{i \in \mathcal{I}_{1}} \Biggl(\frac{I(d,z)(Y_i-\mu_d(Z_i,X_i)) }{\hat \pi_d (Z_i,X_i)^{\mathcal{I}_{2}} \hat \lambda_z(X_i)^{\mathcal{I}_{2}} \pi_d(Z_i,X_i)}(\pi_d(Z_i,X_i) -  \hat \pi_d(Z_i,X_i)^{\mathcal{I}_{2}}) \Biggr) \bigg | \mathcal{I}_{2} \Biggr]\Biggr] \\
    &=\frac{1}{|\mathcal{I}_{1}|}\E\Biggl[\Var\Biggl[ \frac{I(d,z)(Y_i-\mu_d(Z_i,X_i))}{\hat \pi_d (Z_i,X_i)^{\mathcal{I}_{2}} \hat \lambda_z(X_i)^{\mathcal{I}_{2}} \pi_d(Z_i,X_i)}(\pi_d(Z_i,X_i) -  \hat \pi_d(Z_i,X_i)^{\mathcal{I}_{2}}) \bigg | \mathcal{I}_{2} \Biggr]\Biggr] \\
    &=\frac{1}{|\mathcal{I}_{1}|}\E\Biggl[\E\Biggl[ \Biggl(\frac{I(d,z)(Y_i-\mu_d(Z_i,X_i))}{\hat \pi_d (Z_i,X_i)^{\mathcal{I}_{2}} \hat \lambda_z(X_i)^{\mathcal{I}_{2}} \pi_d(Z_i,X_i)}(\pi_d(Z_i,X_i) -  \hat \pi_d(Z_i,X_i)^{\mathcal{I}_{2}})  \Biggr)^2 \bigg |\mathcal{I}_{2} \Biggr]\Biggr] \\
    &\leq \frac{1}{\kappa^3 |\mathcal{I}_{1}|} (1-\kappa)^2\E\Biggl[\E\Biggl[(Y_i-\mu_d(Z_i,X_i))^2(\pi_d(Z_i,X_i) -  \hat \pi_d(Z_i,X_i)^{\mathcal{I}_{2}})^2\bigg | \mathcal{I}_{2} \Biggr]\Biggr] \\
    &= \frac{1}{\kappa^3 |\mathcal{I}_{1}|} (1-\kappa)^2\E\Biggl[(Y_i-\mu_d(Z_i,X_i))^2(\pi_d(Z_i,X_i) -  \hat \pi_d(Z_i,X_i)^{\mathcal{I}_{2}})^2 \Biggr] \\
    &\leq \frac{1}{\kappa^3 |\mathcal{I}_{1}|} (1-\kappa)^2\epsilon_d\E\Biggl[(\pi_d(Z_i,X_i) -  \hat \pi_d(Z_i,X_i)^{\mathcal{I}_{2}})^2 \Biggr] = \frac{o_p(1)}{N}
\end{align*}
Again, the second equality follows because the summands are mean-zero and independent. The last two inequalities follow from Assumption \ref{assumption_consistency_CBGATE} and \ref{assumption_boundness_CBGATE}, the fact that the MSE for the inverse weights decays at the same rate as the MSE for the propensities and the fact that $|\mathcal{I}_{1}| = N/2$. Hence, Part $2$ is $o_p(1/ \sqrt{N})$.

Similarly, using the squared $L_2$-norm of Part $3$:
\begin{align*}
    &\E\Biggl[\Biggl( \frac{1}{\vert \mathcal{I}_{1} \vert}\sum_{i \in \mathcal{I}_{1}} \Biggl(\frac{I(d,z)(Y_i-\mu_d(Z_i,X_i))}{ \pi_d (Z_i,X_i) \hat \lambda_z(X_i)^{\mathcal{I}_{2}} \lambda_z(X_i)}(\lambda_z(X_i) -  \hat \lambda_z(X_i)^{\mathcal{I}_{2}})\Biggr)\Biggr)^2 \Biggr] \\
    &=\E\Biggl[\E\Biggl[ \Biggl( \frac{1}{\vert \mathcal{I}_{1} \vert}\sum_{i \in \mathcal{I}_{1}} \Biggl(\frac{I(d,z)(Y_i-\mu_d(Z_i,X_i))}{ \pi_d (Z_i,X_i) \hat \lambda_z(X_i)^{\mathcal{I}_{2}} \lambda_z(X_i)}(\lambda_z(X_i) -  \hat \lambda_z(X_i)^{\mathcal{I}_{2}})\Biggr) \Biggr)^2 \bigg | I_2 \Biggr]\Biggr] \\
    &=\E\Biggl[\Var\Biggl[\frac{1}{|\mathcal{I}_{1}|} \sum_{i \in \mathcal{I}_{1}} \Biggl(\frac{I(d,z)(Y_i-\mu_d(Z_i,X_i))}{ \pi_d (Z_i,X_i) \hat \lambda_z(X_i)^{\mathcal{I}_{2}} \lambda_z(X_i)}(\lambda_z(X_i) -  \hat \lambda_z(X_i)^{\mathcal{I}_{2}})\Biggr) \bigg | \mathcal{I}_{2} \Biggr]\Biggr] \\
    &=\frac{1}{|\mathcal{I}_{1}|}\E\Biggl[\Var\Biggl[\frac{I(d,z)(Y_i-\mu_d(Z_i,X_i))}{ \pi_d (Z_i,X_i) \hat \lambda_z(X_i)^{\mathcal{I}_{2}} \lambda_z(X_i)}(\lambda_z(X_i) -  \hat \lambda_z(X_i)^{\mathcal{I}_{2}})\bigg | \mathcal{I}_{2} \Biggr]\Biggr] \\
    &=\frac{1}{|\mathcal{I}_{1}|}\E\Biggl[\E\Biggl[ \Biggl(\frac{I(d,z)(Y_i-\mu_d(Z_i,X_i))}{ \pi_d (Z_i,X_i) \hat \lambda_z(X_i)^{\mathcal{I}_{2}} \lambda_z(X_i)}(\lambda_z(X_i) -  \hat \lambda_z(X_i)^{\mathcal{I}_{2}})  \Biggr)^2 \bigg | \mathcal{I}_{2} \Biggr]\Biggr] \\
    &\leq \frac{1}{\kappa^3 |\mathcal{I}_{1}|} (1-\kappa)^2\E\Biggl[\E\Biggl[(Y_i-\mu_d(Z_i,X_i))^2(\lambda_z(X_i) -  \hat \lambda_z(X_i)^{\mathcal{I}_{2}})^2 \bigg | \mathcal{I}_{2} \Biggr]\Biggr] \\
    &= \frac{1}{\kappa^3 |\mathcal{I}_{1}|} (1-\kappa)^2\E\Biggl[(Y_i-\mu_d(Z_i,X_i))^2(\lambda_z(X_i) -  \hat \lambda_z(X_i)^{\mathcal{I}_{2}})^2 \Biggr] \\
    &\leq \frac{1}{\kappa^3 |\mathcal{I}_{1}|} (1-\kappa)^2\epsilon_d\E\Biggl[(\lambda_z(X_i) -  \hat \lambda_z(X_i)^{\mathcal{I}_{2}})^2 \Biggr] = \frac{o_p(1)}{N}
\end{align*}
Again, the second equality follows because the summands are mean-zero and independent. The last two inequalities follow from Assumption \ref{assumption_consistency_CBGATE} and \ref{assumption_boundness_CBGATE}, the fact that the MSE for the inverse weights decays at the same rate as the MSE for the propensities and the fact that $|\mathcal{I}_{1}| = N/2$. Hence, Part $3$ is $o_p(1/ \sqrt{N})$.

Last, using the $L_1$-norm of Part $4$:
\begin{align*}
 &\E\Biggl[\bigg| \frac{1}{\vert \mathcal{I}_{1} \vert}\sum_{i \in \mathcal{I}_{1}} \Biggl( I(d,z)(\hat \mu_d(Z_i,X_i)^{\mathcal{I}_{2}}  -\mu_d(Z_i,X_i)) \left(\frac{1}{ \pi_d (Z_i,X_i) \lambda_z(X_i)} - \frac{1}{ \hat \pi_d (Z_i,X_i)^{\mathcal{I}_{2}} \hat \lambda_z(X_i)^{\mathcal{I}_{2}}} \right)\Biggr) \bigg| \Biggr] \\
 =  &\E\Biggl[\bigg| \frac{1}{\vert \mathcal{I}_{1} \vert}\sum_{i \in \mathcal{I}_{1}} \Biggl( \frac{I(d,z)}{\pi_d(Z_i,X_i) \hat \lambda_z(X_i)^{\mathcal{I}_{2}}}(\hat \mu_d(Z_i,X_i)^{\mathcal{I}_{2}}  -\mu_d(Z_i,X_i)) \\ & \quad \left(\hat \lambda_z(X_i)^{\mathcal{I}_{2}} - \lambda_z(X_i) + \hat \pi_d(Z_i,X_i)^{\mathcal{I}_{2}} - \pi_d(Z_i,X_i)\right)\Biggr) \bigg| \Biggr] \\
 \leq  &\E\Biggl[ \frac{1}{\vert \mathcal{I}_{1} \vert}\sum_{i \in \mathcal{I}_{1}} \bigg| \frac{I(d,z)}{\pi_d(Z_i,X_i) \hat \lambda_z(X_i)^{\mathcal{I}_{2}}}\bigg|\bigg|\hat \mu_d(Z_i,X_i)^{\mathcal{I}_{2}}  -\mu_d(Z_i,X_i)\bigg| \\ & \quad \bigg|\hat \lambda_z(X_i)^{\mathcal{I}_{2}} - \lambda_z(X_i) + \hat \pi_d(Z_i,X_i)^{\mathcal{I}_{2}} - \pi_d(Z_i,X_i)\bigg| \Biggr] \\
=  &\frac{1}{|\mathcal{I}_{1}|} \sum_{i \in \mathcal{I}_{1}} \E\Biggl[ \bigg| \frac{I(d,z)}{\pi_d(Z_i,X_i) \hat \lambda_z(X_i)^{\mathcal{I}_{2}}}\bigg| \bigg|\hat \mu_d(Z_i,X_i)^{\mathcal{I}_{2}}  -\mu_d(Z_i,X_i)\bigg| \\ & \qquad \qquad \quad  \bigg|\hat \lambda_z(X_i)^{\mathcal{I}_{2}} - \lambda_z(X_i) + \hat \pi_d(Z_i,X_i)^{\mathcal{I}_{2}} - \pi_d(Z_i,X_i)\bigg|  \Biggr] \\
 \leq &\frac{1}{|\mathcal{I}_{1}|} \sum_{i \in \mathcal{I}_{1}} \E\Biggl[ \bigg| \frac{I(d,z)}{\pi_d(Z_i,X_i) \hat \lambda_z(X_i)^{\mathcal{I}_{2}}}\bigg|\bigg|\hat \mu_d(Z_i,X_i)^{\mathcal{I}_{2}}  -\mu_d(Z_i,X_i)\bigg| \\ & \qquad \qquad \quad \left( \bigg|\hat \lambda_z(X_i)^{\mathcal{I}_{2}} - \lambda_z(X_i) \bigg| + \bigg|\hat \pi_d(Z_i,X_i)^{\mathcal{I}_{2}} - \pi_d(Z_i,X_i)\bigg| \right) \Biggr] \\
 =  &\E\Biggl[ \bigg| \frac{I(d,z)}{\pi_d(Z_i,X_i) \hat \lambda_z(X_i)^{\mathcal{I}_{2}}}\bigg|\bigg|\hat \mu_d(Z_i,X_i)^{\mathcal{I}_{2}}  -\mu_d(Z_i,X_i)\bigg| \\ & \quad \left(\bigg|\hat \lambda_z(X_i)^{\mathcal{I}_{2}} - \lambda_z(X_i) \bigg| + \bigg|\hat \pi_d(Z_i,X_i)^{\mathcal{I}_{2}} - \pi_d(Z_i,X_i)\bigg| \right) \Biggr] \\
\leq  &\frac{(1-\kappa)^2}{\kappa^2}  \E\Biggl[\bigg|\hat \mu_d(Z_i,X_i)^{\mathcal{I}_{2}}  -\mu_d(Z_i,X_i)\bigg| \left(\bigg|\hat \lambda_z(X_i)^{\mathcal{I}_{2}} - \lambda_z(X_i) \bigg| + \bigg|\hat \pi_d(Z_i,X_i)^{\mathcal{I}_{2}} - \pi_d(Z_i,X_i)\bigg| \right) \Biggr] \\
 \leq &\E\Biggl[ \left(\hat \mu_d(Z_i,X_i)^{\mathcal{I}_{2}}  -\mu_d(Z_i,X_i)\right)^2 \Biggr]^{1/2}\\ & \quad  \left( \E\Biggl[\left(\hat \lambda_z(X_i)^{\mathcal{I}_{2}} - \lambda_z(X_i)\right)^2\Biggr]^{1/2} + \E\Biggl[\left(\hat \pi_d(Z_i,X_i)^{\mathcal{I}_{2}} - \pi_d(Z_i,X_i)\right)^2 \Biggr]^{1/2}\right) \\ &= \frac{o_p(1)}{\sqrt{N}}
\end{align*}
The first inequality follows from Cauchy-Schwarz, the fourth line from the triangle inequality, the sixth line from Assumption \ref{assumption_overlap_CBGATE} and the last equality from Assumption \ref{assumption_consistency_CBGATE} and \ref{assumption_risk_decay_CBGATE} and the fact that $|\mathcal{I}_{1}| = N/2$. Hence, Part $4$ is $o_p(1/\sqrt{N})$.

Hence, we have shown that:
\begin{align*}
    \frac{1}{|\mathcal{I}_{1}|} \sum_{i \in \mathcal{I}_{1}} \hat{\zeta}_{d,z}^{\mathcal{I}_{1}} - \frac{1}{|\mathcal{I}_{1}|} \sum_{i \in \mathcal{I}_{1}}  \zeta_{d,z}^{\star \mathcal{I}_{1}}= o_p\left(\frac{1}{\sqrt{N}}\right)	
\end{align*}

Putting all the parts together, we can conclude that:
\begin{align*}
    \sqrt{N}(\hat \zeta_{d,z} - \zeta_{d,z}) &\\
    &= \underbrace{\sqrt{N} \biggl(\frac{\vert \mathcal{I}_{11} \vert}{N} \hat \zeta_{d,z}^{\mathcal{I}_{1}} -\frac{\vert \mathcal{I}_{1} \vert}{N} \zeta_{d,z}^{\star, \mathcal{I}_{1}} \biggr)}_{=o_p(1)} + 
    \underbrace{\sqrt{N} \biggl( \frac{\vert \mathcal{I}_{2} \vert}{N} \hat \zeta_{d,z}^{\mathcal{I}_{2}} - \frac{\vert \mathcal{I}_{2} \vert}{N} \zeta_{d,z}^{\star, \mathcal{I}_{2}} \biggr)}_{=o_p(1)} + \underbrace{\sqrt{N} (\zeta_{d,z}^{\star} - \zeta_{d,z})}_{\xrightarrow{d} N(0, V^\star)}
\end{align*}

Hence, the estimator $\hat \theta_{l,m,u,v}^{\Delta C}$ is $\sqrt{N}$-consistent and asymptotically normal.

\newpage
\section{Appendix: Estimation Procedures} \label{Estimation_procedures}

\subsection{Algorithms}
Algorithm \ref{alg:BGATE-Learner_ap} shows the DML algorithm to estimate the $\Delta$BGATE. 

\begin{algorithm}[H]
    \setstretch{1}
\caption{\textsc{DML for $\Delta$BGATE}}\label{alg:BGATE-Learner_ap}
\SetKwInOut{Input}{Input}
\SetKwInOut{Output}{Output}
\SetKwBlock{Beginn}{begin}{end}
\Input{Data: $h_i = \{x_i, z_i, d_i, y_i\}$} 
\Output{$\Delta$BGATE = $\hat{\theta}_{l,m,u,v}^{\Delta B}$}
\Beginn{
\textbf{create folds:} Split sample into K random folds $(S_k)_{k=1}^K$ of observations $\{1, \dots, \frac{N}{K}\}$ with size of each fold $\frac{N}{K}$. Define $S_k^c \coloneqq \{1, \dots, N\} \backslash \{S_k\}$\\
\For{k \textnormal{in} $\{1, \dots,K\}$}{
    \For{d \textnormal{in} $\{l, m\}$}{
        \textsc{Response Functions}: \\
        estimate: $\hat \mu_d(z_i,x_i) = \hat \E[Y_i |D_i = d_i, X_i = x_i, Z_i = z_i]$ in $\{x_i, y_i, z_i\}_{i \in S_k^c, d_i = d}$ \\
        \textsc{Propensity Score}: \\
        estimate: $\hat \pi_d(x_i,z_i) = \hat P(D_i = d_i | X_i = x_i, Z_i = z_i)$ in $\{x_i, d_i, z_i\}_{i \in S_k^c}$ \\}
    \textsc{Pseudo-outcome}: \\
    estimate: $\hat{\delta}_{l,m}(h_i) = \hat{\mu}_l(z_i,x_i) - \hat{\mu}_m(z_i, x_i) + \frac{I(l)(y_i - \hat{\mu}_l(z_i, x_i))}{\hat{\pi}_l(x_i,z_i)} - \frac{I(m)(y_i - \hat{\mu}_m(z_i,x_i))}{\hat{\pi}_m(x_i,z_i)}$ in $\{h_i\}_{i \in S_k}$ \\
    \textbf{create folds:} Split sample $S_k$ into J random folds $(S_j)_{j=1}^J$ of observations $\{1, \dots, \frac{N}{K \cdot J}\}$ with size of each fold $\frac{N}{K \cdot J}$. Define $S_{j}^c \coloneqq \{1, \dots, \frac{N}{K}\} \backslash \{S_j\}$\\
    \For{j \textnormal{in} $\{1, \dots,J\}$}{
        \For{z \textnormal{in} $\{v, u\}$}{
            \textsc{Pseudo-outcome regression}: \\
            estimate: $\hat g_z(w_i) = \hat \E[\hat{\delta}_{l,m}(h_i) | Z_i =z_i, W_i = w_i]$ in $\{h_i\}_{i \in S_j^c, z_i = z}$ \\
            \textsc{Propensity score}: \\
            estimate: $\hat \lambda_z(w_i) = \hat P(Z_i =z_i | W_i = w_i)$ in  $\{w_i, z_i\}_{i \in S_j^c}$}
    \textsc{$\Delta$BGATE Function}: \\
    \textsc{Effect}: \\
    estimate: $\hat{\theta}_{j,k,l,m,u,v}^{\Delta B}= \frac{K \cdot J}{N}\sum_{i \in S_j} \Bigl[\hat{g}_u(W_i) - \hat{g}_v(W_i) + \frac{I(u)(\hat{\delta}_{l,m}(H_i) - \hat{g}_u(W_i))}{\hat{\lambda}_u(W_i)} -\frac{I(v)(\hat{\delta}_{l,m}(H_i) - \hat{g}_v(W_i))}{\hat{\lambda}_v(W_i)} \Bigr]$ \\
    \textsc{Standard errors}: \\
    estimate: $\hat{\theta}_{j,k,l,m,u,v}^{\Delta B_{SE}}= \frac{K \cdot J}{N}\sum_{i \in S_j} \Bigl[\Bigl(\hat{g}_u(w_i) - \hat{g}_v(w_i) + \frac{I(u)(\delta_{l,m}(h_i) - \hat{g}_u(w_i))}{\hat{\lambda}_u(w_i)} -\frac{I(v)(\hat{\delta}_{l,m}(h_i) - \hat{g}_v(w_i))}{\hat{\lambda}_v(W_i)} \Bigr)^2\Bigr]$\\}}
    
    estimate effect: $\hat{\theta}_{l,m,u,v}^{\Delta B} = \frac{1}{J\cdot K}\sum_{j = 1}^J \sum_{k = 1}^K \hat{\theta}^{\Delta B}_{j,k,l,m,u,v}$ \\
    estimate standard errors: $SE(\hat{\theta}_{l,m,u,v}^{\Delta B}) = \sqrt{\frac{1}{J \cdot K}\sum_{j = 1}^J \sum_{k = 1}^K \hat{\theta}_{j,k,l,m,u,v}^{\Delta B_{SE}} - \left(\hat{\theta}_{j,k,l,m,u,v}^{\Delta B}\right)^2}$}
\end{algorithm}

When $N/K$ is not an integer, meaning the total number of observations N cannot be evenly divided into K folds, the extra observations are distributed among the folds starting from the first fold. For the implementation of the DML-estimator, the weights (e.g.,$\frac{z}{\lambda_1(w)}$) are normalized to ensure that they do not explode. Furthermore, the weights are truncated such that the weight of each observation is no more than five per cent of the sum of all weights. In a third step, we renormalize the (possibly) truncated weights so that they sum to one \citep*{Huber:2013}. For propensity scores too close to 0 or 1, unnormalized and untruncated weights can lead to implausibly large effect estimates \citep*{Busso:2014}.

Algorithm \ref{alg:normalization} shows how a propensity score can be truncated and normalized for better finite sample properties. This algorithm can be used for all propensity scores in this paper, namely $\lambda_z(w)$, $\lambda_z(x)$, $\pi_d(z,x)$ and $\omega_{d,z}(x)$. From now on, the propensity score is called $\pi_d(z,x)$. Be aware that the score function has to be adapted because weight$_{norm, i}$ replaces $\frac{I(d)}{\pi_d(z,x)}$ and not only $\pi_d(z,x)$. 
\begin{algorithm}[!h]
    \setstretch{1.2}
\caption{\textsc{Normalization and truncation of propensity score weights}}\label{alg:normalization}
\SetKwInOut{Input}{Input}
\SetKwInOut{Output}{Output}
\SetKwBlock{Beginn}{begin}{end}
\Input{Data: $\{d_i\}$ \\
Propensity score: $\pi_d(z_i,x_i)$} 
\Output{Normalized and truncated weights:  weight$_{norm,i}$}
\Beginn{
    Number of observations: $N$\\
\For{i = 1, \dots, N}{
    \textsc{Truncation}: \\ $\pi_d(z_i,x_i) = \max(\pi_d(z_i,x_i),0.0001)$ \\
\textsc{Definition weights}: \\
$weight = \frac{I(d_i = d)}{\pi_d(z_i,x_i)}$ \\}
\textbf{Define:} Sum of the weights: $weight_{sum} = \sum_{i = 1}^N weight_i$ \\
\For{i = 1, \dots, N}{
    \textsc{Normalization}: \\
$weight_{norm,i} = \frac{weight_i}{weight_{sum}}$\\
\textsc{Truncation}: \\ $weight_{norm,i} = \min(weight_{norm,i}, 0.05)$ \\}
\textbf{Define:} Sum of the new weights: $weight_{norm,sum} = \sum_{i = 1}^N weight_{norm,i}$ \\
\For{i = 1, \dots, N}{
\textsc{Normalization}: \\
$weight_{norm,i} = \frac{weight_{norm,i}}{weight_{norm,sum} } \cdot N$\\}
}
\end{algorithm} 

Algorithm \ref{alg:BGATE-Learner_auto} shows the Auto-DML algorithm to estimate the $\Delta$BGATE.

\begin{algorithm}[H]
    \setstretch{1.2}
\caption{\textsc{Auto-DML for $\Delta$BGATE}}\label{alg:BGATE-Learner_auto}
\SetKwInOut{Input}{Input}
\SetKwInOut{Output}{Output}
\SetKwBlock{Beginn}{begin}{end}
\Input{Data: $h_i = \{x_i, z_i, d_i, y_i\}$} 
\Output{$\Delta$BGATE = $\hat{\theta}_{l,m,u,v}^{\Delta B}$}
\Beginn{
\textbf{create folds:} Split sample into K random folds $(S_k)_{k=1}^K$ of observations $\{1, \dots, \frac{N}{K}\}$ with size of each fold $\frac{N}{K}$. Define $S_k^c \coloneqq \{1, \dots, N\} \backslash \{S_k\}$\\
\For{k \textnormal{in} $\{1, \dots,K\}$}{
    \For{d \textnormal{in} $\{l, m\}$}{
        \textsc{Response Functions}: \\
        estimate: $\hat \mu_d(z_i,x_i) = \hat \E[Y_i |D_i = d_i, X_i = x_i, Z_i = z_i]$ in $\{x_i, y_i, z_i\}_{i \in S_k^c, d_i = d}$ \\
        \textsc{Riesz Representer}: \\
        estimate: $\hat \alpha_d(z_i,x_i)$ in $\{x_i, z_i\}_{i \in S_k^c}$ \\}
    \textsc{Pseudo-outcome}: \\
    estimate: $\hat{\delta}_{l,m}(h_i) = \hat{\mu}_l(z_i,x_i) - \hat{\mu}_m(z_i, x_i) + \hat{\alpha}_d(z_i, x_i)(y - \hat{\mu}_d(z_i, x_i))$ in $\{h_i\}_{i \in S_k}$ \\
    \textbf{create folds:} Split sample $S_k$ into J random folds $(S_j)_{j=1}^J$ of observations $\{1, \dots, \frac{N}{K \cdot J}\}$ with size of each fold $\frac{N}{K \cdot J}$. Define $S_{j}^c \coloneqq \{1, \dots, \frac{N}{K}\} \backslash \{S_j\}$\\
    \For{j \textnormal{in} $\{1, \dots,J\}$}{
        \For{z \textnormal{in} $\{v, u\}$}{
            \textsc{Pseudo-outcome regression}: \\
            estimate: $\hat g_z(w_i) = \hat \E[\hat{\delta}_{l,m}(h_i) | Z_i =z_i, W_i = w_i]$ in $\{h_i\}_{i \in S_j^c, z_i = z}$ \\
            \textsc{Riesz Representer}: \\
            estimate: $\hat \alpha_z(w_i)$ in  $\{w_i, z_i\}_{i \in S_j^c}$}
    \textsc{$\Delta$BGATE Function}: \\
    \textsc{Effect}: \\
    estimate: $\hat{\theta}_{k,j,l,m,u,v}^{\Delta B}= \frac{K \cdot J}{N}\sum_{i \in S_j} \Bigl[\hat{g}_u(W_i) - \hat{g}_v(W_i) + \hat{\alpha}_z(W_i)(\hat \delta_{l,m}(H_i) - \hat g_z(W_i)) \Bigr]$ \\
    \textsc{Standard errors}: \\
    estimate: $\hat{\theta}_{k,j,l,m,u,v}^{\Delta B_{SE}}= \frac{K \cdot J}{N}\sum_{i \in S_j} \Bigl[\Bigl(\hat{g}_u(w_i) - \hat{g}_v(w_i) + \hat{\alpha}_z(W_i)(\hat \delta_{l,m}(H_i) - \hat g_z(W_i)) \Bigr)^2\Bigr]$\\}}
    estimate effect: $\hat{\theta}_{l,m,u,v}^{\Delta B} = \frac{1}{J\cdot K}\sum_{j = 1}^J \sum_{k = 1}^K \hat{\theta}^{\Delta B}_{j,k,l,m,u,v}$ \\
    estimate standard errors: $SE(\hat{\theta}_{l,m,u,v}^{\Delta B}) = \sqrt{\frac{1}{J \cdot K}\sum_{j = 1}^J \sum_{k = 1}^K \hat{\theta}_{j,k,l,m,u,v}^{\Delta B_{SE}} - \left(\hat{\theta}_{j,k,l,m,u,v}^{\Delta B}\right)^2}$}
\end{algorithm}

\newpage
Algorithm \ref{alg:BGATE-Learner_resampling} shows the algorithm for reweighting the data in such a way that estimating a $\Delta$GATE equals estimating a $\Delta$BGATE.

\begin{algorithm}[!h]
    \setstretch{1.35}
    \caption{\textsc{Reweighting procedure}}\label{alg:BGATE-Learner_resampling}
    \SetKwInOut{Input}{Input}
    \SetKwInOut{Output}{Output}
    \SetKwBlock{Beginn}{begin}{end}
    
    \Input{Data: $h_i = \{x_i, z_i, d_i, y_i\}$}
    \Output{Data: $h_{i}^{balanced} = \{x_{i}^{balanced}, z_{i}^{balanced}, d_{i}^{balanced}, y_{i}^{balanced}\}$}
    \Beginn{number of observations: $N$ \\ 
    original indices: $index$ \\
    balancing variables: $w_i$ \\

        \textbf{duplicate observations:} Duplicate each observation in $h_i$ and create $h_{i, 1}^{balanced}$ and $h_{i, 0}^{balanced}$\\
        
        \textbf{assign moderator values:} Set z values in $h_{i,1}^{balanced}$ to 1 and in $h_{i,0}^{balanced}$ to 0. Merge $h_{i,1}^{balanced}$ and $h_{i,0}^{balanced}$ to $h_{i}^{balanced}$  that has $2 \times N$ rows. \\
    
        \textbf{compute inverse of covariance matrix of balancing variables:} $\text{cov}_{inv} = \text{cov}(w_i)^{-1}$
        
        \ForEach{$index$ in $1, \dots, 2 \times N$}{
            \textbf{retrieve current balancing vector:} \\
            Define $w_i[z_{i}^{balanced}]$ as the subset of $w_i$ corresponding to $z = z_{i}^{balanced}$\\

            \textbf{calculate distances:} \\
            compute difference vector: \\
            \( \text{diff} = w_i[z_{i}^{balanced}] - w_i^{balanced}[index, :] \) \\
            calculate Mahalanobis distance:
            \begin{equation*}
                \text{distance} = 
            \begin{cases}
                \sum((\text{diff} \cdot \text{cov}_{inv}) \cdot \text{diff}) & \text{if dim}(w_i) > 1 \\
                \text{diff}^2 & \text{otherwise}
            \end{cases}
            \end{equation*}
            
            \textbf{find nearest neighbor:}\\
            Select row with smallest distance as the match for $index$.\\
            Copy matched row values to $h_{i}^{balanced}[index]$.\\
        }
        \Return $h_{i}^{balanced}$
    }
    \end{algorithm}
\newpage
\defcitealias{Chernozhukov:2018}{Chernozhukov et al. (2018)}
Algorithm \ref{alg:DR-Learner_GATE} shows the procedure for estimating ${\theta}^{\Delta G}$ with DML by \citetalias{Chernozhukov:2018}. Summarized, start with estimating an average treatment effect (ATE) in groups $Z_i = 1$ and $Z_i = 0$ separately. Then take the difference of those two effects to receive $\hat{\theta}^{\Delta G}$.

\begin{algorithm}[!h]
    \setstretch{1.35}
\caption{\textsc{DML for $\Delta$GATE}}\label{alg:DR-Learner_GATE}
\SetKwInOut{Input}{Input}
\SetKwInOut{Output}{Output}
\SetKwBlock{Beginn}{begin}{end}
\Input{Data: $h_i = \{x_i, d_i, z_i, y_i\}$} 
\Output{$\Delta$GATE: $\hat{\theta}^{\Delta G}$}
\Beginn{
\For{z \textnormal{in} $\{v, u\}$}{
\textbf{create folds:} Split sample into K random folds $(S_k)_{k=1}^K$ of observations $\{1, \dots, \frac{N}{K}\}$ with size of each fold $\frac{N}{K}$. Also define $S_k^c \coloneqq \{1, \dots, N\} \backslash S_k$\\
\For{k \textnormal{in} $\{1, \dots,K\}$}{
    \For{d \textnormal{in} $\{l, m\}$}{
            \textsc{Response Functions}: \\
            estimate: $\hat \mu_d(z_i, x_i) = \hat \E[Y_i | D_i = d_i, X_i = x_i, Z_i = z_i]$ in $\{x_i, z_i, y_i\}_{i \in S_k^c, d_i = d}$ \\
            \textsc{Propensity Score}: \\
            estimate: $\hat \pi_d(z_i, x_i) = \hat P(D_i = d_i | X_i = x_i, Z_i = z_i)$ in $\{x_i, d_i, z_i \}_{i \in S_k^c}$\\
    }
    \textsc{(G)ATE Function}: \\
    \textsc{Effect}: \\
    estimate: $\hat{\theta}_{k,l,m}=\frac{K}{N}\sum_{i \in S_k} \Bigl[\hat{\mu}_l(z_i, x_i) - \hat{\mu}_m(z_i, x_i) + \frac{\mathcal{I}(D_i = l)(Y_i - \hat{\mu}_l(z_i, x_i))}{\hat{\pi}_l(z_i, x_i)} - \frac{\mathcal{I}(D_i = m)(Y_i - \hat{\mu}_m(z_i, x_i))}{\hat{\pi}_m(z_i, x_i)}\Bigr]$ \\
    \textsc{Standard errors}: \\
    estimate: $\hat{\theta}_{k,l,m}^{SE}=\frac{K}{N}\sum_{i \in S_k} \Bigl[\Bigl(\hat{\mu}_l(z_i, x_i) - \hat{\mu}_m(z_i, x_i) + \frac{\mathcal{I}(d_i = l)(y_i - \hat{\mu}_l(z_i, x_i))}{\hat{\pi}_l(z_i, x_i)} - \frac{\mathcal{I}(d_i = m)(y_i - \hat{\mu}_m(z_i, x_i))}{(\hat{\pi}_m(z_i, x_i))}\Bigr)^2\Bigr]$}
    estimate effect: $\hat{\theta}_{l,m}(z) = \frac{1}{K}\sum_{k = 1}^K \hat{\theta}_{k,l,m}$ \\
    estimate standard errors: $\Var(\hat{\theta}_{l,m}(z)) = \frac{1}{K}\sum_{k = 1}^K \hat{\theta}_{k,l,m}^{SE} - \hat \theta_{l,m}(z)^2$ \\}
    calculate effect: $\hat{\theta}_{l,m,u,v}^{\Delta G} = \hat{\theta}_{l,m}(u) - \hat{\theta}_{l,m}(v)$ \\
    calculate standard errors: $SE(\hat{\theta}_{l,m,u,v}^{\Delta G}) = \sqrt{\Var(\hat{\theta}_{l,m}(u)) + \Var(\hat{\theta}_{l,m}(v))}$
}
\end{algorithm}
\newpage
Algorithm \ref{alg:DDR-Learner} shows the suggested estimation procedure for the $\Delta$CBGATE. It is similar to the estimation of an ATE but with a different score function.
\begin{algorithm}[!h]
    \setstretch{1.35}
\caption{\textsc{DML for $\Delta$CBGATE}}\label{alg:DDR-Learner}
\SetKwInOut{Input}{Input}
\SetKwInOut{Output}{Output}
\SetKwBlock{Beginn}{begin}{end}
\Input{Data: $h_i = \{x_i, z_i, d_i, y_i\}$} 
\Output{$\Delta$CBGATE = $\hat{\theta}^{\Delta C}$}
\Beginn{
\textbf{create folds:} Split sample into K random folds $(S_k)_{k=1}^K$ of observations $\{1, \dots, \frac{N}{K}\}$ with size of each fold $\frac{N}{K}$. Also define $S_k^c \coloneqq \{1, \dots, N\} \backslash S_k$\\
\For{k \textnormal{in} $\{1, \dots,K\}$}{
    \For{d \textnormal{in} $\{l, m\}$}{
        \For{z \textnormal{in} $\{v, u\}$}{
            \textsc{Response Functions}: \\
            estimate: $\hat \mu_{d}(z_i, x_i) = \hat \E[Y_i | X_i = x_i, D_i = d_i, Z_i = z_i]$ in $\{x_i, y_i, z_i\}_{i \in S_k^c, d_i = d, z_i = z}$ \\
            \textsc{Propensity Score}: \\
            \textsc{\textbf{Version 1:}}\\
            estimate: $\hat \omega_{d,z}(x_i) = \hat P(D_i = d_i, Z_i = z_i | X_i = x_i)$ in $\{x_i, d_i, z_i\}_{i \in S_k^c}$ \\
            \textsc{\textbf{Version 2:}}\\
            estimate: $\hat \pi_d(x_i,z_i) = \hat P(D_i = d_i | X_i = x_i, Z_i = z_i)$ in $\{x_i, d_i, z_i\}_{i \in S_k^c}$ \\
            estimate: $\hat \lambda_z(x_i) = \hat P(Z_i = z_i | X_i = x_i)$ in $\{x_i, d_i, z_i \}_{i \in S_k^c}$ \\
            calculate: $\hat \omega_{d,z}(x_i) = \hat \pi_d(x_i,z_i) \cdot \hat \lambda_z(x_i)$ \\
        }
    }
    \textsc{$\Delta$CBGATE Function}: \\
    \textsc{Effect}: \\
    estimate: $\hat{\theta}_{k,l,m,u,v}^{\Delta C}= \frac{K}{N}\sum_{i \in S_k} \Bigl[\hat{\mu}_{l}(v,x_i) - \hat{\mu}_{m}(v,x_i) - \hat{\mu}_{l}(u,x_i) + \hat{\mu}_{m}(u,x_i) + \frac{I(l,v)(y_i - \hat{\mu}_{l}(v,x_i))}{\hat{\omega}_{l,v}(x_i)} - \frac{I(m,v)(y_i - \hat{\mu}_{m}(v,x_i))}{\hat{\omega}_{m,v}(x_i)} -\frac{I(l,u)(y_i - \hat{\mu}_{l}(u,x_i))}{\hat{\omega}_{l,u}(x_i)} + \frac{I(m,u)(y_i - \hat{\mu}_{m}(u,x_i))}{\hat{\omega}_{m,u}(x_i)}\Bigr]$ \\
    \textsc{Standard errors}: \\
    estimate: $\hat{\theta}_{k,l,m,u,v}^{\Delta C, SE}= \frac{K}{N}\sum_{i \in S_k} \Bigl[\Bigl(\hat{\mu}_{l}(v,x_i) - \hat{\mu}_{m}(v,x_i) - \hat{\mu}_{l}(u,x_i) + \hat{\mu}_{m}(u,x_i) + \frac{I(l,v)(y_i - \hat{\mu}_{l}(v,x_i))}{\hat{\omega}_{l,v}(x_i)} - \frac{I(m,v)(y_i - \hat{\mu}_{m}(v,x_i))}{\hat{\omega}_{m,v}(x_i)} -\frac{I(l,u)(y_i - \hat{\mu}_{l}(u,x_i))}{\hat{\omega}_{l,u}(x_i)} + \frac{I(m,u)(y_i - \hat{\mu}_{m}(u,x_i))}{\hat{\omega}_{m,u}(x_i)}\Bigr)^2\Bigr]$\\}
    
    estimate effect: $\hat{\theta}_{l,m,u,v}^{\Delta C} = \frac{1}{K}\sum_{k = 1}^K \hat{\theta}_{k,l,m,u,v}^{\Delta C}$ \\
    estimate standard errors: $SE(\hat{\theta}^{\Delta C}) = \sqrt{\frac{1}{K}\sum_{k = 1}^K \hat{\theta}_{k,l,m,u,v}^{\Delta C, SE} - \left(\hat{\theta}_{l,m,u,v}^{\Delta C}\right)^2}$
}
\end{algorithm}

\subsection{Variance Derivation for Reweighting Approach} \label{Variance_estimation_resampling}
Define an estimator which is a weighted mean of $N$ independent realizations of the i.i.d. random variables $Y$, $y_i$, $\theta = \sum_i \alpha_i y_i$, where $\alpha_i$ denote fixed, positive weights that sum up to one. The procedure defined in Subsection \ref{2_resampling} leads to such weights. The variance of this estimator is given by:

\begin{align*}
    \Var(\hat \theta) &= \Var \left(\sum_{i = 1}^Q a_i y_i\right) = \sum_{i =1}^Q \Var(a_i y_i) =  \sum_{i =1}^Q a_i^2\Var(y_i) = \left(\sum_{i =1}^Q a_i^2\right)\Var(Y)
\end{align*}

Suppose now that the reweighting approach is implemented (as proposed in Subsection \ref{2_resampling}) as a resampling scheme. In this case some observations will appear once, other appear a couple of times. The estimator is applied to such an artificial sample without adjustments. However, the variance needs to be adjusted for the fact that some observations appear several time (i.e. they have a weight larger than $1/N$). Assuming that $s_i$ is the number of duplicates for each observation and $Q$ the number of unique observations, the weights can be written as $a_i = \frac{1 + s_i}{\sum_{i = 1}^Q (1 + s_i)}$ and the variance can be written as follows: 
\begin{align*}
    \Var(\hat \theta) = \left(\sum_{i =1}^Q a_i^2\right)\Var (Y) = \sum_{i =1}^Q\left( \frac{1 + s_i}{\sum_{i = 1}^Q(1 + s_i)}\right)^2\Var(Y) 
    = \sum_{i =1}^Q \frac{(1 + s_i)^2}{\left(\sum_{i = 1}^Q(1 + s_i)\right)^2}\Var(Y)
\end{align*}

\clearpage

\section{Appendix: Monte Carlo Study} \label{Appendix_C}

This section explains details of the Monte Carlo study conducted for the $\Delta$BGATE. For simplicity, the simulation covers the case where $Z_i$ and $D_i$ are binary variables.

\subsection{Performance Measures} \label{performance_measures}

For the evaluation of the different estimators, different measures are considered. First, we describe the measures to evaluate the estimation of the effects. The performance measures are shown for $\hat{\theta}_{l,m,u,v}^{\Delta B}$ only, but the same performance measures are also used for $\hat{\theta}_{l,m,u,v}^{\Delta G}$.
\begin{align*}
    &\text{Bias: }bias(\hat{\theta}_{l,m,u,v}^{\Delta B}) = \frac{1}{R}\sum_{r = 1}^R\hat{\theta}_{r,l,m,u,v}^{\Delta B} - \theta_{l,m,u,v}^{\Delta B} \\
    &\text{Absolut Bias: }|bias(\hat{\theta}_{l,m,u,v}^{\Delta B})| = \frac{1}{R}\sum_{r = 1}^R | \hat{\theta}_{r,l,m,u,v}^{\Delta B} - \theta_{l,m,u,v}^{\Delta B}| \\
    &\text{Standard deviation: }std(\hat{\theta}_{l,m,u,v}^{\Delta B}) = \sqrt{\frac{1}{R}\sum_{r = 1}^R(\hat{\theta}_{r,l,m,u,v}^{\Delta B} - \frac{1}{R}\sum_{r = 1}^R \hat{\theta}_{r,l,m,u,v}^{\Delta B})^2} \\
    &\text{Root mean sqared error: }rmse(\hat{\theta}_{l,m,u,v}^{\Delta B}) = \sqrt{\frac{1}{R} \sum_{r = 1}^R(\hat{\theta}_{r,l,m,u,v}^{\Delta B}- \theta_{l,m,u,v}^{\Delta B})^2} \\
    &\text{Skewness: }skew(\hat{\theta}_{l,m,u,v}^{\Delta B}) = \frac{1}{R} \sum_{i = 1}^R \left( \frac{\hat{\theta}_{l,m,u,v}^{\Delta B} - \frac{1}{R}\sum_{r = 1}^R \hat{\theta}_{r,l,m,u,v}^{\Delta B}}{SD(\hat{\theta}_{l,m,u,v}^{\Delta B})}\right)^3 \\
    &\text{Excess kurtosis: }ex.kurt(\hat{\theta}_{l,m,u,v}^{\Delta B}) = \frac{1}{R} \sum_{i = 1}^R \left( \frac{\hat{\theta}_{r,l,m,u,v}^{\Delta B} - \frac{1}{R}\sum_{r = 1}^R \hat{\theta}_{r,l,m,u,v}^{\Delta B}}{SD(\hat{\theta}_{l,m,u,v}^{\Delta B})}\right)^4 - 3
\end{align*}
Furthermore, the following performance measures are used to evaluate the inference procedures:
\begin{align*}
    &\text{Bias standard error: }bias(\widehat{se}(\hat{\theta}_{l,m,u,v}^{\Delta B})) = \frac{1}{R} \sum_{i = 1}^R SE^r(\hat{\theta}_{r,l,m,u,v}^{\Delta B}) - SD(\hat{\theta}_{l,m,u,v}^{\Delta B}) \\
    &\text{Coverage Probability: }cov(\alpha) = P\Biggl[\left(\theta_{l,m,u,v}^{\Delta B} > \frac{1}{R}\sum_{i = 1}^r(\hat{\theta}_{r,l,m,u,v}^{\Delta B} - Z_{\frac{\alpha}{2}}SE(\hat{\theta}_{r,l,m,u,v}^{\Delta B})) \right) \\ &\cup \left(\theta_{l,m,u,v} < \frac{1}{R}\sum_{i = 1}^r(\hat{\theta}_{r,l,m,u,v}^{\Delta B} + Z_{1 -\frac{\alpha}{2}}SE(\hat{\theta}_{r,l,m,u,v}^{\Delta B}))\right)\Biggr]
\end{align*}

\subsection{Details for $\Delta$BGATE} \label{DGP BGATE non-linear}

\subsubsection{Random Forest Tuning Parameters} \label{RF tuning}
The optimal hyperparameters for the different random forests, namely the tree depth and the minimum leaf size, are tuned by grid search (maximum tree depth: [2, 3, 5, 10, 20], minimum leaf size: [5, 10, 15, 20, 30, 50]) using a random forest with 1000 trees and two folds. They are tuned using twenty different data draws and then fixed for the simulation. The optimal combination of maximal tree depth and minimum leaf size is chosen by taking the combination that appears most often in the twenty draws. Table \ref{Optimal tuning parameters} shows the optimal hyperparameters for the different DGPs, effects of interest and sample sizes.

\begin{table}[!h]
    \centering
    \caption{Simulation Study: Optimal hyperparameters for random forests}
\label{Optimal tuning parameters}
    \begin{adjustbox}{max width=\textwidth}
    \begin{threeparttable}
    \begin{tabular}{llrrrrrr|rrrrrr|rrrrrr|rrrrrr} \toprule 
        &&\multicolumn{6}{c}{N = 1,250} &\multicolumn{6}{c}{N = 2,500} &\multicolumn{6}{c}{N = 5,000} &\multicolumn{6}{c}{N = 10,000}\\ \midrule
        && $\mu_1$ & $\mu_0$ & $\pi_d$ & $g_1$ & $g_0$ & $\lambda_z$ & $\mu_1$ & $\mu_0$ & $\pi_d$ & $g_1$ & $g_0$ & $\lambda_z$ & $\mu_1$ & $\mu_0$ & $\pi_d$ & $g_1$ & $g_0$ & $\lambda_z$ & $\mu_1$ & $\mu_0$ & $\pi_d$ & $g_1$ & $g_0$ & $\lambda_z$\\ \midrule
        \multirow{2}{*}{$\theta^{\Delta B}_{X_0}$} & Max tree depth & 20 & 2 & 10 & 2 & 2 & 2 & 20 & 3 & 5& 2& 2&2 & 10 & 3 & 10 & 2 &2 & 2 & 10 &5 &  10& 2& 2&2\\
        &Min leaf size & 5 & 5 & 10 & 5 & 50 & 50 & 5 & 5 & 10& 10& 50& 50& 5 & 5 & 20 & 5 &50 & 50 & 5 & 5& 15 & 5& 50&50\\ \midrule
        \multirow{2}{*}{$\theta^{\Delta B}_{X_2}$} & Max tree depth &  20 & 2 & 10 & 2 & 2 & 2 &  20 & 3 & 5&2 &2 &2 & 10 & 3 &10  & 2 &2 & 2 &  10& 5& 10 & 2&2 &2\\
        & Min leaf size &  5 & 5 & 10 & 5 & 50 & 50 & 5 & 5 & 10 & 5&5 &50 & 5 & 5 & 20 & 5 & 5& 5 & 5 &5 & 15 & 5& 5&50\\ \midrule
        \multirow{2}{*}{$\theta^{\Delta G}$} & Max tree depth & 20 & 2 &10 & - & -& -& 20 & 2 & 5 & - & - & -& 10 & 3 & 10 &  -& -& - & 10 &3 & 10 & -& -&-\\
        & Min leaf size & 5 & 30 & 10&-&- & -& 5 & 50 & 10 &-   &- & -&5 & 50 & 30 &-  & -& - & 10 & 5& 50 & -& -&-\\  \bottomrule
    \end{tabular}
    \begin{tablenotes}
     \item \textit{Note:} This table depicts the optimal hyperparameters for the random forests. The grid values are as follows: maximum tree depth: [2, 3, 5, 10, 20], minimum leaf size: [5, 10, 15, 20, 30, 50]. Note that in the second step highly shallow trees are used. This might be explained by the fact that we only include one balancing variable. Hence, the tree splits on this single variable only.
    \end{tablenotes}
\end{threeparttable}
\end{adjustbox}
\end{table}

\subsubsection{Implementation of Auto-DML by RieszNet} \label{autoDML RieszNet}

The Auto-DML algorithm uses a neural net as suggested in \cite*{Chernozhukov:2022c}. As shown in Algorithm \ref{alg:BGATE-Learner_auto}, the Auto-DML framework is used twice to estimate the $\Delta$BGATE (same principle as DML for $\Delta$BGATE). Hence, two networks are trained. The first network is used to estimate the response functions $\hat{\mu}_d(z_i, x_i)$ and the Riesz representer $\hat{\alpha}_d(z_i, x_i)$. The second network is used to estimate the pseudo-outcome $\hat{\delta}_{l,m}(h_i)$ and the Riesz representer $\hat{\alpha}_z(w_i)$. 

The network's architecture has four main components: a shared feature representation layer, two regression heads for the conditional expectations, and one head for the Riesz representer. The first neural net takes the covariates $x_i$ as input, the treatment $d_i$, and the moderator $z_i$. First, the input passes through a shared dense layer. Namely, a linear transformation layer that maps the input to a higher-dimensional feature space. After passing through the common layer, the feature representation branches into two regression networks: one predicting the outcome under treatment and one under no treatment. The regression heads consist of a hidden layer and a linear output layer. An Exponential Linear Unit (ELU) activation function is used. The Riesz representer $\alpha$ is also estimated using a linear output layer. Finally, the three outputs are combined into a final prediction output representing the score function $\delta(h_i)$. The second neural net takes as an input the balancing variables $w_i$ and the moderator $z_i$. Afterwards, the neural net is built the same way as the first one, with the final output being the score function $\phi_{Riesz}^{\Delta B}$. Figure \ref{fig:riesznet} shows the architecture of the RieszNet. 

\begin{figure}[h!]
    \centering
    \begin{minipage}{0.8\textwidth}
        \caption{RieszNet architecture}
        \label{fig:riesznet}
        \begin{tikzpicture}[node distance=1.8cm and 2cm]
            \node (H) {$H_i$};
            \node[draw, rectangle, right=2cm of H, minimum width=1cm, minimum height=4cm, align=center] (layer2) {};
            \node[draw, rectangle, right=2cm of layer2, yshift=1.5cm, minimum width=1cm, minimum height=0.9cm, align=center] (layer3) {};
            \node[draw, rectangle, right=1cm of layer3, minimum width=1cm, minimum height=0.9cm, align=center] (layer4) {};
            \node[draw, rectangle, right=2cm of layer2, yshift=0.5cm, minimum width=1cm, minimum height=0.9cm, align=center] (layer5) {};
            \node[draw, rectangle, right=1cm of layer5, minimum width=1cm, minimum height=0.9cm, align=center] (layer6) {};
            \node[draw, rectangle, right=2cm of layer2, yshift=-1cm, minimum width=1cm, minimum height=1.8cm, align=center] (layer7) {};
            \node[right=1cm of layer4] (g1) {$\mu_1(\cdot)$};
            \node[right=1cm of layer6] (g0) {$\mu_0(\cdot)$};
            \node[right=1cm of layer7] (alpha) {$\alpha_d(\cdot)$};
        
            \draw[->] (H) -- (layer2);
            \draw[->] (layer2) -- (layer3);
            \draw[->] (layer3) -- (layer4);
            \draw[->] (layer2) -- (layer5);
            \draw[->] (layer5) -- (layer6);
            \draw[->] (layer2) -- (layer7);
            \draw[->] (layer4) -- (g1);
            \draw[->] (layer6) -- (g0);
            \draw[->] (layer7) -- (alpha);
        
        \end{tikzpicture}    \\
    \scriptsize \textit{Note:} The figure depicts the architecture of the RieszNet. The input $H_i$ is passed through a common layer and branches into two regression heads and one Riesz representer. The architecture is shown for the first stage. The second stage looks the same but with different estimands.
    \end{minipage}
\end{figure}

The RieszNet has been trained with the following loss function:
\begin{align*}
    &\min_{w_{1:d}, \beta, \epsilon} \text{REGloss}(w_{1:d}) + \lambda_1 \text{RRloss}(w_{1:k}) + \lambda_2 \text{TMLEloss}(w_{1:d}, \epsilon) \\
&\text{RRloss}(w_{1:k}) = \E_n \left[ \alpha(z_i, x_i; w_{1:k})^2 - 2m(h_i; \alpha(z_i, x_i; w_{1:k})) \right] \\
&\text{REGloss}(w_{1:d}) = \E_n \left[ \left( Y_i - \mu(z_i, x_i; w_{1:d}) \right)^2 \right] \\
&\text{TMLEloss}(w_{1:d}, \epsilon) = \E_n \left[ \left( Y_i - \mu(z_i, x_i; w_{1:d}) - \epsilon  \alpha(z_i, x_i; w_{1:k})\right)^2  \right]
\end{align*}
with $w_{1:k}$ being the weights with $k$ hidden layers, $w_{1:d}$ being weights with $d$ hidden layers ($k<d$), a function $m(h_i, f) = f(1,z_i, x_i) - f(0, z_i, x_i)$.
For more details to the RieszNet we refer to \cite{Chernozhukov:2022c}. The hyperparameters used are summarized in Table \ref{tab:hyperparameters}. 
\begin{table}[h!]
    \centering
    \caption{Summary of hyperparameters for RieszNet}
    \label{tab:hyperparameters}
    \begin{adjustbox}{max width=0.43\textwidth}
        \begin{threeparttable}
        \begin{tabular}{l|cc} \toprule
            \textbf{Hyperparameter} & \textbf{First Stage} & \textbf{Second Stage} \\
            \hline
            neurons common layer & 200 & 600 \\
            neurons head layer & 100 & 300 \\
            learning rate & $1 \times 10^{-4}$ & $1 \times 10^{-4}$ \\
            patience & 10 & 70 \\
            min $\delta$ & $1 \times 10^{-4}$ & $1 \times 10^{-4}$ \\
            epochs & 600 & 1800 \\
            $\lambda_1$ & 0.1 & 0.1 \\
            $\lambda_2$ & 1 & 1 \\
            number of folds & 2 & 2 \\
            \hline
        \end{tabular}
    \begin{tablenotes}[flushleft]
        \small
     \item \textit{Note:} This table depicts the hyperparameters chosen for the first and second neural net.
    \end{tablenotes}
\end{threeparttable}
\end{adjustbox}
\end{table}

The number of neurons represents the dimensionality of the layer, the learning rate is the step size for the optimization algorithm and controls how much the weights are updated in the learning process, the patience is the number of epochs to wait for improvement before stopping, the minimum $\delta$ is the minimum improvement in validation performance to count as an improvement, the epochs are the maximum number of epochs, $\lambda_1$ is the regularization parameter for the Riesz representer loss and $\lambda_2$ is the regularization parameter for the TMLE loss. The number of folds is the number of folds used for cross-validation.
\clearpage
\subsubsection{Results} \label{Results BGATE non-linear}
Table \ref{Simulation_results_extensive} shows the results of the different simulations. In comparison to the results shown in the main text of the paper, additional performance measures are depicted.
\begin{table}[h!]
    \centering
    \caption{Extensive simulation results}
    \label{Simulation_results_extensive}
    \begin{adjustbox}{max width=0.9\textwidth}
        \begin{threeparttable}
        \begin{tabular}{llrrrrrrrr} \toprule
        &&\multicolumn{6}{l}{Estimation of effects} &\multicolumn{2}{l}{Estimation of std. errors} \\  \midrule
        & Estimator &  bias & |bias| & std & rmse & Skew & ex.kurt& bias(se) & cov(95) \\  \midrule
        &\multicolumn{9}{c}{\textbf{N = 1,250}} \\  \midrule
        $\theta^{\Delta B}(X_0)$ & DML & 0.0072 & 0.1388 & 0.1741 & 0.1743 & -0.0300 & 0.0038 & -0.0075 & 0.9365\\
$\theta^{\Delta B}(X_0)$ & Auto-DML & 0.0171 & 0.1408 & 0.1741 & 0.1750 & -0.0012 & -0.2633 & -0.0225 & 0.9050\\
$\theta^{\Delta B}(X_0)$ & Resampling & -0.0552 & 0.1481 & 0.1772 & 0.1856 & 0.0096 & -0.2091 & 0.0131 & 0.9550\\
  \midrule
        $\theta^{\Delta B}(X_2)$ & DML & -0.0054 & 0.1226 & 0.1541 & 0.1542 & 0.0083 & 0.1055 & -0.0037 & 0.9430\\
$\theta^{\Delta B}(X_2)$ & Auto-DML & 0.0086 & 0.1244 & 0.1546 & 0.1548 & -0.0362 & -0.2190 & -0.0149 & 0.9240\\
$\theta^{\Delta B}(X_2)$ & Resampling & -0.0180 & 0.1336 & 0.1665 & 0.1674 & 0.0030 & -0.0995 & 0.0138 & 0.9655\\
  \midrule
        $\theta^{\Delta G}$ & DML & -0.0083 & 0.1196 & 0.1499 & 0.1501 & 0.0331 & 0.0189 & -0.0000 & 0.9440\\
$\theta^{\Delta G}$ & Auto-DML & 0.0045 & 0.1253 & 0.1572 & 0.1573 & 0.0484 & -0.0701 & -0.0560 & 0.7835\\
$\theta^{\Delta G}$ & Resampling & -0.0079 & 0.1204 & 0.1500 & 0.1502 & -0.0038 & -0.0398 & -0.0001 & 0.9460\\
 \midrule
        &\multicolumn{9}{c}{\textbf{N = 2,500}} \\  \midrule
        $\theta^{\Delta B}(X_0)$ & DML & -0.0234 & 0.0961 & 0.1185 & 0.1208 & 0.0277 & 0.2001 & -0.0004 & 0.9430\\
$\theta^{\Delta B}(X_0)$ & Auto-DML & -0.0010 & 0.0975 & 0.1228 & 0.1228 & 0.0010 & 0.3707 & -0.0164 & 0.9220\\
$\theta^{\Delta B}(X_0)$ & Resampling & -0.0378 & 0.1049 & 0.1253 & 0.1309 & 0.0027 & -0.0191 & 0.0095 & 0.9570\\
  \midrule
        $\theta^{\Delta B}(X_2)$ & DML & -0.0072 & 0.0838 & 0.1048 & 0.1050 & 0.1000 & 0.0847 & -0.0032 & 0.9440\\
$\theta^{\Delta B}(X_2)$ & Auto-DML & 0.0106 & 0.0830 & 0.1052 & 0.1057 & 0.0614 & 0.3309 & -0.0071 & 0.9410\\
$\theta^{\Delta B}(X_2)$ & Resampling & -0.0127 & 0.0915 & 0.1161 & 0.1168 & -0.0056 & 0.2252 & 0.0111 & 0.9650\\
  \midrule
        $\theta^{\Delta G}$ & DML & -0.0066 & 0.0828 & 0.1053 & 0.1055 & 0.1073 & 0.2341 & -0.0020 & 0.9460\\
$\theta^{\Delta G}$ & Auto-DML & 0.0090 & 0.0852 & 0.1075 & 0.1079 & 0.1310 & 0.2222 & -0.0134 & 0.9100\\
$\theta^{\Delta G}$ & Resampling & -0.0075 & 0.0832 & 0.1052 & 0.1055 & 0.0450 & 0.2575 & -0.0018 & 0.9410\\
  \midrule
        &\multicolumn{9}{c}{\textbf{N = 5,000}} \\  \midrule
        $\theta^{\Delta B}(X_0)$ & DML & -0.0239 & 0.0640 & 0.0763 & 0.0799 & -0.0986 & -0.1976 & 0.0032 & 0.9480\\
$\theta^{\Delta B}(X_0)$ & Auto-DML & -0.0350 & 0.0714 & 0.0819 & 0.0891 & -0.0150 & -0.1345 & -0.0074 & 0.9020\\
$\theta^{\Delta B}(X_0)$ & Resampling & -0.0306 & 0.0734 & 0.0885 & 0.0936 & -0.0466 & 0.1273 & 0.0085 & 0.9560\\
  \midrule
        $\theta^{\Delta B}(X_2)$ & DML & -0.0078 & 0.0554 & 0.0687 & 0.0691 & 0.0114 & 0.0037 & 0.0028 & 0.9540\\
$\theta^{\Delta B}(X_2)$ & Auto-DML & 0.0059 & 0.0559 & 0.0699 & 0.0702 & 0.0763 & 0.1414 & -0.0014 & 0.9500\\
$\theta^{\Delta B}(X_2)$ & Resampling & -0.0083 & 0.0618 & 0.0769 & 0.0773 & 0.0541 & -0.1154 & 0.0145 & 0.9820\\
  \midrule
        $\theta^{\Delta G}$ & DML & -0.0015 & 0.0557 & 0.0696 & 0.0696 & 0.0131 & -0.1161 & 0.0026 & 0.9580\\
$\theta^{\Delta G}$ & Auto-DML & 0.0051 & 0.0582 & 0.0722 & 0.0724 & 0.0413 & 0.0426 & -0.0010 & 0.9440\\
$\theta^{\Delta G}$ & Resampling & -0.0015 & 0.0550 & 0.0688 & 0.0688 & -0.0422 & 0.1370 & 0.0033 & 0.9640\\
  \midrule
        &\multicolumn{9}{c}{\textbf{N = 10,000}} \\  \midrule
        $\theta^{\Delta B}(X_0)$ & DML & -0.0176 & 0.0458 & 0.0560 & 0.0587 & -0.2807 & -0.0509 & -0.0015 & 0.9160\\
$\theta^{\Delta B}(X_0)$ & Auto-DML & -0.0307 & 0.0523 & 0.0596 & 0.0671 & -0.2488 & 0.0748 & -0.0075 & 0.8640\\
$\theta^{\Delta B}(X_0)$ & Resampling & -0.0248 & 0.0541 & 0.0641 & 0.0687 & -0.1833 & 0.0186 & 0.0036 & 0.9400\\
  \midrule
        $\theta^{\Delta B}(X_2)$ & DML & -0.0030 & 0.0419 & 0.0513 & 0.0514 & -0.2290 & -0.2410 & -0.0013 & 0.9400\\
$\theta^{\Delta B}(X_2)$ & Auto-DML & 0.0000 & 0.0429 & 0.0530 & 0.0530 & -0.2735 & -0.1250 & -0.0052 & 0.9280\\
$\theta^{\Delta B}(X_2)$ & Resampling & -0.0019 & 0.0445 & 0.0560 & 0.0560 & -0.1730 & -0.1960 & 0.0080 & 0.9640\\
  \midrule
        $\theta^{\Delta G}$ & DML & -0.0009 & 0.0419 & 0.0516 & 0.0516 & -0.2911 & -0.2492 & -0.0013 & 0.9280\\
$\theta^{\Delta G}$ & Auto-DML & 0.0040 & 0.0438 & 0.0529 & 0.0531 & -0.3094 & -0.3732 & -0.0021 & 0.9520\\
$\theta^{\Delta G}$ & Resampling & -0.0008 & 0.0426 & 0.0527 & 0.0527 & -0.2862 & -0.2982 & -0.0024 & 0.9440\\
  \midrule
    \end{tabular}
    \begin{tablenotes}[flushleft]
     \item \textit{Note:} This table shows simulation results for all different sample sizes, estimators and effects. Column (1) shows the effect estimated, and column (2) shows the estimator used. The remaining eight columns depict performance measures explained in \ref{performance_measures}.
    \end{tablenotes}
\end{threeparttable}
\end{adjustbox}
\end{table}

\clearpage
\section{Appendix: Empirical Example} \label{appendix Empiric example}
\subsection{Data Descriptives} \label{data descriptives}

Table \ref{emp_descriptives_both} shows the mean and standard deviation of the covariates included in the analysis by treatment and moderator status. In addition to the covariates in this table, we add caseworkers' fixed effects. For a more detailed description of the data and how the dataset was constructed, please see \cite{Knaus:2022}. The data can be accessed on swissubase.ch for research purposes. Table \ref{emp_descriptives} shows the mean and standard deviation of the covariates included in the analysis by treatment status and Table \ref{emp_descriptives_moderator} by moderator status.

\begin{table}[h!]
    \centering
    \caption{Empirical analysis: Balance table for treatment and moderator variable (nationality)}
    \label{emp_descriptives_both}
    \begin{adjustbox}{max width=0.9\textwidth}
    \begin{threeparttable}
\begin{tabular}{lrrrrr}
\toprule
&\textbf{Treated} & \textbf{Treated} &\textbf{Non-treated} & \textbf{Non-treated} \\ [1ex]
&\textbf{Non-Swiss} & \textbf{Swiss} &\textbf{Non-Swiss} & \textbf{Swiss} \\ [1ex]
Variable &Mean &Mean & Mean& Mean \\ \midrule
Age & 35.62 & 38.09 & 36.42 & 36.69\\
French speaking canton & 0.10 & 0.07 & 0.28 & 0.24\\
German speaking canton & 0.86 & 0.91 & 0.62 & 0.70\\
Italian speaking canton & 0.04 & 0.02 & 0.10 & 0.07\\
Mother tongue in cantonal language & 0.26 & 0.05 & 0.20 & 0.06\\
Lives in big city & 0.21 & 0.14 & 0.19 & 0.15\\
Lives in medium city & 0.17 & 0.15 & 0.15 & 0.12\\
Lives in contryside & 0.62 & 0.71 & 0.67 & 0.73\\
Age of caseworker & 43.70 & 44.54 & 44.27 & 44.46\\
Caseworkers cooperation & 0.51 & 0.48 & 0.50 & 0.47\\
Caseworkers education: above vocational training & 0.45 & 0.45 & 0.43 & 0.44\\
Caseworkers education: tertiary level & 0.20 & 0.22 & 0.25 & 0.24\\
Caseworker female & 0.42 & 0.47 & 0.38 & 0.42\\
Caseworker missing & 0.04 & 0.04 & 0.04 & 0.04\\
Caseworker own unemployment experience & 0.63 & 0.63 & 0.63 & 0.63\\
Caseworker job tenure & 5.54 & 5.55 & 5.90 & 5.85\\
Caseworker education: vocational degree & 0.25 & 0.27 & 0.22 & 0.23\\
Fraction months employed last 2 years & 0.82 & 0.84 & 0.77 & 0.81\\
No employment spells last 5 years & 1.05 & 0.97 & 1.39 & 1.21\\
Employability & 1.94 & 2.00 & 1.94 & 2.01\\
Female & 0.40 & 0.47 & 0.41 & 0.45\\
Cantonal GDP per capita & 0.51 & 0.51 & 0.49 & 0.49\\
Married & 0.67 & 0.36 & 0.70 & 0.35\\
Mother tongue not Swiss language & 0.65 & 0.10 & 0.67 & 0.10\\
Past annual income & 41703 & 47916 & 38057 & 43941\\
Previous job: manager & 0.05 & 0.09 & 0.04 & 0.09\\
Previous job: primary sector & 0.07 & 0.05 & 0.13 & 0.08\\
Previous job: secondary sector & 0.18 & 0.15 & 0.14 & 0.13\\
Previous job: tertiary sector & 0.54 & 0.67 & 0.48 & 0.65\\
Previous job: missing sector & 0.21 & 0.13 & 0.26 & 0.15\\
Previous job: self-employed & 0.00 & 0.00 & 0.01 & 0.01\\
Previous job: skilled worker & 0.47 & 0.72 & 0.45 & 0.70\\
Previous job: unskilled worker & 0.47 & 0.16 & 0.49 & 0.17\\
Qualification: some degree & 0.35 & 0.73 & 0.31 & 0.72\\
Qualification: semiskilled & 0.18 & 0.13 & 0.21 & 0.13\\
Qualification: unskilled & 0.40 & 0.13 & 0.40 & 0.12\\
Qualification: skilled without degree & 0.07 & 0.02 & 0.08 & 0.03\\
Allocation to caseworker: by industry & 0.63 & 0.67 & 0.53 & 0.54\\
Allocation to caseworker: by occupation & 0.56 & 0.59 & 0.56 & 0.56\\
Allocation to caseworker: by age & 0.04 & 0.04 & 0.03 & 0.03\\
Allocation to caseworker: by employability & 0.07 & 0.07 & 0.06 & 0.07\\
Allocation to caseworker: by region & 0.09 & 0.09 & 0.12 & 0.12\\
Allocation to caseworker: other & 0.06 & 0.07 & 0.07 & 0.08\\
No of unemployment spells last 2 years & 0.54 & 0.35 & 0.80 & 0.53\\
Cantonal unemployment rate & 3.71 & 3.60 & 3.83 & 3.70\\
 \midrule
Number of observations & 4,423 & 8,575& 28,169 &  43,415\\ \bottomrule
\end{tabular}
\begin{tablenotes}[flushleft]
    \footnotesize
 \item \textit{Note:} This table shows the mean of some covariates included in the analysis. Column (1) and (2) show it for treated individuals, column (3) and (4) for non-treated individuals. Column (1) and (3) show it for non-Swiss individuals, column (2) and (4) for Swiss individuals.
\end{tablenotes}
\end{threeparttable}
\end{adjustbox}
\end{table}

\begin{table}[h!]
    \centering
    \caption{Empirical analysis: Balance table for treatment and moderator variable (employability)}
    \label{emp_descriptives_both_emp}
    \begin{adjustbox}{max width=0.9\textwidth}
    \begin{threeparttable}
\begin{tabular}{lrrrrr}
\toprule
&\textbf{Treated} & \textbf{Treated} &\textbf{Non-treated} & \textbf{Non-treated} \\ [1ex]
&\textbf{Hard} & \textbf{Easy} &\textbf{Hard} & \textbf{Easy} \\ [1ex]
Variable &Mean &Mean & Mean& Mean \\ \midrule
Age & 36.98 & 39.23 & 36.27 & 38.48\\
French speaking canton & 0.08 & 0.05 & 0.27 & 0.16\\
German speaking canton & 0.89 & 0.92 & 0.65 & 0.77\\
Italian speaking canton & 0.03 & 0.03 & 0.08 & 0.07\\
Mother tongue in cantonal language & 0.12 & 0.13 & 0.12 & 0.09\\
Lives in big city & 0.17 & 0.15 & 0.17 & 0.15\\
Lives in medium city & 0.15 & 0.17 & 0.13 & 0.16\\
Lives in contryside & 0.68 & 0.68 & 0.70 & 0.69\\
Age of caseworker & 44.17 & 44.87 & 44.25 & 45.21\\
Caseworkers cooperation & 0.49 & 0.50 & 0.48 & 0.50\\
Caseworkers education: above vocational training & 0.45 & 0.46 & 0.43 & 0.44\\
Caseworkers education: tertiary level & 0.22 & 0.16 & 0.25 & 0.21\\
Caseworker female & 0.46 & 0.45 & 0.41 & 0.39\\
Caseworker missing & 0.04 & 0.05 & 0.04 & 0.05\\
Caseworker own unemployment experience & 0.63 & 0.65 & 0.63 & 0.64\\
Caseworker job tenure & 5.52 & 5.68 & 5.86 & 5.95\\
Caseworker education: vocational degree & 0.27 & 0.21 & 0.23 & 0.19\\
Fraction months employed last 2 years & 0.84 & 0.81 & 0.80 & 0.73\\
No employment spells last 5 years & 0.97 & 1.18 & 1.25 & 1.45\\
Employability & 2.12 & 1.00 & 2.14 & 1.00\\
Female & 0.45 & 0.43 & 0.44 & 0.45\\
Cantonal GDP per capita & 0.52 & 0.50 & 0.49 & 0.48\\
Married & 0.46 & 0.53 & 0.48 & 0.55\\
Mother tongue not Swiss language & 0.28 & 0.36 & 0.31 & 0.43\\
Past annual income & 46459 & 41129 & 42802 & 34523\\
Previous job: manager & 0.08 & 0.05 & 0.08 & 0.04\\
Previous job: primary sector & 0.06 & 0.07 & 0.10 & 0.09\\
Previous job: secondary sector & 0.16 & 0.15 & 0.13 & 0.13\\
Previous job: tertiary sector & 0.63 & 0.62 & 0.58 & 0.57\\
Previous job: missing sector & 0.15 & 0.16 & 0.19 & 0.21\\
Previous job: self-employed & 0.00 & 0.01 & 0.01 & 0.01\\
Previous job: skilled worker & 0.65 & 0.53 & 0.62 & 0.47\\
Previous job: unskilled worker & 0.24 & 0.39 & 0.27 & 0.46\\
Qualification: some degree & 0.62 & 0.47 & 0.59 & 0.39\\
Qualification: semiskilled & 0.14 & 0.18 & 0.15 & 0.20\\
Qualification: unskilled & 0.21 & 0.31 & 0.21 & 0.36\\
Qualification: skilled without degree & 0.03 & 0.04 & 0.05 & 0.05\\
Allocation to caseworker: by industry & 0.67 & 0.59 & 0.53 & 0.55\\
Allocation to caseworker: by occupation & 0.59 & 0.54 & 0.56 & 0.55\\
Allocation to caseworker: by age & 0.03 & 0.06 & 0.03 & 0.04\\
Allocation to caseworker: by employability & 0.06 & 0.11 & 0.06 & 0.09\\
Allocation to caseworker: by region & 0.09 & 0.11 & 0.12 & 0.12\\
Allocation to caseworker: other & 0.07 & 0.08 & 0.07 & 0.08\\
No of unemployment spells last 2 years & 0.37 & 0.72 & 0.58 & 0.99\\
Cantonal unemployment rate & 3.66 & 3.53 & 3.79 & 3.55\\
 \midrule
Number of observations & 11,396 & 1,602& 61,411 &  10,173\\ \bottomrule
\end{tabular}
\begin{tablenotes}[flushleft]
    \footnotesize
 \item \textit{Note:} This table shows the mean of some covariates included in the analysis. Column (1) and (2) show it for treated individuals, column (3) and (4) for non-treated individuals. Column (1) and (3) show it for hard to employ individuals, column (2) and (4) for easy to employ individuals.
\end{tablenotes}
\end{threeparttable}
\end{adjustbox}
\end{table}

\begin{table}[h!]
    \centering
    \caption{Empirical analysis: Balance table for treatment variable (participation in the program)}
    \label{emp_descriptives}
    \begin{adjustbox}{max width=0.9\textwidth}
    \begin{threeparttable}
\begin{tabular}{lrrrrr}
\toprule
& \multicolumn{2}{c}{\textbf{Treated}} & \multicolumn{2}{c}{\textbf{Control}} &  \\[1ex]
Covariates & Mean & Std. Dev. &  Mean & Std. Dev. & Std. Diff. \\ \midrule
Age & 37.25 & 8.78 & 36.59 & 8.65 & 7.64\\
French speaking canton & 0.08 & 0.27 & 0.25 & 0.44 & 47.96\\
German speaking canton & 0.89 & 0.31 & 0.67 & 0.47 & 57.08\\
Italian speaking canton & 0.03 & 0.16 & 0.08 & 0.27 & 23.99\\
Mother tongue in cantonal language & 0.12 & 0.33 & 0.11 & 0.32 & 3.37\\
Lives in big city & 0.17 & 0.37 & 0.17 & 0.37 & 0.15\\
Lives in medium city & 0.16 & 0.36 & 0.13 & 0.34 & 6.90\\
Lives in contryside & 0.68 & 0.47 & 0.70 & 0.46 & 5.12\\
Age of caseworker & 44.26 & 11.65 & 44.38 & 11.59 & 1.10\\
Caseworkers cooperation & 0.49 & 0.50 & 0.48 & 0.50 & 1.84\\
Caseworkers education: above vocational training & 0.45 & 0.50 & 0.44 & 0.50 & 3.25\\
Caseworkers education: tertiary level & 0.21 & 0.41 & 0.24 & 0.43 & 6.44\\
Caseworker female & 0.45 & 0.50 & 0.41 & 0.49 & 9.95\\
Caseworker missing & 0.04 & 0.20 & 0.04 & 0.20 & 0.28\\
Caseworker own unemployment experience & 0.63 & 0.48 & 0.63 & 0.48 & 0.72\\
Caseworker job tenure & 5.54 & 3.23 & 5.87 & 3.31 & 9.94\\
Caseworker education: vocational degree & 0.26 & 0.44 & 0.22 & 0.42 & 8.27\\
Fraction months employed last 2 years & 0.83 & 0.22 & 0.79 & 0.25 & 18.21\\
No employment spells last 5 years & 0.99 & 1.27 & 1.28 & 1.51 & 20.66\\
Employability & 1.98 & 0.48 & 1.98 & 0.51 & 0.70\\
Female & 0.45 & 0.50 & 0.44 & 0.50 & 1.27\\
Foreigner with permit B & 0.11 & 0.31 & 0.14 & 0.35 & 10.16\\
Foreigner with permit C & 0.23 & 0.42 & 0.25 & 0.43 & 4.63\\
Cantonal GDP per capita & 0.51 & 0.09 & 0.49 & 0.09 & 21.93\\
Married & 0.47 & 0.50 & 0.49 & 0.50 & 3.78\\
Mother tongue not Swiss language & 0.29 & 0.45 & 0.32 & 0.47 & 7.69\\
Past annual income & 45802 & 20194 & 41625 & 20566 & 20.49\\
Previous job: manager & 0.08 & 0.27 & 0.07 & 0.26 & 2.79\\
Previous job: primary sector & 0.06 & 0.23 & 0.10 & 0.30 & 14.85\\
Previous job: secondary sector & 0.16 & 0.37 & 0.13 & 0.34 & 8.24\\
Previous job: tertiary sector & 0.63 & 0.48 & 0.58 & 0.49 & 9.85\\
Previous job: missing sector & 0.15 & 0.36 & 0.19 & 0.39 & 9.94\\
Previous job: self-employed & 0.00 & 0.06 & 0.01 & 0.08 & 4.43\\
Previous job: skilled worker & 0.63 & 0.48 & 0.60 & 0.49 & 6.51\\
Previous job: unskilled worker & 0.26 & 0.44 & 0.29 & 0.46 & 7.08\\
Qualification: some degree & 0.60 & 0.49 & 0.56 & 0.50 & 7.43\\
Qualification: semiskilled & 0.15 & 0.35 & 0.16 & 0.37 & 3.41\\
Qualification: unskilled & 0.22 & 0.41 & 0.23 & 0.42 & 2.64\\
Qualification: skilled without degree & 0.03 & 0.18 & 0.05 & 0.21 & 6.78\\
Swiss & 0.66 & 0.47 & 0.61 & 0.49 & 11.06\\
Allocation to caseworker: by industry & 0.66 & 0.47 & 0.53 & 0.50 & 25.37\\
Allocation to caseworker: by occupation & 0.58 & 0.49 & 0.56 & 0.50 & 4.28\\
Allocation to caseworker: by age & 0.04 & 0.19 & 0.03 & 0.17 & 3.57\\
Allocation to caseworker: by employability & 0.07 & 0.25 & 0.07 & 0.25 & 0.00\\
Allocation to caseworker: by region & 0.09 & 0.28 & 0.12 & 0.33 & 10.35\\
Allocation to caseworker: other & 0.07 & 0.25 & 0.07 & 0.26 & 2.02\\
No of unemployment spells last 2 years & 0.41 & 1.00 & 0.64 & 1.28 & 19.36\\
Cantonal unemployment rate & 3.64 & 0.77 & 3.75 & 0.86 & 13.95\\
 \midrule
Number of observations & \multicolumn{2}{c}{12,998} & \multicolumn{2}{c}{71,584} \\ \bottomrule
\end{tabular}
\begin{tablenotes}[flushleft]
    \footnotesize
 \item \textit{Note:} This table shows the mean and standard deviation of the covariates included in the analyzes. Columns (1) and (2) show it for the treated group, columns (3) and (4) for the control group. The last column shows the standardized difference between the two groups. The standardized difference is calculated as $ SD=\frac{\left|\bar{X}_\text{treated}-\bar{X}_\text{control}\right|}{\sqrt{1 / 2\left(\operatorname{Var}\left(\bar{X}_\text{treated}\right)+\operatorname{Var}\left(\bar{X}_\text{control}\right)\right)}} \cdot 100$ where $\bar{X}_\text{treated}$ and $\bar{X}_\text{control}$ indicate the sample mean of the treatment and control group, respectively.
\end{tablenotes}
\end{threeparttable}
\end{adjustbox}
\end{table}

\begin{table}[h!]
    \centering
    \caption{Empirical analysis: Balance table for moderator variable (nationality)}
    \label{emp_descriptives_moderator}
    \begin{adjustbox}{max width=0.9\textwidth}
    \begin{threeparttable}
\begin{tabular}{lrrrrr}
\toprule
& \multicolumn{2}{c}{\textbf{Non Swiss}} & \multicolumn{2}{c}{\textbf{Swiss}} &  \\[1ex]
Covariates & Mean & Std. Dev. &  Mean & Std. Dev. & Std. Diff. \\ \midrule
Age & 36.31 & 8.23 & 36.92 & 8.94 & 7.09\\
French speaking canton & 0.26 & 0.44 & 0.21 & 0.41 & 11.88\\
German speaking canton & 0.65 & 0.48 & 0.73 & 0.44 & 17.66\\
Italian speaking canton & 0.09 & 0.29 & 0.06 & 0.24 & 11.81\\
Lives in big city & 0.19 & 0.39 & 0.15 & 0.36 & 10.31\\
Lives in medium city & 0.15 & 0.36 & 0.13 & 0.33 & 7.16\\
Lives in contryside & 0.66 & 0.47 & 0.72 & 0.45 & 13.76\\
Age of caseworker & 44.20 & 11.68 & 44.47 & 11.55 & 2.36\\
Caseworkers cooperation & 0.50 & 0.50 & 0.47 & 0.50 & 6.45\\
Caseworkers education: above vocational training & 0.43 & 0.50 & 0.44 & 0.50 & 2.58\\
Caseworkers education: tertiary level & 0.24 & 0.43 & 0.23 & 0.42 & 2.06\\
Caseworker female & 0.39 & 0.49 & 0.43 & 0.49 & 7.51\\
Caseworker missing & 0.04 & 0.20 & 0.04 & 0.19 & 2.17\\
Caseworker own unemployment experience & 0.63 & 0.48 & 0.63 & 0.48 & 0.29\\
Caseworker job tenure & 5.85 & 3.37 & 5.80 & 3.26 & 1.56\\
Caseworker education: vocational degree & 0.22 & 0.42 & 0.23 & 0.42 & 2.94\\
Fraction months employed last 2 years & 0.77 & 0.26 & 0.81 & 0.24 & 15.43\\
No employment spells last 5 years & 1.35 & 1.59 & 1.17 & 1.40 & 11.69\\
Employability & 1.94 & 0.51 & 2.01 & 0.50 & 13.22\\
Female & 0.41 & 0.49 & 0.46 & 0.50 & 8.96\\
Cantonal GDP per capita & 0.49 & 0.09 & 0.50 & 0.09 & 2.51\\
Married & 0.70 & 0.46 & 0.35 & 0.48 & 73.36\\
Mother tongue not Swiss language & 0.66 & 0.47 & 0.10 & 0.30 & 142.02\\
Past annual income & 38552 & 18648 & 44596 & 21353 & 30.15\\
Previous job: manager & 0.05 & 0.21 & 0.09 & 0.29 & 18.27\\
Previous job: primary sector & 0.12 & 0.32 & 0.07 & 0.26 & 15.81\\
Previous job: secondary sector & 0.14 & 0.35 & 0.13 & 0.34 & 3.01\\
Previous job: tertiary sector & 0.49 & 0.50 & 0.65 & 0.48 & 33.03\\
Previous job: missing sector & 0.25 & 0.43 & 0.15 & 0.35 & 26.47\\
Previous job: self-employed & 0.01 & 0.07 & 0.01 & 0.08 & 2.68\\
Previous job: skilled worker & 0.45 & 0.50 & 0.70 & 0.46 & 51.65\\
Previous job: unskilled worker & 0.48 & 0.50 & 0.17 & 0.37 & 71.63\\
Qualification: some degree & 0.32 & 0.47 & 0.72 & 0.45 & 89.15\\
Qualification: semiskilled & 0.20 & 0.40 & 0.13 & 0.33 & 20.33\\
Qualification: unskilled & 0.40 & 0.49 & 0.12 & 0.33 & 67.12\\
Qualification: skilled without degree & 0.08 & 0.27 & 0.03 & 0.16 & 23.62\\
Allocation to caseworker: by industry & 0.54 & 0.50 & 0.56 & 0.50 & 3.85\\
Allocation to caseworker: by occupation & 0.56 & 0.50 & 0.57 & 0.50 & 1.75\\
Allocation to caseworker: by age & 0.03 & 0.17 & 0.03 & 0.18 & 2.38\\
Allocation to caseworker: by employability & 0.06 & 0.24 & 0.07 & 0.26 & 4.72\\
Allocation to caseworker: by region & 0.12 & 0.32 & 0.11 & 0.32 & 0.94\\
Allocation to caseworker: other & 0.07 & 0.26 & 0.08 & 0.27 & 2.31\\
No of unemployment spells last 2 years & 0.77 & 1.40 & 0.50 & 1.12 & 21.55\\
Cantonal unemployment rate & 3.82 & 0.83 & 3.69 & 0.85 & 15.12\\
 \midrule
Number of observations & \multicolumn{2}{c}{32,592} & \multicolumn{2}{c}{51,990} \\ \bottomrule
\end{tabular}
\begin{tablenotes}[flushleft]
    \footnotesize
 \item \textit{Note:} This table shows the mean and standard deviation of some covariates included in the analysis. Column (1) and (2) show it for Non Swiss individuals, column (3) and (4) for Swiss individuals. The last column shows the standardized difference between the two groups. The standardized difference is calculated as $ SD=\frac{\left|\bar{X}_\text{Swiss}-\bar{X}_\text{non-Swiss}\right|}{\sqrt{1 / 2\left(\operatorname{Var}\left(\bar{X}_\text{Swiss}\right)+\operatorname{Var}\left(\bar{X}_\text{non-Swiss}\right)\right)}} \cdot 100$ where $\bar{X}_\text{Swiss}$ and $\bar{X}_\text{non-Swiss}$ indicate the sample mean of the Non-Swiss and the Swiss individuals, respectively.
\end{tablenotes}
\end{threeparttable}
\end{adjustbox}
\end{table}
\newpage

\begin{table}[h!]
    \centering
    \caption{Empirical analysis: Balance table for moderator variable (employability)}
    \label{emp_descriptives_moderator_employability}
    \begin{adjustbox}{max width=0.9\textwidth}
    \begin{threeparttable}
\begin{tabular}{lrrrrr}
\toprule
& \multicolumn{2}{c}{\textbf{Easy to employ}} & \multicolumn{2}{c}{\textbf{Hard to employ}} &  \\[1ex]
Covariates & Mean & Std. Dev. &  Mean & Std. Dev. & Std. Diff. \\ \midrule
Age & 36.38 & 8.58 & 38.58 & 9.02 & 24.94\\
French speaking canton & 0.24 & 0.43 & 0.14 & 0.35 & 24.66\\
German speaking canton & 0.69 & 0.46 & 0.79 & 0.41 & 23.44\\
Italian speaking canton & 0.07 & 0.26 & 0.07 & 0.25 & 2.32\\
Lives in big city & 0.17 & 0.37 & 0.15 & 0.36 & 4.99\\
Lives in medium city & 0.13 & 0.34 & 0.16 & 0.37 & 9.29\\
Lives in contryside & 0.70 & 0.46 & 0.69 & 0.46 & 3.18\\
Age of caseworker & 44.24 & 11.56 & 45.16 & 11.83 & 7.90\\
Caseworkers cooperation & 0.48 & 0.50 & 0.50 & 0.50 & 3.72\\
Caseworkers education: above vocational training & 0.44 & 0.50 & 0.44 & 0.50 & 0.76\\
Caseworkers education: tertiary level & 0.24 & 0.43 & 0.21 & 0.40 & 8.30\\
Caseworker female & 0.41 & 0.49 & 0.40 & 0.49 & 2.70\\
Caseworker missing & 0.04 & 0.19 & 0.05 & 0.22 & 6.20\\
Caseworker own unemployment experience & 0.63 & 0.48 & 0.64 & 0.48 & 2.56\\
Caseworker job tenure & 5.80 & 3.32 & 5.92 & 3.20 & 3.43\\
Caseworker education: vocational degree & 0.24 & 0.42 & 0.19 & 0.39 & 10.38\\
Fraction months employed last 2 years & 0.81 & 0.24 & 0.74 & 0.28 & 23.92\\
No employment spells last 5 years & 1.21 & 1.46 & 1.41 & 1.58 & 13.12\\
Female & 0.44 & 0.50 & 0.44 & 0.50 & 1.01\\
Cantonal GDP per capita & 0.50 & 0.09 & 0.48 & 0.09 & 16.13\\
Married & 0.48 & 0.50 & 0.55 & 0.50 & 14.66\\
Mother tongue not Swiss language & 0.30 & 0.46 & 0.42 & 0.49 & 24.55\\
Past annual income & 43374 & 20502 & 35422 & 19607 & 39.64\\
Previous job: manager & 0.08 & 0.27 & 0.04 & 0.20 & 15.56\\
Previous job: primary sector & 0.09 & 0.29 & 0.09 & 0.28 & 1.00\\
Previous job: secondary sector & 0.14 & 0.34 & 0.13 & 0.34 & 1.12\\
Previous job: tertiary sector & 0.59 & 0.49 & 0.57 & 0.49 & 3.23\\
Previous job: missing sector & 0.18 & 0.39 & 0.20 & 0.40 & 5.73\\
Previous job: self-employed & 0.01 & 0.08 & 0.01 & 0.09 & 2.55\\
Previous job: skilled worker & 0.63 & 0.48 & 0.48 & 0.50 & 29.62\\
Previous job: unskilled worker & 0.26 & 0.44 & 0.45 & 0.50 & 40.41\\
Qualification: some degree & 0.59 & 0.49 & 0.40 & 0.49 & 38.51\\
Qualification: semiskilled & 0.15 & 0.36 & 0.19 & 0.40 & 11.30\\
Qualification: unskilled & 0.21 & 0.41 & 0.35 & 0.48 & 32.38\\
Qualification: skilled without degree & 0.04 & 0.21 & 0.05 & 0.21 & 1.20\\
Foreigner & 0.37 & 0.48 & 0.45 & 0.50 & 15.65\\
Allocation to caseworker: by industry & 0.55 & 0.50 & 0.55 & 0.50 & 0.41\\
Allocation to caseworker: by occupation & 0.57 & 0.50 & 0.55 & 0.50 & 3.52\\
Allocation to caseworker: by age & 0.03 & 0.17 & 0.04 & 0.20 & 7.44\\
Allocation to caseworker: by employability & 0.06 & 0.24 & 0.10 & 0.30 & 12.99\\
Allocation to caseworker: by region & 0.12 & 0.32 & 0.12 & 0.32 & 0.77\\
Allocation to caseworker: other & 0.07 & 0.26 & 0.08 & 0.27 & 2.09\\
No of unemployment spells last 2 years & 0.54 & 1.16 & 0.95 & 1.61 & 28.84\\
Cantonal unemployment rate & 3.77 & 0.84 & 3.55 & 0.84 & 25.80\\
 \midrule
Number of observations & \multicolumn{2}{c}{72,807} & \multicolumn{2}{c}{11,775} \\ \bottomrule
\end{tabular}
\begin{tablenotes}[flushleft]
    \footnotesize
 \item \textit{Note:} This table shows the mean and standard deviation of some covariates included in the analysis. Column (1) and (2) show it for easy to employ individuals, column (3) and (4) for hard to employ individuals. The last column shows the standardized difference between the two groups. The standardized difference is calculated as $ SD=\frac{\left|\bar{X}_\text{high}-\bar{X}_\text{low}\right|}{\sqrt{1 / 2\left(\operatorname{Var}\left(\bar{X}_\text{high}\right)+\operatorname{Var}\left(\bar{X}_\text{low}\right)\right)}} \cdot 100$ where $\bar{X}_\text{high}$ and $\bar{X}_\text{low}$ indicate the sample mean of the hard and the easy to employ individuals, respectively.
\end{tablenotes}
\end{threeparttable}
\end{adjustbox}
\end{table}
\newpage

\end{appendices}
\end{document}